\documentclass[a4paper,12pt]{article}
\usepackage{changepage} 

\usepackage{xcolor}
\AtBeginEnvironment{equation}{\par\nobreak\noindent}
\usepackage[utf8]{inputenc}
\usepackage{makecell}
\usepackage{natbib}
\usepackage{utfsym}
\usepackage{graphicx}
\usepackage{amsmath}
\usepackage{titlesec}
\usepackage{color}
\usepackage{algorithm}
\usepackage{algpseudocode}
\usepackage{mathrsfs}
\usepackage{appendix}
\usepackage{soul}
\usepackage{pdflscape}
\usepackage{subcaption}
\usepackage{adjustbox}
\usepackage{subcaption}
\usepackage{booktabs}
\usepackage{lipsum}
\usepackage{float}
\usepackage[font=small,labelfont=bf]{caption}
\usepackage[shortlabels]{enumitem}
\usepackage[top=2cm,bottom=2cm,left=3cm,right=3cm,marginparwidth=1.75cm]{geometry}

\newcommand{\Ind}{\mathbb{I}}          
\usepackage{amsmath}
\usepackage{graphicx}
\usepackage{comment}
\usepackage[colorlinks=true, allcolors=blue]{hyperref}
\usepackage{amsmath, amsthm, amssymb,nccmath}
\usepackage[status=draft,inline,index]{fixme}
\fxsetup{theme=color,margin=false,mode=multiuser}
\FXRegisterAuthor{ra}{rae}{RA}
\FXRegisterAuthor{dn}{dne}{\color{magenta}DN}
\FXRegisterAuthor{uo}{uoe}{\color{cyan}UO}
\FXRegisterAuthor{ms}{mse}{\color{orange}MS}
\newtheorem{Proposition}{Proposition}
\newtheorem{Assumption}{Assumption}
\newtheorem*{Assumption*}{Assumption}

\newtheorem*{remark*}{Remark}
\newtheorem{lemma}{Lemma}
\usepackage{caption}
\usepackage{subcaption}
\usepackage{calc}

\makeatletter
\def\textSq#1{%
\begingroup
\setlength{\fboxsep}{0.3ex}
\setbox1=\hbox{#1}
\setlength{\@tempdima}{\maxof{\wd1}{\ht1+\dp1}}
\setlength{\@tempdimb}{(\@tempdima-\ht1+\dp1)/2}
\raise-\@tempdimb\hbox{\fbox{\vbox to \@tempdima{%
  \vfil\hbox to \@tempdima{\hfil\copy1\hfil}\vfil}}}%
\endgroup%
}
\def\Sq#1{\textSq{\ensuremath{#1}}}%
\makeatother
\usepackage{tikz}
\usetikzlibrary{arrows.meta,shapes}
\usetikzlibrary{arrows,shapes.arrows,shapes.geometric,shapes.multipart,decorations.pathmorphing,positioning,shapes.swigs}
\usetikzlibrary{shapes,decorations,arrows,calc,arrows.meta,fit,positioning}
\tikzset{
    -Latex,auto,node distance =1 cm and 1 cm,semithick,
    state/.style ={ellipse, draw, minimum width = 0.7 cm},
    point/.style = {circle, draw, inner sep=0.04cm,fill,node contents={}},
    bidirected/.style={Latex-Latex,dashed},
    el/.style = {inner sep=2pt, align=left, sloped}
}
\usepackage{float}
\def\Sq#1{\textSq{\ensuremath{#1}}}%

\title{Nonparametric bounds for vaccine effects in randomized trials}

\author{
Rachel Axelrod$^{1}$\thanks{Corresponding author: rachel.xlrod@gmail.com},
Uri Obolski$^{2}$,
Daniel Nevo$^{1}$
}
\date{
\small
${}^1$Department of Statistics and Operations Research, Faculty of Exact Sciences, Tel Aviv University, Tel Aviv, Israel\\
${}^2$Department of Epidemiology and Preventive Medicine, School of Public Health, Gray Faculty of Medical and Health Sciences, Tel Aviv University, Tel Aviv, Israel
}
\begin{document}
\maketitle
\begin{abstract}
Randomized controlled vaccine trials are typically designed to be blinded, ensuring that the estimated vaccine efficacy (VE) reflects the immunological effect of vaccination. When blinding is broken, however, the estimated VE may reflect not only the immunological effect but also behavioral effects stemming from participants' awareness of their treatment status. Recent work has proposed alternative causal estimands to the standard VE to address this issue, but their point identification results require a strong assumption: the absence of unmeasured common causes of infection risk and participants' belief about whether they received the vaccine. Personality traits, for example, may plausibly violate this assumption. We relax this assumption and derive nonparametric causal bounds for different types of VE. We construct these bounds using two approaches: linear programming-based and monotonicity-based methods. We further consider several possible causal structures for vaccine trials and show how the nonparametric bounds differ across these scenarios. Finally, we illustrate the performance of the proposed bounds using fully synthetic data and a semi-synthetic data example based on a COVID-19 vaccine trial.
\end{abstract}

\section{Introduction}
\label{sec:introduction}

In vaccine trials, \textit{broken blinding}, i.e., when participants become aware of their treatment assignment, has recently been recognized as a factor that may cause the immunological effect, typically the target of the trial, to be non-identifiable \citep{gabriel2025elucidating}. Moreover, the estimated vaccine efficacy (VE) in such trials has been recently shown to be a biased estimator of the true immunological effect \citep{stensrud2024distinguishing, obolski2024call}. In the presence of broken blinding,  the estimated VE reflects not only the immunological effect but also behavioral changes induced by participants’ awareness of their treatment status. For example, after COVID-19 vaccination, individuals who felt more protected were found to increase social contacts, reduce mask use, and attend larger gatherings, potentially elevating their exposure to SARS-CoV-2 \citep{trogen2021risk,buckell2023covid}.

Since examples of broken blinding in vaccine trials have been documented \citep{lazarus2021safety}, it is crucial to discuss how the VE can be estimated. As reported in \cite{lazarus2021safety}, a possible reason for the blinding process to fail is that the active vaccine causes more local reaction than placebo, which could unmask the active vaccine versus placebo allocation.

Alternative causal estimands have recently been proposed to study behavioral and immunological effects arising from broken blinding in vaccine trials. One such estimand corresponds to the VE that would be observed had participants known whether they were vaccinated or not \citep{stensrud2024distinguishing,obolski2024call}. Importantly, such an estimand is not only relevant for broken-blinding scenarios but also reflects vaccine effectiveness in real-world settings, where people are aware of their vaccination status and may adjust their behavior accordingly. The point-identification strategy of the suggested estimands builds on the idea of measuring trial participants \textit{belief} about their vaccination status.

However, these point-identification results require an arguably strong assumption: that there are no unmeasured common causes of infection risk and of belief about the treatment received. A key example of such a common cause in vaccine studies is personality traits, which may violate this assumption. For instance, optimism may lead participants to believe they received the vaccine even when they did not (``wishful thinking'', \citealp{bang2016random}), and the same optimistic disposition may also make them more outgoing, thereby increasing their chances of exposure to infectious agents.

We develop in this paper nonparametric causal bounds for different types of VE, that avoid making the assumption of no unmeasured common causes. Nonparametric bounds have been proposed in a variety of contexts: \citet{robins1989analysis} and \citet{manski1990nonparametric} independently derived the first informative nonparametric bounds on causal effects in studies with imperfect compliance, where the standard randomization assumption cannot be invoked. Later, \citet{horowitz2000nonparametric} derived nonparametric bounds for randomized experiments with missing data, relying only on observed confounders and making no assumptions about the missingness mechanism. Other contexts in which nonparametric bounds have been used include, for example, mediation analysis \citep{robins2010alternative,vanderweele2010bias,gabriel2023sharp} and outcome-dependent sampling with unmeasured confounding \citep{gabriel2022causal}.

We construct nonparametric bounds using two different approaches: linear programming (LP)-based and monotonicity-based methods. The LP-based approach originates in the seminal paper of \citet{balke1994counterfactual}, whose framework enabled bounding the average treatment effect in studies with imperfect compliance. They recognized that deriving such bounds is equivalent to solving an LP optimization problem, with the target causal parameter and the constraints implied by observed probabilities expressed as linear combinations of the same underlying parameters. In the binary treatment and binary outcome setting, Balke and Pearl further derived closed-form symbolic solutions, providing analytical expressions without the need for iterative optimization. Their results were later extended to the more general instrumental variables case (see \citealp{swanson2018partial}, for a review). The original Balke--Pearl bounds can be further tightened by imposing monotonicity assumptions \citep{balke1994counterfactual,kuroki2008formulating}, and more recent work has generalized the framework to accommodate multiple scenarios \citep{gabriel2023nonparametric,sachs2023general}.

While LP-based bounds are sharp (i.e., the tightest possible bounds under the stated assumptions), they can be wide in practice, especially when the data provide limited information about the joint distribution of unmeasured variables. As an alternative, we consider a second approach based on monotonicity assumptions, namely that both the belief variable and the outcome are monotonically related to the unmeasured variable. To this end, we build on ideas similar to those of \citet{vanderweele2008sign}, who established sign determination results for causal effects in the presence of unmeasured confounding in observational studies.

We consider several possible causal structures for a vaccine trial. Each structure consists of particular assumptions about the relationships between key trial variables, which in turn yield different nonparametric bounds for the different VEs. We begin with a baseline setting that assumes no unmeasured common causes between belief and risk of infection, thereby permitting point identification \citep{stensrud2024distinguishing}. Next, we relax this assumption by allowing an unmeasured common cause of belief and risk of infection. We further relax the assumptions by permitting the unmeasured common cause to also directly affect adverse events (AEs). For each structure, we first provide real-world examples, supported by existing studies and then present the corresponding LP-based and monotonicity-based bounds for the different VEs.

As in illustrative example, we consider the ENSEMBLE2 trial, a randomized, double-blind, placebo-controlled study evaluating a COVID-19 vaccine in adults ($\geq18$ years) across multiple international sites \citep{hardt2022efficacy}. In a prespecified safety subset, solicited local and systemic AEs occurred in 57.3\% of vaccine recipients versus 22.5\% of placebo recipients, raising concerns that blinding may have been compromised. Consequently, as previously noted \citep{stensrud2024distinguishing,obolski2024call}, the estimated VE in this trial may not reflect the causal immunological effects of the vaccine. At the same time, it is not plausible to assume the absence of common causes influencing both belief and risk of infection (e.g., personality traits). Thus, the point identification approach of \citet{stensrud2024distinguishing} is not feasible here, motivating instead a partial identification strategy based on nonparametric bounds for VEs.

The remainder of the paper is structured as follows. In Section \ref{sec:review}, we define the notation, describe the causal estimands of interest, and review the assumptions under which these estimands can be identified from the observed data. In Section \ref{sec:motivation}, we briefly present the scenarios in which the identification assumptions are not satisfied, motivating the derivation of nonparametric bounds for the VEs. In Sections \ref{sec:fig2}--\ref{sec:viol.Y.S.dis}, we develop LP-based and monotonicity-based bounds for the VEs under these scenarios. 
Section \ref{sec:sim} provides a numerical example illustrating the performance of the proposed bounds. Finally, in Section \ref{sec:real_data_example}, we return to the illustrative example of the ENSEMBLE2 trial and apply our methods to estimate both LP-based and monotonicity-based bounds.

\section{Review of the existing framework}
\label{sec:review}

\subsection{Notation}

We begin by reviewing the framework of \citet{stensrud2024distinguishing}. Let $A$ denote the randomized vaccine assignment, with $A=1$ for active vaccine and $A=0$ for placebo. In this framework, a key component is the introduction of the \textit{message} $M$. Let $M \in \{-1,0,1\}$ indicate that vaccine status is blinded $(M=-1)$, that the individual received the message they are vaccinated $(M=1)$, or that they received the message they are unvaccinated $(M=0)$. Unlike a real-world setting, where $A = M$, blinding in a randomized trial implies that $M$ is fixed to $-1$.  

A second key component is the \textit{belief} $B \in \{0,1\}$, which indicates whether the individual believes they were vaccinated $(B=1)$ or not $(B=0)$. Additionally, let $S \in \{0,1\}$ be an indicator of an AE, and let $E$ represent the level of exposure to the infectious agent. The variable $E$ is introduced only to illustrate causal scenarios; we do not assume $E$ is measured, and it will not be used in the bound calculations. While the causal directed acyclic graphs (DAGs) \citep{pearl2009causality} presented later on clarify the ordering of these variables in the scenarios we analyze, we note here that $A$ and $M$ are determined first, followed by $S$, then $B$, and finally the outcome $Y$.  

We use superscripts to denote potential outcomes. Let $Y^{a,m}$ denote the potential disease status had a participant been assigned treatment $A=a$ and received message $M=m$. Similarly, $S^{a,m}$ and $B^{a,m}$ denote the potential AE and belief under treatment $A=a$ and message $M=m$. We assume consistency with respect to all potential outcomes and their observed counterparts; for example, if $A=a$ and $M=m$ then $Y=Y^{a,m}$.  

For the observed data distribution, we use the following notation:
\begin{align*}
 p_{y.a}   &= \Pr(Y=y \mid A=a),\\
 p_{y.ab}  &= \Pr(Y=y \mid A=a,B=b),\\
 p_{y.abs} &= \Pr(Y=y \mid A=a,B=b,S=s),\\
 p_{yb.a}  &= \Pr(Y=y,B=b \mid A=a),\\
 p_{yb.as} &= \Pr(Y=y,B=b \mid A=a,S=s).
\end{align*}
Additionally, let $\pi_{b.a} = \Pr(B=b \mid A=a)$, $\pi_{b.as} = \Pr(B=b \mid A=a,S=s)$, and $\gamma_{s.a} = \Pr(S=s \mid A=a)$. Note that $p_{yb.a} = p_{y.ab}\pi_{b.a}$ and $p_{yb.as} = p_{y.abs}\pi_{b.as}$.

\subsection{Causal estimands}
\label{SubSec:estimands}

In a blinded two-arm vaccine trial, henceforth denoted by $\mathcal{T}_{II}$, the standard VE under blinding ($m=-1$) is
\begin{equation}
\label{eq:blind_effect}
VE(-1) = 1 - \frac{E(Y^{a=1,m=-1})}{E(Y^{a=0,m=-1})}.
\end{equation}
This quantity is identifiable by replacing $E(Y^{a=1,m=-1})$ and $E(Y^{a=0,m=-1})$ with the corresponding sample infection rates in the active vaccine and placebo groups, $E(Y \mid A=1,M=-1)$ and $E(Y \mid A=0,M=-1)$, respectively. However, when blinding is broken and participants’ behavior depends on their perceived treatment status, the observed data instead involve quantities such as $E(Y \mid A=1,B=1,M=-1)$ and $E(Y \mid A=0,B=0,M=-1)$. Without additional assumptions, these sample quantities no longer identify the causal quantities $E(Y^{a=1,m=-1})$ and $E(Y^{a=0,m=-1})$. Moreover, even under perfect blinding, one may argue that the effect in \eqref{eq:blind_effect} is not the target estimand for real-world vaccine deployment, where individuals know their treatment status ($A=M$ with probability one). These limitations motivate defining new VE measures that account not only for treatment assignment ($A$) but also for the message received ($M$). Such estimands allow, for example, estimation of the VE that would have been observed had all participants known whether they were vaccinated or not.

To formalize this idea, consider a hypothetical four-arm trial, denoted by $\mathcal{T}_{IV}$. In this trial, each participant is assigned both to a blinded treatment ($A$) and to a message ($M$) indicating, possibly contrary to fact, that they received either placebo ($M=0$) or active vaccine ($M=1$). Thus, in $\mathcal{T}_{IV}$ it is possible that $A \ne M$. In addition, the belief $B$ is measured. By design, such a trial permits identification of
\begin{equation}\label{eq:gen.con}
    VE(m)=1-\frac{E(Y^{a=1,m})}{E(Y^{a=0,m})},
\end{equation}
and $VE_M(a)=1-\frac{E ( Y^{a,m=1})}{E ( Y^{a,m=0})}$ for $a,m\in\{0,1\}^2$. These estimands correspond to the immunological effect of the vaccine under the message $m$ and the behavioral effect of the message at treatment level $a$, respectively. In practice, such four-arm trials have rarely been conducted, likely because it is ethically problematic to mislead participants about their vaccination status in ways that might influence their behavior and increase risk. For this reason, \citet{stensrud2024distinguishing} presented conditions under which data from a standard two-arm trial, augmented with measurements of $B$, can be used to identify and interpret causal contrasts as $VE(m)$ and $VE_M(a)$. We review the details in Section \ref{SubSec:PointIdent}.

In addition, such a four-arm trial would enable measurement of the following \textit{total effect},
\begin{equation}\label{eq:tot_effect}
   VE_T=1-\frac{E(Y^{a=1,m=1})}{E(Y^{a=0,m=0})}. 
\end{equation}
This effect is directly relevant in real-world settings, where individuals know whether they have been vaccinated and may change their behavior accordingly. Policymakers may therefore be interested in comparing the infection risks that would be observed if all individuals were vaccinated versus unvaccinated, both under awareness of their vaccination status.

In addition to the four-arm trial, we will also consider a six-arm trial, denoted by $\mathcal{T}_{VI}$, which combines the arms of the two- and four-arm trials introduced so far. Lastly, throughout the paper we use subscripts to indicate probabilities, expectations, and (conditional) independence statements under a particular vaccine trial. For example, $E_{IV}(\cdot)$ is the expected value under the observed data distribution in the four-arm trial $\mathcal{T}_{IV}$. Probabilities, expectations, and  independence statements without subscripts are with respect to the observed data from the two-arm trial, $\mathcal{T}_{II}$. In addition, we assume that $E(Y^{a,m})=E_{\mathcal{T}_{IV}}(Y^{a,m})=E_{\mathcal{T}_{VI}}(Y^{a,m})$, meaning that expectations of potential outcomes are the same under all trials. This assumption formalizes that all trials, hypothetical or observed, sample participants from the same underlying population. 

Throughout this paper, we present the results for $VE(m),\; m\in\{-1,0,1\}$ and $VE_T$. In Appendix~\ref{app:diff_scale} we present theoretical results for other causal effects, including $VE_M(a)$, the behavioral effect on the VE scale, as well as causal contrasts on the difference scale. 

\subsection{Point identification}
\label{SubSec:PointIdent}
\cite{stensrud2024distinguishing} presented assumptions under which $VE(m), \; m\in\{-1,0,1\}$, and $VE_T$ are identifiable from a two-arm trial $\mathcal{T}_{II}$ that also includes the belief $B$. 

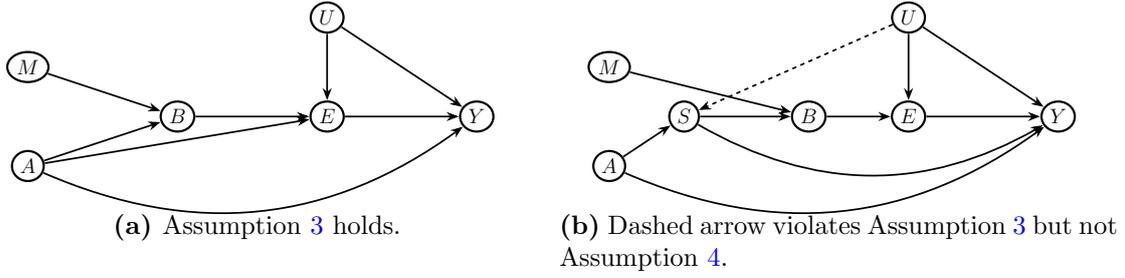
\begin{figure}[t]
\centering

\begin{subfigure}[t]{0.49\textwidth}
\centering
\resizebox{\linewidth}{!}{%
\begin{tikzpicture}[line width=1.5pt, baseline=(current bounding box.north), scale=1.2, transform shape]
  \path[use as bounding box] (-7.0,-1.6) rectangle (4.2,2.6);

  \tikzset{ell/.style={draw,fill=white,inner sep=2pt,line width=1.5pt}}
  \node (A) [ell,ellipse] at (-6,-1) {$A$};
  \node (M) [ell,ellipse] at (-6, 1) {$M$};
  \node (B) [ell,ellipse] at (-3, 0) {$B$};
  \node (E) [ell,ellipse] at ( 0, 0) {$E$};
  \node (U) [ell,ellipse] at ( 0, 2) {$U$};
  \node (Y) [ell,ellipse] at ( 3, 0) {$Y$};

  \begin{scope}[>={Stealth[black]}, every edge/.style={draw=black,very thick}]
    \path[->] (A) edge (E) (A) edge (B) (M) edge (B) (B) edge (E)
              (U) edge (E) (U) edge (Y) (E) edge (Y)
              (A) edge[bend left=-30] (Y);
  \end{scope}
\end{tikzpicture}}
\caption{Assumption \ref{assump:y.dismissible} holds.}
\label{fig:DAG_full_identification1}
\end{subfigure}
\hfill
\begin{subfigure}[t]{0.49\textwidth}
\centering
\resizebox{\linewidth}{!}{%
\begin{tikzpicture}[line width=1.5pt, baseline=(current bounding box.north), scale=1.2, transform shape]
  \path[use as bounding box] (-7.0,-1.6) rectangle (4.2,2.6);

  \tikzset{ell/.style={draw,fill=white,inner sep=2pt,line width=1.5pt}}
 \node (A) [ell,ellipse] at (-6,-1) {$A$};
\node (S) [ell,ellipse] at (-4.5, 0) {$S$};  
\node (B) [ell,ellipse] at (-2, 0) {$B$};
\node (E) [ell,ellipse] at ( 0, 0) {$E$};
\node (U) [ell,ellipse] at ( 0, 2) {$U$};
\node (Y) [ell,ellipse] at ( 3, 0) {$Y$};
\node (M) [ell,ellipse] at (-6, 1) {$M$};

  \begin{scope}[>={Stealth[black]}, every edge/.style={draw=black,very thick}]
    \path[->] (A) edge (S) (S) edge (B) (M) edge (B) (B) edge (E)
              (U) edge (E) (U) edge[dashed] (S) (U) edge (Y) (E) edge (Y)
              (S) edge[bend left=-30] (Y) (A) edge[bend left=-30] (Y);
  \end{scope}
\end{tikzpicture}}
\caption{Dashed arrow violates Assumption \ref{assump:y.dismissible} but not Assumption \ref{assump:y.s.dis}.}
\label{fig:DAG_full_identification2}
\end{subfigure}

\caption{DAGs describing six-arm trial $\mathcal{T}_{VI}$. In these trials, the VEs are identifiable.}
\label{fig:DAG_full_identification}
\end{figure}

The first assumption implies that when there is no blinding, the belief $B$ is identical to the message $M \in \{0, 1\}$.
\begin{Assumption}\label{assump:B_M}
     In the four-arm trial $\mathcal{T}_{IV}$, $B = M$ with probability 1.
\end{Assumption}
\noindent The second assumption is the following positivity assumption.
 \begin{Assumption}(Positivity)\label{assump:pos1}
    In $\mathcal{T}_{II}$ , $\pi_{b.a}>0.$ for all $a,b\in \{0, 1\}^2$.
\end{Assumption}
\noindent These assumptions are plausible in most studies; for further discussion see \cite{stensrud2024distinguishing}.
The third assumption is inspired by the literature on separable effects \citep{robins2010alternative, stensrud2021generalized,stensrud2023conditional}. 
\begin{Assumption}
\label{assump:y.dismissible}(Y Dismissible component condition).
$Y \perp_{VI} M| A,B$.
\end{Assumption}
\noindent To get more intuition on this assumption, consider the DAG given in Figure \ref{fig:DAG_full_identification1}, which represents a six-arm trial obeying Assumption \ref{assump:y.dismissible}. In the DAG, it can be seen that the causal path from $M$ to $Y$ is blocked by $B$, and the non-causal path $M \rightarrow B \leftarrow A \rightarrow Y$, which is opened by conditioning on $B$, is blocked by $A$.

Under Assumptions \ref{assump:B_M}-\ref{assump:y.dismissible}, Proposition 1 in \cite{stensrud2024distinguishing} asserts that $E(Y^{a,m}), \;a,m\in\{0,1\}$, is identified from the two-arm trial $\mathcal{T}_{II}$ by $E(Y^{a,m})=p_{1.am}$, and therefore, $VE(m),\; m \in \{-1,0,1\}$, and $VE_T$ are identified by
\begin{align*}
VE(m)&=1-\frac{p_{1.1m}}{p_{1.0m}},\\
VE_T&=1-\frac{p_{1.11}}{p_{1.00}}.
\end{align*}
 In some cases, Assumption \ref{assump:y.dismissible} may fail but the VEs can still be point identified with additional data and alternative assumptions. Consider Figure \ref{fig:DAG_full_identification2}, which presents a trial with possible broken blinding. In Figure \ref{fig:DAG_full_identification2}, the occurrence of AEs plays a crucial role in the violation of Assumption \ref{assump:y.dismissible}. This structure is motivated by our illustrative example, the ENSEMBLE2 trial \citep{hardt2022efficacy}, in which AEs are more likely among the treated and experiencing an AE increases the likelihood of believing one received the treatment (the path $A \rightarrow S \rightarrow B$). Moreover, the likelihood of AEs might be affected by an unmeasured variable $U$ that also affects the outcome $Y$ as illustrated by the path $S\leftarrow U \rightarrow Y$. This variable $U$ can represent, for example, general physiological frailty (e.g., immunosuppression) which may increase the risk of AEs and of infection. The path $M \rightarrow B \leftarrow S\leftarrow U \rightarrow Y$ violates Assumption \ref{assump:y.dismissible}.

Point identification in the scenario, is possible under the following less restrictive dismissible component conditions.
\begin{Assumption} \label{assump:y.s.dis} ($Y$, $S$ Dismissible component conditions)
\begin{align}
\begin{split}
&Y \perp_{VI} M| A,S,B\\
&S \perp_{VI} M| A
\end{split}
\end{align}
\end{Assumption}
\noindent In addition,  Assumption \ref{assump:pos1} is replaced with the following assumption.
\begin{Assumption}\label{assump:pos2} (Positivity 2). In $\mathcal{T}_{II}$
\begin{align}
\begin{split}
&\gamma_{s.a} > 0, \text{for all a, s }\in \{0, 1\}^2\\
&\pi_{b.as}> 0, \text{for all a, b, s} \in \{0, 1\}^3.
\end{split}
\end{align}   
\end{Assumption}
\noindent It can be shown \citep[][ Proposition 2]{stensrud2024distinguishing} that under Assumptions \ref{assump:B_M}, \ref{assump:y.s.dis}-\ref{assump:pos2}, the expectations $E(Y^{a,m})$ for $a,m\in\{0,1\}^2$ are identified from the two-arm trial $\mathcal{T}_{II}$ even under broken blinding by $E(Y^{a,m})=\sum_sp_{1.ams}\gamma_{s.a}$. Therefore, $VE(m), \; m \in \{-1,0,1\}$ and $VE_T$ are identified by
\begin{align*}\label{eq:idet_with_s}
    VE(m)&=1-\frac{\sum_sp_{1.1ms}\gamma_{s.1}}{\sum_sp_{1.0ms}\gamma_{s.0}},\\
VE_T&=1-\frac{\sum_sp_{1.11s}\gamma_{s.1}}{\sum_sp_{1.00s}\gamma_{s.0}}.
\end{align*}


\section{Motivation for nonparametric bounds}\label{sec:motivation}

\begin{figure}[t] 
\centering
\begin{minipage}{1\textwidth}
\centering
\resizebox{0.5\textwidth}{!}{ 
\begin{tikzpicture}
\tikzset{line width=1.5pt, outer sep=0pt,ell/.style={draw,fill=white, inner sep=2pt,line width=1.5pt},swig vsplit={gap=5pt,inner line width right=0.5pt}};
\node[name=A,ell,  shape=ellipse] at (-6,-1) {$A$};
\node[name=M,ell,  shape=ellipse] at (-6,1) {$M$};
\node[name=B,ell,  shape=ellipse] at (-3,0) {$B$};
\node[name=E,ell,  shape=ellipse] at (0,0) {$E$};
\node[name=U,ell,  shape=ellipse] at (0,2) {$U$};
\node[name=Y,ell,  shape=ellipse] at (3,0) {$Y$};
\begin{scope}[>={Stealth[black]},every edge/.style={draw=black,very thick}]
\path [->] (E) edge (Y);
\path [->] (A) edge (E);
\path [->] (A) edge (B);
\path [->] (M) edge (B);
\path [->] (B) edge (E);
\path [->] (U) edge (E);
\path [->] (U) edge [dashed](B);
\path [->] (U) edge (Y);
\path [->] (E) edge (Y);
\path [->] (A) edge[bend left=-30] (Y);    
\end{scope}
\end{tikzpicture}}
\caption{A DAG describing a six-arm trial $\mathcal{T}_{VI}$. In this trial the VEs are not identifiable because Assumption \ref{assump:y.dismissible} is violated due to the dashed arrow. }
\label{fig:sub.viol.1}
\end{minipage}
\end{figure}
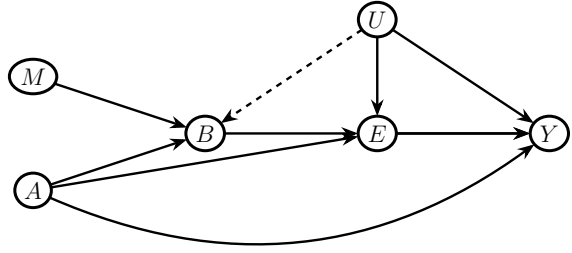

Arguably, the strongest of the identification assumptions described above is Assumption \ref{assump:y.dismissible}. It will fail if there is an unmeasured common cause, $U$, between $B$ and $Y$, leading to the unblocked path $M \rightarrow B\leftarrow U \rightarrow Y$, as illustrated in Figure \ref{fig:sub.viol.1}. These unmeasured variables $U$ can be, for example, psychological factors such as stress or illness anxiety disorder (hypochondria). In the two-arm trial, participants with higher anxiety levels (high $U$) might be more likely to believe they did not receive the vaccine ($B=0$) and are also more likely to obey the public health guidelines and minimize their exposure to the infectious agent ($E$), which can decrease the risk of infection ($Y$). Additionally, elevated anxiety levels can directly influence the immune system and susceptibility to infection through physiological mechanisms such as hormonal changes and immunosuppression \citep{cohen1991psychological,segerstrom2004psychological}.

When Assumption \ref{assump:y.dismissible} fails, the alternative identification strategy is built upon Assumption \ref{assump:y.s.dis}. However, this assumption could also fail; the example in the previous paragraph may still apply if $S$ does not fully mediate the effect of $U$ on $B$. 

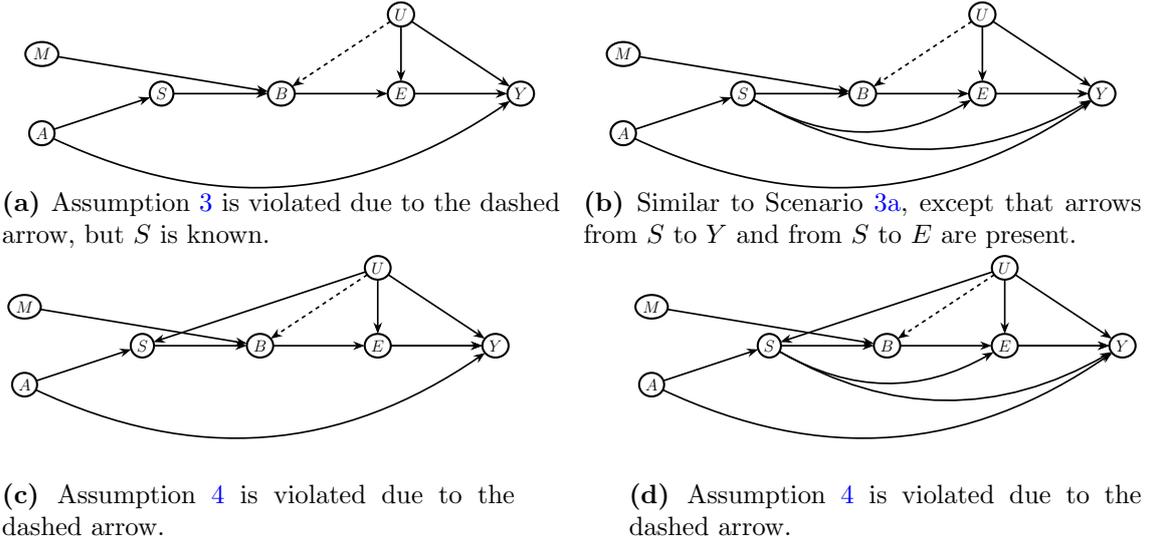
\begin{figure}[t]
  \centering

  \begin{subfigure}[t]{0.49\textwidth}
    \centering
    \resizebox{\linewidth}{!}{%
      \begin{tikzpicture}[baseline=(current bounding box.north)]
        \path[use as bounding box] (-10,-2) rectangle (4,3);

        \tikzset{
          line width=1.5pt, outer sep=0pt,
          ell/.style={draw,fill=white,inner sep=2pt,line width=1.5pt},
          swig vsplit/.style={gap=5pt,inner line width right=0.5pt}
        }

        \node[name=A,ell,ellipse] at (-9,-1) {$A$};
        \node[name=S,ell,ellipse] at (-6, 0) {$S$};
        \node[name=B,ell,ellipse] at (-3, 0) {$B$};
        \node[name=E,ell,ellipse] at ( 0, 0) {$E$};
        \node[name=U,ell,ellipse] at ( 0, 2) {$U$};
        \node[name=Y,ell,ellipse] at ( 3, 0) {$Y$};
        \node[name=M,ell,ellipse] at (-9, 1) {$M$};

        \begin{scope}[>={Stealth[black]}, every edge/.style={draw=black,very thick}]
          \path [->] (E) edge (Y);
          \path [->] (A) edge (S);
          \path [->] (S) edge (B);
          \path [->] (M) edge (B);
          \path [->] (B) edge (E);
          \path [->] (U) edge (E);
          \path [->] (U) edge[dashed] (B);
          \path [->] (U) edge (Y);
          \path [->] (A) edge[bend left=-30] (Y);
        \end{scope}
      \end{tikzpicture}%
    }
    \subcaption{Assumption \ref{assump:y.dismissible} is violated due to the dashed arrow, but $S$ is known.}
    \label{fig:sub.viol.3}
  \end{subfigure}
  \hfill
  \begin{subfigure}[t]{0.49\textwidth}
    \centering
    \resizebox{\linewidth}{!}{%
      \begin{tikzpicture}[baseline=(current bounding box.north)]
        \path[use as bounding box] (-10,-2) rectangle (4,3);

        \tikzset{
          line width=1.5pt, outer sep=0pt,
          ell/.style={draw,fill=white,inner sep=2pt,line width=1.5pt},
          swig vsplit/.style={gap=5pt,inner line width right=0.5pt}
        }

        \node[name=A,ell,ellipse] at (-9,-1) {$A$};
        \node[name=S,ell,ellipse] at (-6, 0) {$S$};
        \node[name=B,ell,ellipse] at (-3, 0) {$B$};
        \node[name=E,ell,ellipse] at ( 0, 0) {$E$};
        \node[name=U,ell,ellipse] at ( 0, 2) {$U$};
        \node[name=Y,ell,ellipse] at ( 3, 0) {$Y$};
        \node[name=M,ell,ellipse] at (-9, 1) {$M$};

        \begin{scope}[>={Stealth[black]}, every edge/.style={draw=black,very thick}]
          \path [->] (E) edge (Y);
          \path [->] (A) edge (S);
          \path [->] (S) edge (B);
          \path [->] (M) edge (B);
          \path [->] (B) edge (E);
          \path [->] (U) edge (E);
          \path [->] (U) edge[dashed] (B);
          \path [->] (U) edge (Y);
          \path [->] (S) edge[bend left=-30] (E);
          \path [->] (S) edge[bend left=-30] (Y);
          \path [->] (A) edge[bend left=-30] (Y);
        \end{scope}
      \end{tikzpicture}%
    }
    \subcaption{Similar to Scenario \ref{fig:sub.viol.3}, except that arrows from $S$ to $Y$ and from $S$ to $E$ are present.}
    \label{fig:sub.viol.4}
  \end{subfigure}

\begin{subfigure}[b]{0.45\textwidth} 
\centering
\resizebox{\linewidth}{!}{ 
\begin{tikzpicture}
\tikzset{line width=1.5pt, outer sep=0pt, ell/.style={draw,fill=white, inner sep=2pt,
line width=1.5pt},
swig vsplit={gap=5pt, inner line width right=0.5pt}};
\node[name=A,ell,  shape=ellipse] at (-9,-1) {$A$};
\node[name=S,ell,  shape=ellipse] at (-6,0) {$S$};
\node[name=B,ell,  shape=ellipse] at (-3,0) {$B$};
\node[name=E,ell,  shape=ellipse] at (0,0) {$E$};
\node[name=U,ell,  shape=ellipse] at (0,2) {$U$};
\node[name=Y,ell,  shape=ellipse] at (3,0) {$Y$};
\node[name=M,ell,  shape=ellipse] at (-9,1) {$M$};
\begin{scope}[>={Stealth[black]},
every edge/.style={draw=black,very thick}]
\path [->] (E) edge (Y);
\path [->] (A) edge (S);
\path [->] (S) edge (B);
\path [->] (M) edge (B);
\path [->] (B) edge (E);
\path [->] (U) edge (E);
\path [->] (U) edge[dashed] (B);
\path [->] (U) edge[] (S);
\path [->] (U) edge (Y);
\path [->] (E) edge (Y);  
\path [->] (A) edge[bend left=-30] (Y);    
\end{scope}
\end{tikzpicture}
}
\subcaption{Assumption \ref{assump:y.s.dis} is violated due to the dashed arrow.}        \label{fig:DAG_with_s_violation2}
\end{subfigure}
\hfill
\begin{subfigure}[b]{0.45\textwidth} 
\centering
\resizebox{\linewidth}{!}{ 
\begin{tikzpicture}
\tikzset{line width=1.5pt, outer sep=0pt, ell/.style={draw,fill=white, inner sep=2pt,
line width=1.5pt},
swig vsplit={gap=5pt, inner line width right=0.5pt}};
\node[name=A,ell,  shape=ellipse] at (-9,-1) {$A$};
\node[name=S,ell,  shape=ellipse] at (-6,0) {$S$};
\node[name=B,ell,  shape=ellipse] at (-3,0) {$B$};
\node[name=E,ell,  shape=ellipse] at (0,0) {$E$};
\node[name=U,ell,  shape=ellipse] at (0,2) {$U$};
\node[name=Y,ell,  shape=ellipse] at (3,0) {$Y$};
\node[name=M,ell,  shape=ellipse] at (-9,1) {$M$};
\begin{scope}[>={Stealth[black]},
every edge/.style={draw=black,very thick}]
\path [->] (E) edge (Y);
\path [->] (A) edge (S);
\path [->] (S) edge (B);
\path [->] (M) edge (B);
\path [->] (B) edge (E);
\path [->] (U) edge (E);
\path [->] (U) edge[dashed] (B);
\path [->] (U) edge[] (S);
\path [->] (U) edge (Y);
\path [->] (E) edge (Y);
\path [->] (S) edge[bend left=-30] (Y);  
\path [->] (S) edge[bend left=-30] (E);
\path [->] (A) edge[bend left=-30] (Y);    
\end{scope}
\end{tikzpicture}
        }
\subcaption{Assumption \ref{assump:y.s.dis} is violated due to the dashed arrow.}
\label{fig:DAG_with_s_violation}
\end{subfigure}
\caption{DAGs describing six-arm trials $\mathcal{T}_{VI}$. In these trials the VEs are not identifiable because Assumptions \ref{assump:y.dismissible} or \ref{assump:y.s.dis} are violated.}
    \label{fig:violation.assump.4}
\end{figure}

Because of the importance of the $VE(m)$,$VE_M(a)$, and $VE_T$, we develop in this paper nonparametric bounds whenever Assumptions \ref{assump:y.dismissible} and/or \ref{assump:y.s.dis} do not hold. We will show that the assumed causal structure may lead to different nonparametric bounds. These scenarios are divided into the following four main groups.
\begin{enumerate}
\item Figures \ref{fig:DAG_full_identification}\ref{fig:DAG_full_identification1}-\ref{fig:DAG_full_identification}\ref{fig:DAG_full_identification2}. The path $M \rightarrow B\leftarrow U \rightarrow Y$ does not exist and thus point identification is possible.
\item Figure \ref{fig:sub.viol.1}.
The path $M \rightarrow B\leftarrow U \rightarrow Y$ exists and thus point identification is not possible. A special case of this scenario captures the scenario described in Figure \ref{fig:DAG_full_identification2}, but because  $S$ is unobserved, point identification is not possible.
\item Figure \ref{fig:sub.viol.3}.
The path $M \rightarrow B\leftarrow U \rightarrow Y$ violates Assumption \ref{assump:y.dismissible} and thus point identification is not possible. The paths $S \rightarrow Y$ and $S\leftarrow U \rightarrow Y$ do not exist. 
\item Figures \ref{fig:sub.viol.4}, \ref{fig:DAG_with_s_violation2} and \ref{fig:DAG_with_s_violation}.
The path $M \rightarrow B \leftarrow U \rightarrow Y$ is present, together with one or both of the following paths between $S$ and $Y$.  If only $S \rightarrow Y$ exists, then Assumption \ref{assump:y.dismissible} is violated (as in Group 3). If $S \leftarrow U \rightarrow Y$ exists, then Assumption \ref{assump:y.s.dis} is violated. In either situation, point identification is not possible.
\end{enumerate}
For Groups 2–4, we derive two types of nonparametric bounds: LP-based bounds and monotonicity-based bounds with the latter relying on monotonicity assumptions. As will be demonstrated in the following sections, the bounds for Groups 2 and 4 are identical, while those for Group 3 may be narrower.

The scenarios above are presented without additional measured baseline covariates beyond $S$. In practice, vaccine trials typically collect baseline covariates (other than $S$), denoted by $L$. In Appendix~\ref{app:sharpening_bounds}, we examine when incorporating $L$ can lead to narrower bounds and provide the corresponding bounds under those settings.

To construct the bounds, we consider the following assumption.
\begin{Assumption}\label{assump:isolation}(M partial isolation).
The only causal paths from $M$ to $Y$ are directed paths intersected by $B$.
\end{Assumption}
\noindent This assumption is arguably quite plausible: it asserts that apart from its effect through belief, $M$ has no direct causal effect on $Y$. It is weaker than Assumptions \ref{assump:y.dismissible} and \ref{assump:y.s.dis}, which additionally require the absence of the path $B\leftarrow U \rightarrow Y$.

As we show in Appendix \ref{app:NPSEM},  under the nonparametric structural equation model with independent errors (NPSEM-IE) \citep{pearl2009causality}, 
Assumption \ref{assump:isolation} implies that in the two-arm trial, the following condition holds for 
$a,a',m \in \{0,1\}^3$:
\begin{equation}
\label{Eq:POidentBm}
\Pr\left(Y^{a',m}=Y^{a',m=-1}\middle|  A=a,B^{a',m=-1}=m,U=u\right)=1.
\end{equation}
In words, since $M$ affects $Y$ only through $B$, the potential outcome $Y^{a',m}$ is identical to $Y^{a',m=-1}$ within the subpopulation of individuals for whom $B^{a',m=-1} = m$. Nevertheless, the implication of Equation \eqref{Eq:POidentBm} is strong: it requires the condition to hold at the individual level and thus constitutes a \textit{cross-world assumption}, involving the distribution of $Y^{a',m}$ conditional on $B^{a',m=-1}$ for $m=0,1$. Nevertheless, as discussed later, our results remain the same even if condition \eqref{Eq:POidentBm} holds 
only in expectation, namely
$E\!\left(Y^{a',m}\mid A=a,B^{a',m=-1}=m,U=u\right)
=E\!\left(Y^{a',m=-1}\mid A=a,B^{a',m=-1}=m,U=u\right).$

The subsequent sections present LP-based and monotonicity-based bounds for Groups 2--4. All proofs are provided in the Appendix. In Sections \ref{sec:sim} and \ref{sec:real_data_example}, we compare the LP-based and monotonicity-based bounds in terms of length, and examine the factors that affect the width of the bounds. For Group 2, we explain in length the methods we leverage in this paper to construct the nonparametric bounds. 

\section{Nonparametric bounds for vaccine effects under Figure \ref{fig:sub.viol.1}} \label{sec:fig2}
LP bounds are based on the potential response approach; see \cite{maclehose2005bounding} for a review of the approach. Compared to existing research, the belief $B$ introduces additional nuances in these definitions. As a result, the standard LP-based results \citep{balke1997bounds} cannot be applied directly. We therefore derive LP-based bounds tailored to our setting, presenting the main ideas in the text and deferring the technical details to Appendix \ref{app:LP1}.

\subsection{LP-based bounds}\label{sec:3a.LP}
Let $r_{y|a, b}$ be the
\textit{potential response type} of $Y^{a,m}$ given $A=a$ and $B^{a,m=-1}=b$ for the different values of $a,m \in \{0,1\}
\times
\{-1,0,1\}$ in the two-arm trial. That is, for each combination of $a$ and $b$, the potential response type is the collection of potential outcomes for all $a$ and $m$ conditionally on $A=a$ and $B^{a,m=-1}=b$. For clarity, when conditioning on $A=a$ we write $B^{a,m=-1}$ simply as $B$, since $B^{a,m=-1}$ is the value of $B$ in $\mathcal{T}_{II}$ for those with vaccine status $A=a$. For example, as can be seen from Table \ref{tab:potential.responses.b0}, $r_{y|a, b=0}=0$ is the subgroup of participants that would not have been infected under any combination of $a$ and $m$, among those with treatment level $A=a$ and belief $B=0$ in trial $\mathcal{T}_{II}$. Note that although in the two-arm trial $M=-1$ with probability one, the potential outcomes $Y^{a,m}$ are well-defined for all $a,m$. For each combination of $a$ and $b$, there are $2^{2\times3}=64$ potential response types. Under Assumption \ref{assump:isolation} and the resulting Equation \eqref{Eq:POidentBm}, the number of possible potential response types is reduced to  $2^{2\times2}=16$. Tables \ref{tab:potential.responses.b0} and \ref{tab:potential.responses.b1} list all $r_{y|a, b}$ values for $a,b = 0,1$. 

For each $i\in \{0,...,15\}$, let $q_{i.ab}=\Pr(r_{y|a, b}=i)$ denote the proportion of the two-arm trial participants with potential response type $i$, among those with $A=a$ and $B=b$. Finally, let $Q_{a,b}(a,m)=\big\{r_{y|a, b}:\{Y^{a,m}|A=a,B=b\big\}=1\}$ be the set of potential response types such that with probability one, the potential outcome $Y^{a,m}$ equals to one conditionally on $A=a$ and $B=b$. 


We derive bounds for each VE by solving an optimization problem that maximizes (upper bound) or minimizes (lower bound) the VE under constraints derived from the observed data distribution. To obtain the bounds, as discussed later, we do not optimize the VE directly but instead its building blocks, namely $E_{IV}(Y^{a,m})=E_{VI}(Y^{a,m})=E(Y^{a,m})$. To begin with, we show how both the target parameters and the constraints can be expressed as linear functions of the $q_{i.ab}$, thereby framing the problem as an LP problem.
 
The target quantity, $E(Y^{a,m})$, can be expressed in terms of $q_{i.ab}$ as follows
\begin{align}
\label{eq:target.function.LP}
\begin{split}
 E(Y^{a,m})&=E(Y^{a,m}|A=a)\\
&=\pi_{0.a}E(Y^{a,m}|A=a,B=0)+\pi_{1.a}E(Y^{a,m}|A=a,B=1)\\
&=\pi_{0.a} \sum_{i \in Q_{a,0}}q_{i.a0}
+\pi_{1.a} \sum_{i \in Q_{a,1}}q_{i.a1},\\
\end{split}
\end{align}
where the first equality follows from randomization of $A$ and the last one by the definition of the potential response types. The constraints are derived from the observed conditional probability $p_{y.ab}$. As in the last line of Equation \eqref{eq:target.function.LP}, $p_{y.ab}$ can be represented as the sum $p_{y.ab}=\sum_{i \in Q_{a,b}(a,m=-1)}q_{i.ab}$. Additional constraints are induced from $q_{i.ab}$ being probabilities that sum to one over the potential response types. Putting it together, for every combination of $A=a$ and $B=b$, the constraints are
\begin{align}
\label{eq:const.LP}
\begin{split}
1.\, &p_{y.ab}= \sum_{i \in Q_{a,b}(a,m=-1)}q_{i.ab} \; \text{for } \; y \in \{0,1\}\\
2.\, &\sum_{i=0}^{15}q_{i.ab}=1\\
3.\,&q_{i.ab}\geq 0, \text{for} \, i=0,...,15
\end{split}
\end{align}

Having expressed both the target quantity in Equation \eqref{eq:target.function.LP} and the constraints in Equation \eqref{eq:const.LP} as linear functions of the $q_{i.ab}$, we now illustrate how this LP formulation can be used to derive bounds for the VEs. In particular, using Equation \eqref{eq:target.function.LP}, we express each of the VE terms as a combination of four conditional expectations $E(Y^{a,m}\mid A=a,B=b)$. Thus, to obtain the upper and lower bounds of the VEs, we solve four separate LP problems. For example, to find an upper bound for $VE(0)=1-\frac{E(Y^{a=1,m=0})}{E(Y^{a=0,m=0})}$, we first write it as
\begin{align}\label{eq:4_LP}
\begin{split}
VE(0)
&=1-\frac{\pi_{0.1}E(Y^{a=1,m=0}|A=1,B=0)
+\pi_{1.1}E(Y^{a=1,m=0}|A=1,B=1)}{\pi_{0.0}E(Y^{a=0,m=0}|A=0,B=0)
+\pi_{1.0}E(Y^{a=0,m=0}|A=0,B=1)},
    \end{split}
\end{align}
and then solve the following four separate LP problems, where the constraints in \eqref{eq:const.LP} are specified for different values of $a$ and $b$ based on the conditioning event:
$$
\min\{E(Y^{a=1,m=0}\mid A=1,B=b)\}
\; \text{for } a=1,\; b\in\{0,1\},
$$
and
$$
\max\{E(Y^{a=0,m=0}\mid A=0,B=b)\}
\; \text{for } a=0,\; b\in\{0,1\}.
$$

Finally, each LP problem can be solved using the program developed by \citet{balke1994counterfactual}, which takes symbolic LP expressions as input and returns symbolic solutions for the bounds, unlike standard LP solvers that provide only numerical values. Following \cite{gabriel2023nonparametric}, we instead use the \texttt{rcdd} R package \citep{Geyer2023} to solve each of the four LP problems. The \texttt{rcdd} R package is more general and suitable for complex scenarios beyond those handled by the original program. The following proposition presents the LP-based bounds under Figure \ref{fig:sub.viol.1}..

\begin{Proposition}
\label{prop:LP1}
 Consider the setting of Figure \ref{fig:sub.viol.1}. The following bounds for $VE(m)$, $m\in \{0,1\}$ and $VE_T$ hold. 
$$
1-\frac{1-p_{00.1}}{p_{10.0}}
\leq
VE(0)
\leq
1-\frac{p_{10.1}}{1-p_{00.0}}
$$ 
$$
1-\frac{1-p_{01.1}}{p_{11.0}}
   \leq
   VE(1)
   \leq
   1-\frac{p_{11.1}}{1-p_{01.0}}
   $$  
   $$
1-\frac{1-p_{01.1}}{p_{10.0}}
\leq
VE_T
\leq
1 - \frac{p_{11.1}}{ 1-p_{00.0}}.
$$
 \end{Proposition}  
The LP bounds in Proposition~\ref{prop:LP1} rely on the observed data and Assumption \ref{assump:isolation}. In Appendix \ref{app:sharp.LP.bounds}, we show that by further imposing individual-level monotonicity assumptions on the effect of $A$ or $M$ on $Y$, the bounds become narrower. For example, assuming $\Pr\left( Y^{a,m=0}\leq Y^{a,m=1}\middle|A=a,B=b,U=u\right)=1$ for all $a,b\in \{0,1\}$ and for all $u$, leads to the bounds for $VE(0)$:
$$
1-\frac{p_{10.1}+p_{1.11}}{1-p_{00.0}}
\leq\; VE(0) \leq
1-\frac{p_{10.1}}{p_{1.01}+p_{1.11}}.
$$
An alternative approach to the LP method to find the VE bounds is to derive the bounds directly from their representation in terms of conditional expectations, such as the representation of $VE(0)$ in Equation~\eqref{eq:4_LP}. In Appendix \ref{app:eq1}, we show that both the numerator and denominator of VE can be decomposed into two components: an identifiable component, equal to $p_{y.ab}$ for the relevant $a,b$, and a non-identifiable component that is bounded between 0 and 1. 
This decomposition makes it possible to bound VE without an optimization algorithm. Importantly, this approach requires Equation~\eqref{Eq:POidentBm} to hold only in expectation:
$$
E\big( Y^{a',m} \mid A=a, B^{a',m=-1}=m, U=u \big)
= E\big( Y^{a',m=-1} \mid A=a, B^{a',m=-1}=m, U=u \big).
$$
In addition, monotonicity assumptions on the effect of $A$ or $M$ on $Y$ can likewise be imposed in expectation, 
for example $E(Y^{a=1,m} \mid A=a, B=b) \leq E(Y^{a=0,m} \mid A=a, B=b)$. The bounds obtained in this way are equivalent to those derived using the LP procedure (Appendix \ref{app:eq1}).

\subsection{Monotonicity-based bounds}\label{sec:mon.2}
In this section, we derive monotonicity-based bounds based on two monotonicity assumptions, different from those just described. The first assumes that both the belief and the outcome are monotonically related to the unmeasured common cause.
\begin{Assumption}\label{assump:mon_in_U_general}(U monotonicity)
    In the setting of Figure \ref{fig:sub.viol.1},
    suppose that $Y\perp_{VI} M|A,B,U$ and one of the following conditions hold:
    \begin{enumerate}
    \item \label{assump:mon_in_U_general_1}$\Pr\left(Y=1|A=a,B=1,U=u\right)$ and $\Pr(B=1|A=a,U=u)$ are both nondecreasing or both nonincreasing in $u$, for all $a$.
    \item \label{assump:mon_in_U_general_2}One of $\Pr\left(Y=1|A=a,B=1,U=u\right)$ or $\Pr(B=1|A=a,U=u)$ is nondecreasing in $u$ and the other one is nonincreasing in $u$, for all $a$.
    \end{enumerate} 
\end{Assumption}
An example of a variable $U$ that might satisfy Assumption \ref{assump:mon_in_U_general}.\ref{assump:mon_in_U_general_1} is optimism. Optimistic participants may engage in ``wishful thinking” and believe they received the vaccine even when they did not \citep{bang2016random}. They may also be more outgoing, thereby increasing their risk of infection. By contrast, pessimism provides an example of the same monotonicity but in the opposite direction: both $\Pr(Y=1\mid A=a,B=1,U=u)$ and $\Pr(B=1\mid A=a,U=u)$ decrease with $u$. An illustration of Assumption \ref{assump:mon_in_U_general}.\ref{assump:mon_in_U_general_2} is neuroticism. Neurotic participants may be more anxious and thus more likely to believe they received the vaccine, consistent with findings that placebo effects are more common among them \citep{kern2020influence}. Such participants often have more introverted personalities and may therefore avoid or minimize social contact, which reduces their risk of infection \citep{glei2023midst}.
  
Unfortunately, under Assumption \ref{assump:mon_in_U_general}, we obtain two-sided bounds only for $VE(0)$. The resulting bounds do not provide information on $VE(1)$ or on an upper bound for $VE_T$ (see Appendix~\ref{app:proof of prop.bounds}). 
To remedy this problem, the second assumption we consider asserts that the sign of the effect of the message $M$ on the mean outcome $Y$ is known. 

\begin{Assumption}\label{assump:non.negative.m.general} (M monotonicity) The effect of $M$ on $Y$ is monotone. Specifically, one of the following conditions hold:
\begin{enumerate}
    \item \label{assump:non.negative.m} Non-negative effect of $M$:
    \begin{equation*}
      E(Y^{a,m=0})\leq E(Y^{a,m=-1})\leq E(Y^{a,m=1}),\, \text{for } \,a\in\{0,1\}
\end{equation*}
\item \label{assump:negative.m} Non-positive effect of $M$:
 \begin{equation*}
      E(Y^{a,m=0})\geq E(Y^{a,m=-1})\geq E(Y^{a,m=1}),\, \text{for } \,a\in\{0,1\}
\end{equation*}
\end{enumerate}
\end{Assumption}
This assumption is plausible under Assumptions \ref{assump:B_M} and \ref{assump:isolation}. For instance, if a participant receives a message indicating vaccination ($M=1$), then by Assumption \ref{assump:B_M} this also implies $B=1$. Assumption \ref{assump:non.negative.m.general}.\ref{assump:non.negative.m} reflects the scenario in which believing that one is vaccinated is likely to increase the risk of infection by encouraging riskier behaviors, such as more frequent contacts or reduced adherence to safety guidelines \citep{trogen2021risk,buckell2023covid}. By contrast, Assumption \ref{assump:non.negative.m.general}.\ref{assump:negative.m} reflects the opposite scenario, where believing one is vaccinated would reduce the risk of infection, a pattern that appears unlikely in practice.

The following proposition builds on Assumptions \ref{assump:mon_in_U_general}- \ref{assump:non.negative.m.general} to provide two-sided bounds for $VE(m)$, $m\in\{0,1\}$, and for $VE_T$.
\begin{Proposition}\label{prop:bounds.two.sided1}
 Consider the setting of Figure \ref{fig:sub.viol.1}. Under Assumptions \ref{assump:pos1},\ref{assump:mon_in_U_general}.\ref{assump:mon_in_U_general_1}, and \ref{assump:non.negative.m.general}.\ref{assump:non.negative.m}, the following bounds for $VE(m), \;m\in\{0,1\}$ and $VE_T$ hold.
$$
1-\frac{p_{1.1}}{p_{1.00}}\le VE(0) \le 1-\frac{p_{1.10}}{p_{1.0}},
$$ 
$$
1-\frac{p_{1.11}}{p_{1.0}}\le VE(1) \le 1-\frac{p_{1.1}}{p_{1.01}},
$$
 $$
1-\frac{p_{1.11}}{p_{1.00}}\le VE_T \le 1-\frac{p_{1.1}}{p_{1.0}}.
  $$
Under Assumptions \ref{assump:pos1}, \ref{assump:mon_in_U_general}.\ref{assump:mon_in_U_general_2} and \ref{assump:non.negative.m.general}.\ref{assump:negative.m} the bounds for $VE(m)$, $m\in\{0,1\}$ and $VE_T$ given in Appendix \ref{app:proof.two.bounds} hold.
\end{Proposition}

\section{Bounds for vaccine effects under Figure \ref{fig:sub.viol.3}}\label{sec:fig3}

Similarly to the scenario illustrated in Figure \ref{fig:sub.viol.1}, Assumption \ref{assump:y.dismissible} is also violated in Figure \ref{fig:sub.viol.3} due to the path $ B \leftarrow U \rightarrow Y $. However, unlike Figure \ref{fig:sub.viol.1}, Figure \ref{fig:sub.viol.3} introduces an additional factor: AEs ($S$) are observed and directly influence the belief ($B$). In vaccine trials, it is more realistic to assume that treatment ($A$) affects belief through AEs, rather than through a direct effect on belief. Consistent with this assumption, empirical evidence shows that participants in vaccine trials often rely on perceived AEs to infer their treatment allocation, leading to unblinding \citep{hrobjartsson2014risk}.

Under Figure \ref{fig:sub.viol.3}, the arrows $U \to S$ and $S \to Y$ are absent. This implies two additional independence assumptions: $S \perp_{\mathcal{T}_{VI}} U$ and $Y \perp_{\mathcal{T}_{VI}} S \mid A, B, M, U$. As we show in this section, these two independence assumptions are exactly what allow us to obtain narrower bounds under both the LP-based and the monotonicity-based approaches.

\subsection{LP-based bounds}\label{sec:3c-d.LP}
Similarly to Proposition \ref{prop:LP1}, we use the potential response approach to construct bounds for the VE under the settings described by Figure \ref{fig:sub.viol.3}. Let $r_{y|a,s,b}$ be the potential response of $Y^{a,m}$ given $A=a$, $S^{a,m=-1}=s$ and $B^{a,m=-1}=b$ for the different values of $a,m\in\{0,1\}^2$ in the two arm trial. As in Section \ref{sec:3a.LP}, and for clarity, when conditioning on $A=a$ we write $S^{a,m=-1}$ and $B^{a,m=-1}$ simply as $S$ and $B$, respectively. Furthermore, let $q_{i.asb}=\Pr(r_{y|a,s,b}=i)$ and define
$$
Q_{a,s,b}(a,m)=\{r_{y|a,s,b}:\{Y^{a,m}|A=a,S=s,B=b\}=1\}.
$$

Using the above results, we now show how the target parameter $E(Y^{a,m})$ and the constraints can be expressed in terms of $q_{i.asb}$. Similarly to Equation \eqref{eq:target.function.LP}, we can express $E(Y^{a,m})$ for all $s\in\{0,1\}$ as
\begin{align}\label{eq:target.function.LP.2c}
\begin{split}
E(Y^{a,m})
&=E(Y^{a,m}|A=a,S=s)\\
&=\pi_{0.as}E(Y^{a,m}|A=a,S=s,B=0)+\pi_{1.as}E(Y^{a,m}|A=a,S=s,B=1)\\
&=\pi_{0.as}\sum_{i \in Q_{a,s,0}}q_{i.as0}
+\pi_{1.as}\sum_{i \in Q_{a,s,1}}q_{i.as1},\\
\end{split}
\end{align}
where the first equality follows from the randomization of $A$ and because $A$ blocks the only backdoor path between $S$ and $Y$ under Figure \ref{fig:sub.viol.3}, so $Y^{a,m}\perp_{\mathcal{T}_{VI}} A,S^{a,m=-1}$ (for a proof see Appendix \ref{app:indepndece.proof}). Moving to the constraints, similarly to Equation \eqref{eq:const.LP}, we obtain the following system of linear constraints for every combination of $A=a$, $S=s$ and $B=b$:
 \begin{align}
 \label{eq:const.LP.s}
 \begin{split}
1.\, &p_{y.asb}= \sum_{i \in Q_{a,s,b}(a,m=-1)}q_{i.asb} \; \text{for } \; y \in \{0,1\}\\
2.\, &\sum_{i=0}^{15}q_{i.asb}=1\\
3.\,&q_{i.asb}\geq 0, \text{for} \, i=0,...,15
 \end{split}
 \end{align}

Now that we have expressed both the target parameter and the constraints in terms of $q_{i.asb}$, we can formulate the corresponding LP problem. Since Equations \eqref{eq:target.function.LP.2c} and \eqref{eq:const.LP.s} must hold for both $s=0$ and $s=1$, the LP can be solved separately within each stratum of $s$. We then select, for the lower and upper bound separately, the value of $s$ that yields the tightest result. This procedure ensures that the bounds presented in the next proposition are narrower than the bounds in Proposition \ref{prop:LP1}.

\begin{Proposition}\label{prop:LP1_withS}
   Consider the setting of Figure \ref{fig:sub.viol.3}. The following bounds for the $VE(m)$, $m \in \{0,1\}$ and $VE_T$ hold.
 $$
    \max_{s_1,s_2\in \{0,1\}}\bigg \{1-\frac{ 1-p_{00.1s_1}}{p_{10.0s_2}}\bigg\}
   \leq
   VE(0)
   \leq
  \min_{s_1,s_2\in \{0,1\}}\bigg \{1-\frac{p_{10.1s_1}}{1-p_{00.0s_2}}\bigg\},
   $$ 
    $$
 \max_{s_1,s_2\in \{0,1\}}\bigg \{1-\frac{1-p_{01.1s_1}}{p_{11.0s_2}}\bigg\}
   \leq
   VE(1)
   \leq
   \min_{s_1,s_2\in \{0,1\}}\bigg \{1-\frac{p_{11.1s_1}}{1-p_{01.0s_2}}\bigg\},
   $$  
   $$
\max_{s_1,s_2\in \{0,1\}}\bigg \{1-\frac{1-p_{01.1s_1}}{p_{10.0s_2}}\bigg\}
        \leq
         VE_T
         \leq
\min_{s_1,s_2\in \{0,1\}}\bigg \{1 - \frac{p_{11.1s_1}}{ 1-p_{00.0s_2}}\bigg\}.
   $$
 \end{Proposition} 

\subsection{Monotonicity-based bounds}
\label{sec:3c-d.mon}
Because under Figure \ref{fig:sub.viol.3} we have $S \perp_{\mathcal{T}_{VI}} Y \mid A,B,M,U$, we can replace Assumption \ref{assump:mon_in_U_general} with the following requirement that conditions on $S$ and imposes monotonicity within each stratum of $S$.

\begin{Assumption}\label{assump:U.mon.S.general}($U,S$ monotonicity)
    Consider the setting of Figure \ref{fig:sub.viol.3}. Suppose that $Y\perp_{\mathcal{T}_{VI}} M|A,B,S,U$ and one of the following conditions holds:
\begin{enumerate}
\item  \label{assump:U.mon.S.versiona} $\Pr\left(Y=1|A=a,B=1,S=s,U=u\right)$ and $\Pr(B=1|A=a,S=s,U=u)$ are both nondecreasing or both nonincreasing in $u$, for all $a$ and $s$.
\item \label{assump:U.mon.S.versionb} One of $\Pr\left(Y=1|A=a,B=1,S=s,U=u\right)$ or $\Pr(B=1|A=a,S=s,U=u)$ is nondecreasing in $u$ and the other one is nonincreasing in $u$, for all $a$ and $s$.
\end{enumerate} 
\end{Assumption}
\noindent As formally presented in the following Proposition, under Assumption \ref{assump:U.mon.S.general}, the expressions $p_{1.ab}$ in Proposition \ref{prop:bounds.two.sided1} can be replaced by $\min_{s\in \{0,1\}}\{p_{1.abs}\}$ or by $\max_{s\in \{0,1\}}=\{p_{1.abs}\}$ depending on whether they appear in the numerator or denominator and whether we are computing an upper or lower bound. 

\begin{Proposition}
\label{prop:bounds.mon.s.iv}
 Consider the setting of Figure \ref{fig:sub.viol.3}. Under Assumptions \ref{assump:pos2}, \ref{assump:non.negative.m.general}.\ref{assump:non.negative.m}, and \ref{assump:U.mon.S.general}.\ref{assump:U.mon.S.versiona}, the following bounds for $VE(m)$, $m\in\{0,1\}$ and $VE_T$ hold.
\begin{alignat*}{3}
\max_{s\in \{0,1\}}\Bigg\{1-\frac{p_{1.1}}{p_{1.00s}}\Bigg\}
&\;\le\;& VE(0) &\;\le\;
\min_{s\in \{0,1\}}\Bigg\{1-\frac{p_{1.10s}}{p_{1.0}}\Bigg\}, \\[1ex]
\max_{s\in \{0,1\}}\Bigg\{1-\frac{p_{1.11s}}{p_{1.0}}\Bigg\}
&\;\le\;& VE(1) &\;\le\;
\min_{s\in \{0,1\}}\Bigg\{1-\frac{p_{1.1}}{p_{1.01s}}\Bigg\}, \\[1ex]
\max_{s_1,s_2\in \{0,1\}}\Bigg\{1-\frac{p_{1.11s_1}}{p_{1.00s_2}}\Bigg\}
&\;\le\;& VE_T &\;\le\;
1-\frac{p_{1.1}}{p_{1.0}}.
\end{alignat*}
Under Assumptions \ref{assump:pos2}, \ref{assump:non.negative.m.general}.\ref{assump:negative.m} and \ref{assump:U.mon.S.general}.\ref{assump:U.mon.S.versionb}, the bounds for $VE(m)$, $m\in\{0,1\}$ and $VE_T$ given in Appendix \ref{app:proof:prop:bounds.mon.s.iv} hold.
\end{Proposition}

The monotonicity-based bounds in Proposition \ref{prop:bounds.mon.s.iv} are guaranteed to be at least as narrow as those in Proposition \ref{prop:bounds.two.sided1}, since they are obtained by additionally conditioning on $S$. As in the LP case, this conditioning allows the bounds to be computed within each stratum of $S$, and then combined by taking the appropriate minimum or maximum, which can only tighten the results.

\section{Bounds for vaccine effects under Figures \ref{fig:sub.viol.4}--\ref{fig:DAG_with_s_violation}}
\label{sec:viol.Y.S.dis}

In this section, we discuss Figures \ref{fig:sub.viol.4}--\ref{fig:DAG_with_s_violation}, where $S$ is measured. As we will show, however, $S$ cannot be used to obtain narrower bounds in Propositions \ref{prop:LP1} and \ref{prop:bounds.two.sided1}. We begin by describing how the role of AEs differs across these figures. 

Figures \ref{fig:sub.viol.3} and \ref{fig:sub.viol.4} differ in whether AEs have a direct effect on the exposure and the outcome. A natural explanation for this difference is the severity of the events. When AEs are mild (e.g., arm pain or low-grade fever), it may be reasonable to assume that $S$ influences $B$ but not $E$ and $Y$, as in Figure \ref{fig:sub.viol.3}. By contrast, when AEs are severe (e.g., limiting the ability to perform day-by-day activities), they may cause individuals to stay home, thereby reducing their exposure ($E$) to pathogens and in turn lowering their risk of infection ($Y$), as represented in Figure \ref{fig:sub.viol.4}. In practice, this pathway from $S$ to $E$ or to $Y$ is likely to have only a minor impact, since AEs are typically short-lived relative to the trial duration and early post-vaccination days are often excluded from outcome assessment. Nevertheless, we include this possibility in Figure \ref{fig:sub.viol.4} to ensure that our framework captures all relevant causal pathways. 

Figures \ref{fig:DAG_with_s_violation2}--\ref{fig:DAG_with_s_violation} violate Assumption \ref{assump:y.s.dis} due to the path $B \leftarrow U \rightarrow Y$. However, unlike Figure \ref{fig:sub.viol.4}, they also include the path $S \leftarrow U \rightarrow Y$. Such scenario is possible, for example, if $U$ influence not only $B$ and $Y$ but also the perception and reporting of AEs ($S$). For instance, anxiety sensitivity and neuroticism have been linked to a heightened tendency to interpret mild symptoms as severe AEs, a phenomenon known as the \textit{nocebo effect} \citep{meijers2022possible,nocebo_vaccine_anxiety}.

\subsection{LP-based bounds}\label{sec:LP.S.not.sharp}
As in Section \ref{sec:3c-d.LP}, the LP-based bounds rely on $r_{y|a,b,s}$, $q_{i.asb}$, and $Q_{a,s,b}(a,m)$. The key difference here is in the representation of $E(Y^{a,m})$. Here, we can no longer argue that $E(Y^{a,m}) = E(Y^{a,m} \mid A=a,S=s)$. However, by the law of total probability, we can still express both the target parameter and the constraints in terms of $q_{i.asb}$. In particular, $E(Y^{a,m})$ can be written as follows:
\begin{align}\label{eq:target.function.LP.U.S}
\begin{split}
E(Y^{a,m})&=E(Y^{a,m}|A=a)\\
&=\sum_{s\in\{0,1\}}\gamma_{s.a}E(Y^{a,m}|A=a,S=s)\\
&=\sum_{b\in\{0,1\}}\sum_{s\in\{0,1\}}\gamma_{s.a}\pi_{b.as}E(Y^{a,m}|A=a,S=s,B=b)\\
&=\sum_{s\in\{0,1\}}\gamma_{s.a}\Big[\pi_{0.as}\sum_{i \in Q_{a,s,0}}q_{i.as0}
+\pi_{1.as}\sum_{i \in Q_{a,s,1}}q_{i.as1}\Big].
\end{split}
\end{align}
The constraints are the same as in Section \ref{sec:3c-d.LP}. This leads to the following proposition.
\begin{Proposition}\label{prop:bounds.4ab.LP}
The bounds for the VE given in Proposition \ref{prop:LP1} hold in the setting of Figures \ref{fig:sub.viol.4}--\ref{fig:DAG_with_s_violation}.
\end{Proposition}

\subsection{Monotonicity-based bounds}
In the case of Figures \ref{fig:sub.viol.4}--\ref{fig:DAG_with_s_violation}, to obtain bounds, in addition to the conditions in Assumption \ref{assump:U.mon.S.general}, we also assume the directions of the effects of $S$ on $B$ and of $S$ on the conditional expectation $E(Y^{a,m}\mid A=a,S=s)$.

\begin{Assumption}\label{assump:U.mon.S.general.SU}($U|S$ monotonicity)
    Consider the setting of Figures \ref{fig:sub.viol.4}--\ref{fig:DAG_with_s_violation}. Suppose that $Y\perp_{\mathcal{T}_{VI}} M|A,B,S,U$, and the following conditions hold:

\begin{enumerate}
\setlength{\itemsep}{0pt}
\setlength{\topsep}{0pt}
\setlength{\parsep}{0pt}
\setlength{\partopsep}{0pt}
    \item  \label{assump:U.mon.S.versiona.SU} $\Pr\left(Y=1|A=a,B=1,S=s,U=u\right)$ and $\Pr(B=1|A=a,S=s,U=u)$ are both nondecreasing or both nonincreasing in $u$, for all $a$ and $s$.
    \item \label{assump:a1} $\Pr(B=1|A=a,S=s)$ is nondecreasing in $s$, for all $a$.
    \item \label{assump:b1} $E(Y^{a,m}|A=a,S=s)$ is nondecreasing in $s$, for all $a$.
\end{enumerate} 
\end{Assumption}
Next, we clarify the additional conditions in Assumption \ref{assump:U.mon.S.general.SU}. Condition \ref{assump:U.mon.S.general.SU}.\ref{assump:a1} captures the idea that individuals who experience AEs are more likely to believe that they received the vaccine, which is typically plausible in vaccine trials. To explain the meaning of Condition \ref{assump:U.mon.S.general.SU}.\ref{assump:b1}, we consider Figure~\ref{fig:DAG_with_s_violation2}, in which the path $S \to Y$ is absent. Under this figure, Condition \ref{assump:U.mon.S.general.SU}.\ref{assump:b1} is implied by two simpler monotonicity assumptions: (i) $\Pr(Y=1 \mid A=a,B=1,U=u)$ is nondecreasing in $u$; and (ii) $\Pr(S=1 \mid U=u)$ is nondecreasing in $u$. Intuitively, Condition \ref{assump:U.mon.S.general.SU}.\ref{assump:b1} means that both the outcome and the occurrence of AEs increase with $U$, so higher values of $S$ correspond to higher values of $E(Y^{a,m})$. Nevertheless, since the arrow $S \to Y$ may exist in general, the simplified monotonicity conditions serve only as intuition, and the analysis still requires Condition \ref{assump:U.mon.S.general.SU}.\ref{assump:b1} itself.

As in the LP-based case, the bounds under Figures \ref{fig:sub.viol.4}--\ref{fig:DAG_with_s_violation}, together with Assumptions \ref{assump:pos2}, \ref{assump:non.negative.m.general} and \ref{assump:U.mon.S.general.SU}, coincide with the bounds presented in Proposition \ref{prop:bounds.two.sided1}.
\begin{Proposition}\label{prop:bounds.4ab.mon}
  Under Assumptions \ref{assump:pos2}, \ref{assump:non.negative.m.general} and \ref{assump:U.mon.S.general.SU}, the bounds for the VE given in Proposition \ref{prop:bounds.two.sided1} hold in the setting of Figures \ref{fig:sub.viol.4}--\ref{fig:DAG_with_s_violation}.
\end{Proposition}
\noindent 
The assumption corresponding to Assumption \ref{assump:U.mon.S.general.SU} that yields reversed bounds is provided in Appendix \ref{app:proof.prop.bounds2.reversed.bounds}.

To emphasize the findings from this section, if either (i) the arrow $S \rightarrow Y$ exists, or (ii)  the arrow $U\rightarrow S$ exists (in addition of being a common cause of $B$ and $Y$), then $S$ cannot be used to improve the width of the bounds, and the resulting bounds (in Propositions \ref{prop:bounds.4ab.LP}--\ref{prop:bounds.4ab.mon}) remain the same as the results without observing $S$ (Propositions~\ref{prop:LP1}--\ref{prop:bounds.two.sided1}). While $B$ is not a treatment variable, these finding are reminiscing of a demand that $S$ has a similar role to an instrumental variable (with respect to $B$). Furthermore, finding (ii) aligns with the results of \cite{jonzon2025adding} (their Section 4.1), who similarly demonstrated that when the covariate is confounded by an unobserved variable, ``covariate averaging does not improve the width of the bounds,'' and the covariate-adjusted and unadjusted bounds coincide.

As a remark, in some applications it may be more realistic to assume distinct unmeasured variables, $U_1$ and $U_2$, rather than a single common cause (Figure~\ref{fig:DAG_violations_two_U}). For example, $U_1$ could represent personality traits that affect both belief and outcome, while $U_2$ could represent pre-existing health conditions, such as chronic inflammation or immunodeficiency, that increase both the likelihood of strong AEs and susceptibility to infection \citep{polack2020safety}. As shown in Appendix~\ref{app:two_u}, however, introducing $U_2$ does not affect the key independence relations and therefore does not alter the resulting bounds: $Y \perp_{\mathcal{T}_{VI}} M \mid A,B,S,U_1$ still holds, while $Y^{a,m} \not\perp_{\mathcal{T}_{VI}} S$, so Propositions~\ref{prop:bounds.4ab.LP} and \ref{prop:bounds.4ab.mon} continue to apply.

\section{Numerical examples}\label{sec:sim}
We conducted a simulation study to examine the behavior of the nonparametric bounds under various settings. The simulation study objectives were: (1) to evaluate how the extent of broken blinding influences the width of the true bounds; (2) to assess the consequences of violations of the monotonicity assumptions, specifically (2a)  Assumption \ref{assump:U.mon.S.general.SU}\ref{assump:U.mon.S.versiona.SU} and (2b) Assumption \ref{assump:non.negative.m.general}\ref{assump:non.negative.m}, on the validity of the bounds; (3) to evaluate the consequences of applying bounds derived under an incorrect causal structure; and (4) to assess the performance of bounds estimated from a finite sample.

To compute two-sided monotonicity-based bounds, the data generating mechanism (DGM) was designed to satisfy Assumption \ref{assump:non.negative.m.general}. For comparison, we also computed the LP-based bounds under the same M-monotonicity assumption, but at the individual level as described in Section \ref{sec:3a.LP}. For clarity, when we state that bounds were computed using Proposition \ref{prop:bounds.4ab.LP}, this refers to calculations carried out under the conditions of Proposition \ref{prop:bounds.4ab.LP} together with the M-monotonicity assumption. The corresponding formulas for the LP-based bounds under this assumption are provided in Appendix \ref{app:sharp.LP.bounds}.


\subsection{DGM}\label{sec:DGM}
We generated a two-arm vaccine trial following the DGM described below. To avoid sampling noise and approximate the true population values of the estimands and their bounds under the DGM, we simulated a large population of size $n = 10^6$. For each participant, we simulated $\{A,U\}$ and then the full set of potential outcomes $\{S^{a},B^{a,m},Y^{a,m}\}$ for $a,m \in \{0,1\}\times \{-1,0,1\}$. First, when $m \in \{0,1\}$, we set $B^{a,m}=m$ and simulated according to the model
\begin{equation}
\begin{aligned}
\begin{split}
\label{eq:sim_PO}
\Pr(A=a)&\sim \mathrm{Bernoulli}(0.5)\\
\Pr(U=u)&\sim \Ind_{u}\,\mathrm{Bernoulli}(0.5)+ \Ind_{u^2}\,\mathrm{N}(0,1)\\
\Pr(S^{a}=1|U=u) &= \text{expit}(\delta_0+\delta_A a + \delta_U u) \\
\Pr(Y^{a,m}=1|U=u) &= \text{expit}(\gamma_0+\gamma_A a + \gamma_{B^a} m +  \gamma_S S^{a} + \Ind_{u}\gamma_U u+\Ind_{u^2}\gamma_U u^2),
\end{split}
\end{aligned}
\end{equation}
where $\mathrm{expit}(x) = \frac{e^x}{1 + e^x}$. Here, $\Ind_{\{\cdot\}}$ denotes a binary switch used to control whether Assumptions~\ref{assump:U.mon.S.general.SU}.\ref{assump:U.mon.S.versiona.SU} hold: with $\Ind_{u}=1$ and $\Ind_{u^2}=0$ the assumption is satisfied, while with $\Ind_{u}=0$ and $\Ind_{u^2}=1$ it is violated (note that $\Ind_{u}$ and $\Ind_{u^2}$ control $U$ variable type but also its association with $Y$). The coefficients $\gamma_{B^1}$ and $\gamma_{B^0}$ represent the effects of $B$ on $Y$ when $a=1$ and $a=0$, respectively. These coefficients are used to control whether Assumptions~\ref{assump:non.negative.m.general}\ref{assump:non.negative.m} hold: when both are positive the assumption is satisfied, whereas setting $\gamma_{B^1}>0$ and $\gamma_{B^0}<0$ corresponds to violation.

Next, when $m=-1$, we simulate $B^{a,m=-1}$ for $a\in\{0,1\}$ by
\begin{align}
\begin{split}
\label{eq:sim_PO_minus_1}
\Pr(B^{a,m=-1}=1|U=u)&=\text{expit}(\beta_0+  \beta_{S} S^{a} +  \Ind_{u}\beta_U u+\Ind_{u^2}\beta_U u^2).
\end{split}
\end{align}
We then set $Y^{a,m=-1}$ according to the value of $B^{a,m=-1}$: if  $B^{a,m=-1}=1$, we take $Y^{a,m=-1}=Y^{a,m=1}$ and if $B^{a,m=-1}=0$ we take $Y^{a,m=-1}=Y^{a,m=0}$. Finally, we set the observed values of $(B,S,Y)$ to their corresponding $m=-1$ potential outcomes. To generate the different scenarios, we used the general model in Equations \eqref{eq:sim_PO}--\eqref{eq:sim_PO_minus_1} and varied the parameter specifications: setting all parameters to positive values except $\delta_U=\gamma_S=0$ corresponded to Figure \ref{fig:sub.viol.3}; setting all parameters to positive values except $\delta_U=0$ corresponded to Figure \ref{fig:sub.viol.4}; setting all parameters to positive values except $\gamma_S=0$ corresponded to Figure \ref{fig:DAG_with_s_violation2}; and setting all parameters to positive values corresponded to Figure \ref{fig:DAG_with_s_violation}.

\subsection{Simulation setup}\label{sec:sim:analysis}

To address each of the objectives, we specified the causal structure under which data were generated (the corresponding figure), the imposed assumptions, and the propositions used to calculate LP-based and monotonicity-based bounds. For each objective, we also calculated the corresponding population-level values of $\mathrm{VE}(0)$, $\mathrm{VE}(1)$, and $\mathrm{VE}_T$, as well as the true bounds, which can be recovered in this large-sample setting where sampling variability is negligible. Table \ref{tab:sim_fig_prop} summarizes these design choices.

For Objective (4), in addition to the large two-arm trial, we also generated $1,000$ smaller samples of size $n=500$ and $n=5,000$. For each such dataset, we constructed 95\% percentile-bootstrap confidence intervals with $B=200$ to evaluate the performance of the bounds under finite samples.

\begin{table}[t]
\caption{Objectives with corresponding simulation figures, assumptions, and propositions.}
\label{tab:sim_fig_prop}
\centering
\begin{tabular}{lllll}
\toprule
Objective & Figure & Assumptions&\makecell{LP- \\ Propositions}  & \makecell{Monotonicity- \\ Propositions} \\
\midrule
(1)&\ref{fig:DAG_with_s_violation}&\ref{assump:non.negative.m.general}.\ref{assump:non.negative.m} and \ref{assump:U.mon.S.general.SU}&\ref{prop:bounds.4ab.LP}&\ref{prop:bounds.4ab.mon}\\
(2a)&  \ref{fig:DAG_with_s_violation}&\ref{assump:non.negative.m.general}.\ref{assump:non.negative.m}, \ref{assump:U.mon.S.general.SU}.\ref{assump:a1} and \ref{assump:U.mon.S.general.SU}.\ref{assump:b1}&\ref{prop:bounds.4ab.LP} &\ref{prop:bounds.4ab.mon}\\
(2b)&  \ref{fig:DAG_with_s_violation}&\ref{assump:U.mon.S.general.SU}&\ref{prop:bounds.4ab.LP}&\ref{prop:bounds.4ab.mon}\\
(3)& \ref{fig:DAG_with_s_violation2} and \ref{fig:DAG_with_s_violation} &\ref{assump:non.negative.m.general}\ref{assump:non.negative.m} and \ref{assump:U.mon.S.general.SU}&\ref{prop:LP1_withS}&\ref{prop:bounds.mon.s.iv}\\
(4a)& 
\ref{fig:DAG_with_s_violation}&\ref{assump:non.negative.m.general}.\ref{assump:non.negative.m} and \ref{assump:U.mon.S.general.SU}&\ref{prop:bounds.4ab.LP}&\ref{prop:bounds.4ab.mon}\\
(4b)& \ref{fig:sub.viol.3}&\ref{assump:non.negative.m.general}.\ref{assump:non.negative.m} and \ref{assump:U.mon.S.general.SU}&\ref{prop:LP1_withS} &\ref{prop:bounds.mon.s.iv}\\
\bottomrule
\end{tabular}
\end{table}

\subsection{Results}
\begin{figure}[t]
    \centering
    \includegraphics[width=1\linewidth]{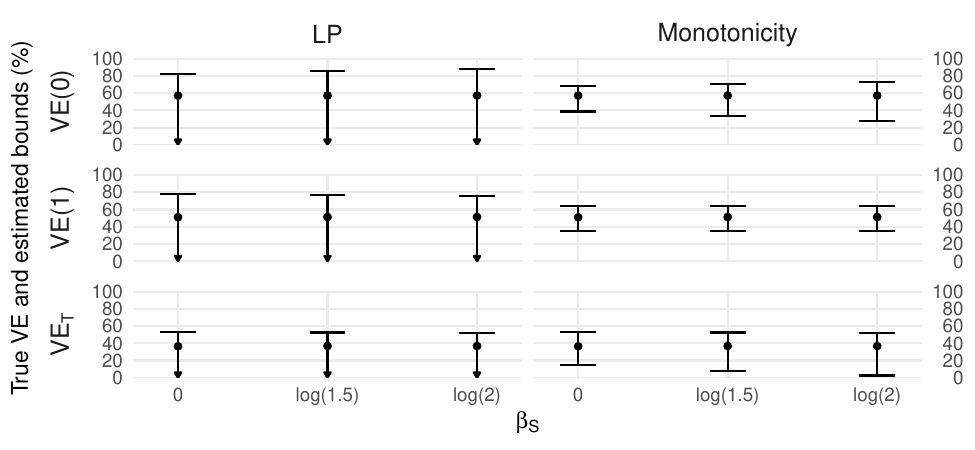}
    \caption{The relationship between magnitude of broken blinding ($\beta_S$) and the width of the bounds for $VE(0)$, $VE(1)$ and $VE_T$ under the setting of Figure~\ref{fig:DAG_with_s_violation}. Assumptions \ref{assump:non.negative.m.general}\ref{assump:non.negative.m} and \ref{assump:U.mon.S.general.SU} are satisfied. A value of $\beta_S=0$ corresponds to the case where the belief was not influenced by AEs, $\beta_S=\log(1.5)$ represented moderate, and $\beta_S=\log(2)$ strong, broken blinding due to AEs. Circles represent the true values. LP: bounds constructed via Proposition~\ref{prop:bounds.4ab.LP} and the results in Appendix~\ref{app:sharp.LP.bounds}; Monotonicity: bounds constructed via Proposition~\ref{prop:bounds.4ab.mon}. Triangles at the lower boundary indicate bounds that extended below 0\% but were truncated for display.
    }
    \label{fig:sim.affect.of.blinding}
\end{figure}

For Objective (1), Figure \ref{fig:sim.affect.of.blinding} displays the resulting bounds for $\mathrm{VE}(0)$, $\mathrm{VE}(1)$, and $\mathrm{VE}_T$ across different levels of the effect of $S$ on $B$ ($\beta_S$). In all scenarios, the monotonicity-based bounds were narrower than those obtained using the LP-based method. As $\beta_S$ increased, the bounds for $\mathrm{VE}(0)$ and $\mathrm{VE}_T$ became wider, while the bounds for $\mathrm{VE}(1)$ became narrower. Importantly, even when $\beta_S=0$, the belief remained affected by $U$, which prevented point identification of the VEs.

\begin{figure}[t]
    \centering
    \includegraphics[width=1\linewidth]{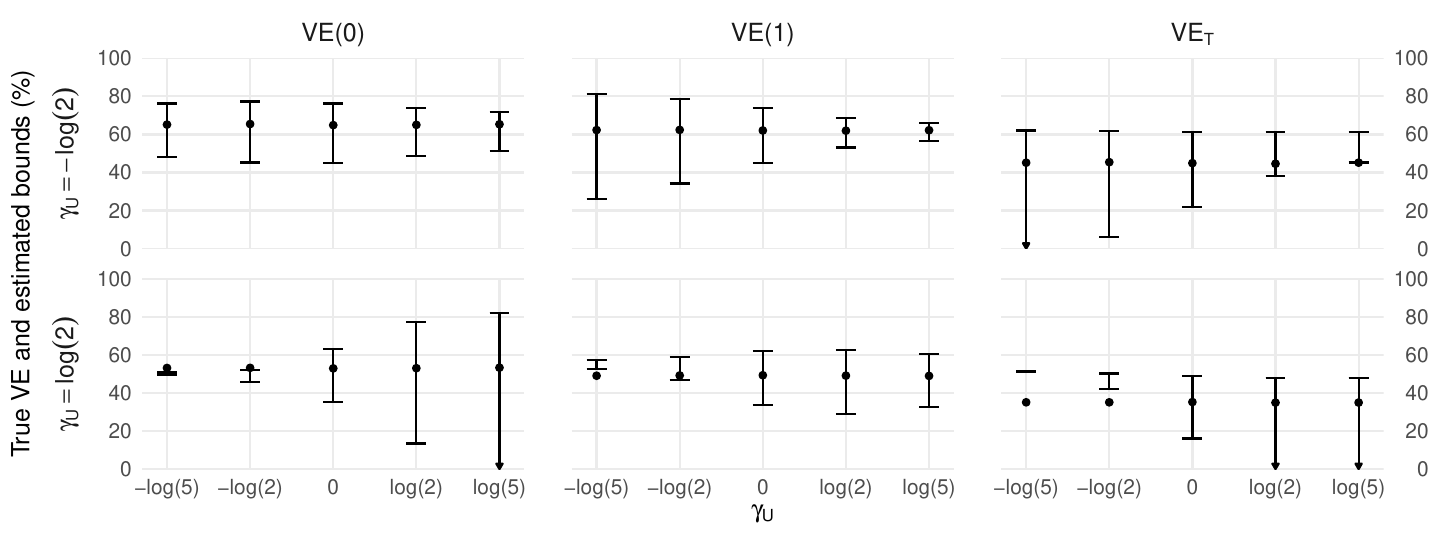}
\caption{Monotonicity-based bounds for \(\mathrm{VE}(0)\), \(\mathrm{VE}(1)\), and \(\mathrm{VE}_T\) for $\gamma_U \in \{-\log(2),\log(2)\}$ and across values of $\beta_U$ when Assumption~\ref{assump:U.mon.S.general} is violated and Assumption~\ref{assump:non.negative.m.general} is maintained. The parameters $\beta_U$ and $\gamma_U$ controlled the monotone relationships between $U$ and $B$ or $Y$, respectively: setting either to zero enforced monotonicity by removing the corresponding non-linear effect of $U$, while increasing their magnitude strengthened violations of the assumption. Circles indicate the true values. The monotonicity-based bounds were calculated using Proposition~\ref{prop:bounds.4ab.mon}. Triangles at the lower boundary indicate bounds that extended below 0\% but were truncated for display.}
    \label{fig:sim.mon_assump}
\end{figure}

To address Objective (2a), Figure \ref{fig:sim.mon_assump} presents the monotonicity-based bounds under violations of Assumption \ref{assump:U.mon.S.general.SU}.\ref{assump:U.mon.S.versiona.SU}. For $\gamma_U>0$, as $\beta_U$ increased the monotonicity-based bounds became wider, whereas for $\gamma_U<0$ the pattern was mirrored -- bounds became wider as $\beta_U$ decreased, with the effect most pronounced for $\mathrm{VE}(1)$ and $\mathrm{VE}_T$. Interestingly, when $\beta_U$ and $\gamma_U$ shared the same sign, the bounds remained valid despite the technical violation of monotonicity, while opposite signs led to clearly invalid bounds, particularly for $\mathrm{VE}_T$, as shown in Figure \ref{fig:sim.mon_assump}. This observation aligns with theoretical results from \citet{vanderweele2008causal}, who showed that the direction and magnitude of bias due to unmeasured confounding depend not only on whether monotonicity holds, but also on whether the confounder affects different pathways in opposing directions. When $\beta_U$ and $\gamma_U$ share the same sign, the effects of $U$ on $B$ and $Y$ act in the same direction, preserving an approximately monotonic total effect of $U$. In contrast, when $\beta_U$ and $\gamma_U$ have opposite signs, the effects of $U$ on $B$ and $Y$ tend to cancel each other, creating non-monotonic behavior that violates Assumption \ref{assump:U.mon.S.general.SU}.\ref{assump:U.mon.S.versiona.SU}. 

For Objectives (2b) and (3), Figures~\ref{fig:app:M_mon_assump}--\ref{fig:app:SY_SU} examine the robustness of the bounds under different misspecifications. In some settings, particularly under stronger violations, the estimated lower bound exceeded the upper bound, indicating that the assumed causal structure or underlying assumptions (e.g., Assumption \ref{assump:non.negative.m.general}.\ref{assump:non.negative.m}) were incompatible with the DGM. In general, the monotonicity-based bounds were less robust to misspecification, as they were narrower. When Assumption \ref{assump:non.negative.m.general}.\ref{assump:non.negative.m} was violated, both LP-based and monotonicity-based bounds were robust for mild violation ($\gamma_{B1}=-\log(1.5)$) but became invalid for moderate to large violations ($\gamma_{B1}\in{-\log(3), -\log(5)}$) (Figure~\ref{fig:app:M_mon_assump}). When the $S \rightarrow U$ path was incorrectly assumed to be absent, both the LP-based and monotonicity-based bounds remained valid for all values of $\delta_U$. However, the lower bound of the monotonicity-based interval for $\mathrm{VE}_T$ was very close to being invalid (Figure~\ref{fig:app:SY}). When both $S \rightarrow U$ and $S \rightarrow Y$ arrows were incorrectly assumed to be absent, only the LP-based bounds for $\mathrm{VE}(0)$ remained valid across all values of $\delta_U$, whereas all other bounds were either invalid or very close to becoming invalid (Figure~\ref{fig:app:SY_SU}).

Finally, we turn to Objective (4). In practice, the bounds we developed in this paper must be estimated from finite samples. Tables \ref{app:tab:coverage_lp} and \ref{app:tab:coverage_mon} summarize the empirical coverage rates of the LP-based and monotonicity-based bounds. Overall, most intervals achieve coverage very close to the nominal 95\% level, with only a few lower or upper limits showing slight under- or over-coverage, especially for the smaller sample size. For further discussion of the bootstrap CI coverage, see Section~\ref{sec:Discussion}.

\section{Real data example}\label{sec:real_data_example}

The ENSEMBLE2 study is a randomized, double-blind, placebo-controlled, phase 3 trial of COVID-19 vaccines. We analyzed the data via the YODA platform \citep{krumholz2016yale}. The dataset available to us included a modestly larger cohort than the one reported in \citet{hardt2022efficacy}, since additional participants were enrolled after the primary analysis was completed. As a result, the numbers reported in this section slightly differ from those in \citet{hardt2022efficacy}. We considered semi-synthetic analysis to illustrate how this trial could have been analyzed and bounds could have been calculated had the belief was measured. 
 
A total of 31,704 individuals participated in the trial, of whom 16,083 (51\%) were randomized to receive three doses (including a booster) of the COVID-19 vaccine, and 15,621 (49\%) were randomized to receive three doses (including a booster) of saline placebo. The solicited local AE was assessed in a safety subset containing 6,184 randomly selected participants from the full cohort. To avoid complexities introduced by dropouts and arm-switching, we chose as our primary endpoint the VE during days 14-56 following the first vaccination, and the solicited AE after the first dose. In the full cohort, the estimated VE after the first vaccination was $\widehat{VE}(-1)=53.8\%$, while the corresponding estimate in the safety subset was $\widehat{VE}(-1)=39.3\%$. The sample means of the solicited local AE were $\widehat{\Pr}(S=1|A=1)=57.3\%$ in the vaccinated group and $\widehat{\Pr}(S=1|A=0)=22.5\%$ in the placebo group, indicating that AEs are far more likely for those treated with the vaccine, thus compromising the blinding process in this trial.

Since $B$ was not measured in this trial, we generated it to reflect two considerations. First, individuals who experienced AEs ($S=1$) were assumed to be more likely to believe they had received the vaccine. Second, we aimed to align $B$ with Assumption~\ref{assump:non.negative.m.general}.\ref{assump:non.negative.m}, which states that receiving a message of vaccination ($M=1$) increases the risk of infection ($Y=1$). As discussed in Section~\ref{sec:mon.2}, this assumption implies that believing one is vaccinated ($B=1$) is also associated with a higher risk of infection ($Y=1$). Because $B$ was unobserved, we generated it in the opposite direction, individuals with a higher infection risk were made more likely to believe they were vaccinated. The full probabilities used to generate $B$ are provided in Appendix \ref{app:prob.B.real.data}.

\begin{figure}[t]
    \centering
    \includegraphics[width=1\linewidth]{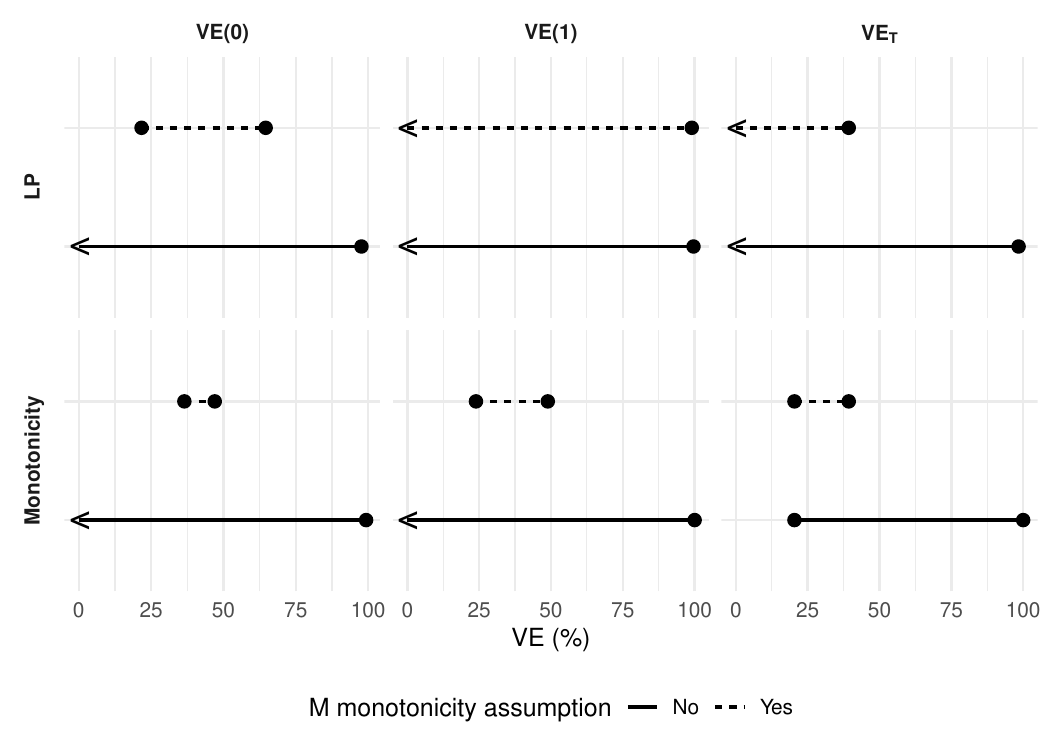}
    \caption{Estimated LP-based and monotonicity-based bounds under Figure \ref{fig:DAG_with_s_violation}. 
Each horizontal line shows the identified interval for $VE(0)$, $VE(1)$, and $VE_{T}$, stratified by method (LP vs. Monotonicity) and whether Assumption \ref{assump:non.negative.m.general} \ref{assump:non.negative.m} was imposed (solid = no, dashed = yes). 
Triangles at the left boundary indicate bounds that extended below 0\% but were truncated for display.}
    \label{fig:numerical_example}
\end{figure}

We first computed the point estimates of the VEs under Assumptions~\ref{assump:B_M}--\ref{assump:y.dismissible}. We obtained 
$\widehat{VE}(0) = 44.5\%$ (95\% CI: 1.0\%, 71.5\%), 
$\widehat{VE}(1) = 36.0\%$ (95\% CI: -70.0\%, 75.0\%), 
and $\widehat{VE_T} = 20.4\%$ (95\% CI: -52.1\%, 64.1\%). Next, we estimated the LP-based and monotonicity-based bounds for $VE(0)$, $VE(1)$ and $VE_T$ plugging the sample proportions instead of the corresponding probabilities. Figure \ref{fig:numerical_example} presents the bounds under Figure \ref{fig:DAG_with_s_violation}, both with and without Assumption \ref{assump:non.negative.m.general}.\ref{assump:non.negative.m}. The monotonicity-based bounds under Assumption~\ref{assump:non.negative.m.general}.\ref{assump:non.negative.m} were the most informative. These bounds were notably narrower than the LP-based bounds and included the corresponding point estimates within their ranges. The bounds for $VE(0)$, (36.5\%, 47.0\%), suggest that the VE among participants who received the message $m=0$ could be even higher than the estimated $\widehat{VE}(-1) = 39.3\%$ in the safety subset, reaching up to 47.0\%. The bounds for $VE(1)$, (23.9\%, 48.9\%), indicate that the VE for those who received the message $m=1$ could be stronger than $VE(0)$ (and hence stronger than $VE(-1)$), but could also be considerably lower than the point estimate $\widehat{VE}(1) = 36.0\%$ and $\widehat{VE}(-1)$. Finally, the bounds for $VE_T$, (20.4\%, 39.3\%), suggest that the maximum total VE is similar to the published estimate, implying that the overall efficacy remains modest.

Figure \ref{app:fig:bounds_under_figure3} and Table~\ref{app:tab:bounds-figure3} present the LP-based and the monotonicity-based bounds under Figure~\ref{fig:sub.viol.3}, both with and without imposing Assumption~\ref{assump:non.negative.m.general}.\ref{assump:non.negative.m}. Under Assumption~\ref{assump:non.negative.m.general}.\ref{assump:non.negative.m}, in most cases we obtained upper bounds that were lower than the corresponding lower bounds, indicating that, under Figure~\ref{fig:sub.viol.3}, at least one of the Assumptions~\ref{assump:non.negative.m.general}.\ref{assump:non.negative.m} or~\ref{assump:U.mon.S.general}.\ref{assump:U.mon.S.versiona} must be violated. Under both Figures \ref{fig:sub.viol.3} and \ref{fig:DAG_with_s_violation}, when Assumption~\ref{assump:non.negative.m.general}.\ref{assump:non.negative.m} was not imposed, both the LP-based and monotonicity-based bounds were wide and largely non-informative, with upper limits close to 100\% and negative lower limits. This finding reflects that, without additional assumptions on how $M$ affects $Y$, we cannot determine whether the vaccine provides protection in this setting (Tables~\ref{app:tab:bounds-figure4}--\ref{app:tab:bounds-figure3}). All 95\% bootstrap confidence intervals are reported in Table~\ref{app:tab:bounds-figure3} for Figure~\ref{fig:sub.viol.3} and in Table~\ref{app:tab:bounds-figure4} for Figure~\ref{fig:DAG_with_s_violation}.

\section{Discussion}\label{sec:Discussion}
In this paper, we address the challenge of disentangling the immunological and behavioral components of vaccine effects when the identification assumptions proposed by \citet{stensrud2024distinguishing} do not hold. Our work is motivated by growing evidence of compromised blinding in vaccine trials, such as the ENSEMBLE2 study, where unblinding may be possible due to differences in AE rates. In such cases, the standard VE estimate from the trial may not reflect the true immunological effect. However, the methodology proposed by \citet{stensrud2024distinguishing} to identify the immunological effect in such cases is not applicable here, as the unobserved common cause may affect belief, AEs, and the outcome simultaneously, violating key assumptions.

We propose large-sample bounds using two complementary approaches: LP-based bounds and monotonicity-based bounds. We study three groups of plausible causal DAGs: (i) when $S$ is unmeasured; (ii) when $U$ affects both $B$ and $Y$; and (iii) when $U$ affects $B$, $S$, and $Y$, deriving bounds for each setting.


Our numerical study reveals that when the monotonicity assumptions hold, monotonicity-based bounds are typically narrower than LP-based bounds. This is expected, as stronger assumptions generally lead to narrower bounds. When Assumption~\ref{assump:U.mon.S.general.SU} was violated, the resulting bounds were invalid. By contrast, when Assumption~\ref{assump:non.negative.m.general} was violated, the bounds were largely robust under mild to moderate violations. Misspecifying the $S \rightarrow U$ path did not affect validity under our numerical example, though this need not hold in general. In contrast, misspecifying the $S \rightarrow Y$ path led to invalid bounds, particularly for $\mathrm{VE}_T$. Due to the latter case, unless domain knowledge justifies the assumptions, we recommend using the bounds derived under the most conservative DAG in Figure~\ref{fig:DAG_with_s_violation} rather than the narrower bounds from Figure~\ref{fig:sub.viol.3}.

To construct confidence intervals, we used percentile bootstrap-based intervals, as  proposed in recent work applying LP bounds \citep{gabriel2023nonparametric, gabriel2023sharp}. However, under DAGs like Figure \ref{fig:sub.viol.3}, where bounds involve the minimum and maximum of observed probabilities (observed up to estimation), standard bootstrap procedures may yield inconsistent intervals. As noted by \citet{chernozhukov2013intersection}, such nonparametric bootstrap intervals can substantially underestimate the width of the true bounds, leading to undercoverage. This aligns with our simulations, where the empirical coverage of bootstrap intervals was often around 90\% instead of 95\%. While more sophisticated corrections and exact inference methods exist \citep{Haile2003, Kreider2007, chernozhukov2013intersection}, we did not apply these in the present study. Since the true DGM is unknown, we recommend computing the bounds under the structure shown in Figure~\ref{fig:DAG_with_s_violation}, under which the bootstrap coverage was closest to 95\%.

In summary, this study illustrates that even when the immunological and behavioral VEs cannot be separately point identified,  information about these quantities can still be extracted using nonparametric bounds across a range of plausible DGMs.

\section{Software}
\label{sec5}

Code and reproducibility materials for the numerical examples and the real-data analysis, with a synthetic dataset provided instead of the original data, are available at \url{https://github.com/xlrod1/vaccine_for_publication}.

\section*{Acknowledgments}

We thank Mats Stensrud for very helpful discussions that contributed to the development of this paper.

Section \ref{sec:real_data_example}, carried out under YODA Project \#2024-0972, used data obtained from the Yale University Open Data Access (YODA) Project, which has an agreement with Janssen Research \& Development, L.L.C. The interpretation and reporting of research using this data are solely the responsibility of the authors and do not necessarily represent the official views of the Yale University Open Data Access Project or Janssen Research \& Development, L.L.C. The original proposal is available at: \url{https://yoda.yale.edu/data-request/2024-0972/}. {\it Conflict of Interest}: None declared.

\bibliographystyle{apalike}
\bibliography{vaccine.bib}

@article{stensrud2024distinguishing,
  title={Distinguishing immunologic and behavioral effects of vaccination},
  author={Stensrud, Mats J and Nevo, Daniel and Obolski, Uri},
  journal={Epidemiology},
  volume={35},
  number={2},
  pages={154--163},
  year={2024},
  publisher={LWW}
}

@article{esary1967association,
  title={Association of random variables, with applications},
  author={Esary, James D and Proschan, Frank and Walkup, David W},
  journal={The Annals of Mathematical Statistics},
  volume={38},
  number={5},
  pages={1466--1474},
  year={1967},
  publisher={Institute of mathematical statistics}
}

@article{vanderweele2008sign,
  title={The sign of the bias of unmeasured confounding},
  author={VanderWeele, Tyler J},
  journal={Biometrics},
  volume={64},
  number={3},
  pages={702--706},
  year={2008},
  publisher={Wiley Online Library}
}

@article{stensrud2021generalized,
  title={A generalized theory of separable effects in competing event settings},
  author={Stensrud, Mats J and Hern{\'a}n, Miguel A and Tchetgen Tchetgen, Eric J and Robins, James M and Didelez, Vanessa and Young, Jessica G},
  journal={Lifetime data analysis},
  volume={27},
  number={4},
  pages={588--631},
  year={2021},
  publisher={Springer}
}

@article{kern2020influence,
  title={The influence of personality traits on the placebo/nocebo response: a systematic review},
  author={Kern, Alexandra and Kramm, Christoph and Witt, Claudia M and Barth, J{\"u}rgen},
  journal={Journal of Psychosomatic research},
  volume={128},
  pages={109866},
  year={2020},
  publisher={Elsevier}
}

@article{bang2016random,
  title={Random guess and wishful thinking are the best blinding scenarios},
  author={Bang, Heejung},
  journal={Contemporary clinical trials communications},
  volume={3},
  pages={117--121},
  year={2016},
  publisher={Elsevier}
}

@article{obolski2024call,
  title={A call for blinding assessments in dengue vaccine trials},
  author={Obolski, Uri and Stensrud, Mats J and Nevo, Daniel},
  journal={The Lancet Infectious Diseases},
  volume={24},
  number={1},
  pages={e10},
  year={2024},
  publisher={Elsevier}
}

@article{sachs2023general,
  title={A general method for deriving tight symbolic bounds on causal effects},
  author={Sachs, Michael C and Jonzon, Gustav and Sj{\"o}lander, Arvid and Gabriel, Erin E},
  journal={Journal of Computational and Graphical Statistics},
  volume={32},
  number={2},
  pages={567--576},
  year={2023},
  publisher={Taylor \& Francis}
}

@article{balke1997bounds,
  title={Bounds on treatment effects from studies with imperfect compliance},
  author={Balke, Alexander and Pearl, Judea},
  journal={Journal of the American statistical Association},
  volume={92},
  number={439},
  pages={1171--1176},
  year={1997},
  publisher={Taylor \& Francis}
}

@inproceedings{balke1994counterfactual,
  title={Counterfactual probabilities: Computational methods, bounds and applications},
  author={Balke, Alexander and Pearl, Judea},
  booktitle={Uncertainty in artificial intelligence},
  pages={46--54},
  year={1994},
  organization={Elsevier}
}

@book{pearl2009causality,
  title={Causality},
  author={Pearl, Judea},
  year={2009},
  publisher={Cambridge university press}
}

@article{gabriel2023nonparametric,
  title={Nonparametric bounds for causal effects in imperfect randomized experiments},
  author={Gabriel, Erin E and Sj{\"o}lander, Arvid and Sachs, Michael C},
  journal={Journal of the American Statistical Association},
  volume={118},
  number={541},
  pages={684--692},
  year={2023},
  publisher={Taylor \& Francis}
}

@article{kuroki2008formulating,
  title={Formulating tightest bounds on causal effects in studies with unmeasured confounders},
  author={Kuroki, Manabu and Cai, Zhihong},
  journal={Statistics in medicine},
  volume={27},
  number={30},
  pages={6597--6611},
  year={2008},
  publisher={Wiley Online Library}
}

@article{maclehose2005bounding,
  title={Bounding causal effects under uncontrolled confounding using counterfactuals},
  author={MacLehose, Richard F and Kaufman, Sol and Kaufman, Jay S and Poole, Charles},
  journal={Epidemiology},
  volume={16},
  number={4},
  pages={548--555},
  year={2005},
  publisher={LWW}
}

@misc{Geyer2023,
  author       = {Charles J. Geyer and Glen D. Meeden},
  title        = {Rcdd: Computational Geometry},
  year         = {2023},
  note         = {Incorporates code from cddlib (ver. 0.94f) written by Komei Fukuda},
  url          = {https://doi.org/10.32614/CRAN.package.rcdd},
  urldate      = {2023-09-30}
}

@article{robins2010alternative,
  title={Alternative graphical causal models and the identification of direct effects},
  author={Robins, James M and Richardson, Thomas S},
  journal={Causality and psychopathology: Finding the determinants of disorders and their cures},
  volume={84},
  pages={103--158},
  year={2010},
  publisher={Oxford University Press Oxford}
}

@article{stensrud2023conditional,
  title={Conditional separable effects},
  author={Stensrud, Mats J and Robins, James M and Sarvet, Aaron and Tchetgen Tchetgen, Eric J and Young, Jessica G},
  journal={Journal of the American Statistical Association},
  volume={118},
  number={544},
  pages={2671--2683},
  year={2023},
  publisher={Taylor \& Francis}
}

@article{glei2023midst,
  title={In the midst of a pandemic, more introverted individuals may have a mortality advantage},
  author={Glei, Dana A and Weinstein, Maxine},
  journal={Dialogues in Health},
  volume={2},
  pages={100087},
  year={2023},
  publisher={Elsevier}
}

@article{hrobjartsson2014risk,
  author    = {Asbjørn Hróbjartsson and Kare Birger Hagen and Peter C Gøtzsche and Maria B Ljungberg and Mette H Thomsen and Lene I Sörensen},
  title     = {The risk of unblinding was infrequently and incompletely reported in 300 randomized clinical trial publications},
  journal   = {Journal of Clinical Epidemiology},
  volume    = {67},
  number    = {3},
  pages     = {280-285},
  year      = {2014},
  publisher = {Elsevier},
  doi       = {10.1016/j.jclinepi.2013.07.012}
}

@article{nocebo_vaccine_anxiety,
  author    = {Witthöft, M. and Hiller, W.},
  title     = {The Nocebo Effect in the Context of Vaccination: The Role of Anxiety Sensitivity and Expectancies},
  journal   = {Psychosomatic Medicine},
  year      = {2022},
  volume    = {84},
  number    = {5},
  pages     = {621-628},
  url       = {https://journals.lww.com/psychosomaticmedicine/fulltext/2022/05000/the_nocebo_effect_in_the_context_of_vaccination.7.aspx}
}

@article{polack2020safety,
  author    = {Polack, F. P. and Thomas, S. J. and Kitchin, N.},
  title     = {Safety and efficacy of the BNT162b2 mRNA COVID-19 vaccine},
  journal   = {New England Journal of Medicine},
  year      = {2020},
  volume    = {383},
  pages     = {2603-2615},
  url       = {https://doi.org/10.1056/NEJMoa2034577}
}

@article{vanderweele2008causal,
  title={Causal directed acyclic graphs and the direction of unmeasured confounding bias},
  author={VanderWeele, Tyler J and Hern{\'a}n, Miguel A and Robins, James M},
  journal={Epidemiology},
  volume={19},
  number={5},
  pages={720--728},
  year={2008},
  publisher={LWW}
}

@article{trogen2021risk,
  author       = {Trogen, B. and Caplan, A.},
  title        = {Risk compensation and COVID-19 vaccines},
  journal      = {Annals of Internal Medicine},
  year         = {2021},
  volume       = {174},
  number       = {6},
  pages        = {858--859},
  doi          = {10.7326/M20-8251},
}

@article{buckell2023covid,
  author       = {Buckell, J. and Avery, C. and Hale, T. and others},
  title        = {COVID-19 vaccination, risk‐compensatory behaviours, and contacts in the UK},
  journal      = {Scientific Reports},
  year         = {2023},
  volume       = {13},
  pages        = {8441},
}

@article{lazarus2021safety,
  title={Safety and immunogenicity of concomitant administration of COVID-19 vaccines (ChAdOx1 or BNT162b2) with seasonal influenza vaccines in adults in the UK (ComFluCOV): a multicentre, randomised, controlled, phase 4 trial},
  author={Lazarus, Rajeka and Baos, Sarah and Cappel-Porter, Heike and Carson-Stevens, Andrew and Clout, Madeleine and Culliford, Lucy and Emmett, Stevan R and Garstang, Jonathan and Gbadamoshi, Lukuman and Hallis, Bassam and others},
  journal={The Lancet},
  volume={398},
  number={10318},
  pages={2277--2287},
  year={2021},
  publisher={Elsevier}
}

@article{hardt2022efficacy,
  title={Efficacy, safety, and immunogenicity of a booster regimen of Ad26. COV2. S vaccine against COVID-19 (ENSEMBLE2): results of a randomised, double-blind, placebo-controlled, phase 3 trial},
  author={Hardt, Karin and Vandebosch, An and Sadoff, Jerald and Le Gars, Mathieu and Truyers, Carla and Lowson, David and Van Dromme, Ilse and Vingerhoets, Johan and Kamphuis, Tobias and Scheper, Gert and others},
  journal={The Lancet Infectious Diseases},
  volume={22},
  number={12},
  pages={1703--1715},
  year={2022},
  publisher={Elsevier}
}

@article{segerstrom2004psychological,
  author       = {Segerstrom, Suzanne C. and Miller, Gregory E.},
  title        = {Psychological Stress and the Human Immune System: A Meta‐Analytic Study of 30 Years of Inquiry},
  journal      = {Psychological Bulletin},
  year         = {2004},
  volume       = {130},
  number       = {4},
  pages        = {601--630},
  doi          = {10.1037/0033-2909.130.4.601},
}

@article{cohen1991psychological,
  author       = {Cohen, Sheldon and Tyrrell, David A. J. and Smith, Andrew P.},
  title        = {Psychological Stress and Susceptibility to the Common Cold},
  journal      = {New England Journal of Medicine},
  year         = {1991},
  volume       = {325},
  number       = {9},
  pages        = {606--612},
  doi          = {10.1056/NEJM199108293250903},
}

@article{jonzon2025adding,
  title={Adding covariates to bounds: What is the question?},
  author={Jonzon, Gustav and Gabriel, Erin E and Sj{\"o}lander, Arvid and Sachs, Michael C},
  journal={arXiv preprint arXiv:2502.03156},
  year={2025}
}

@article{gabriel2023sharp,
  title={Sharp nonparametric bounds for decomposition effects with two binary mediators},
  author={Gabriel, Erin E and Sachs, Michael C and Sj{\"o}lander, Arvid},
  journal={Journal of the American Statistical Association},
  volume={118},
  number={544},
  pages={2446--2453},
  year={2023},
  publisher={Taylor \& Francis}
}

@article{chernozhukov2013intersection,
  title={Intersection bounds: Estimation and inference},
  author={Chernozhukov, Victor and Lee, Sokbae and Rosen, Adam M},
  journal={Econometrica},
  volume={81},
  number={2},
  pages={667--737},
  year={2013},
  publisher={Wiley Online Library}
}

@article{Haile2003,
  title={Inference with an incomplete model of English auctions},
  author={Haile, Philip A. and Tamer, Elie},
  journal={Journal of Political Economy},
  volume={111},
  number={1},
  pages={1--51},
  year={2003},
  publisher={University of Chicago Press}
}

@article{Kreider2007,
  title={Disability and employment: Reevaluating the evidence in light of reporting errors},
  author={Kreider, Brent and Pepper, John V.},
  journal={Journal of the American Statistical Association},
  volume={102},
  number={478},
  pages={432--441},
  year={2007},
  publisher={Taylor \& Francis}
}

@article{robins1989analysis,
  title={The analysis of randomized and non-randomized AIDS treatment trials using a new approach to causal inference in longitudinal studies},
  author={Robins, James M},
  journal={Health service research methodology: a focus on AIDS},
  pages={113--159},
  year={1989},
  publisher={US Public Health Service}
}

@article{manski1990nonparametric,
  title={Nonparametric bounds on treatment effects},
  author={Manski, Charles F},
  journal={The American Economic Review},
  volume={80},
  number={2},
  pages={319--323},
  year={1990},
  publisher={JSTOR}
}

@article{horowitz2000nonparametric,
  title={Nonparametric analysis of randomized experiments with missing covariate and outcome data},
  author={Horowitz, Joel L and Manski, Charles F},
  journal={Journal of the American statistical Association},
  volume={95},
  number={449},
  pages={77--84},
  year={2000},
  publisher={Taylor \& Francis}
}

@article{vanderweele2010bias,
  title={Bias formulas for sensitivity analysis for direct and indirect effects},
  author={VanderWeele, Tyler J},
  journal={Epidemiology},
  volume={21},
  number={4},
  pages={540--551},
  year={2010},
  publisher={LWW}
}

@article{swanson2018partial,
  title={Partial identification of the average treatment effect using instrumental variables: review of methods for binary instruments, treatments, and outcomes},
  author={Swanson, Sonja A and Hern{\'a}n, Miguel A and Miller, Matthew and Robins, James M and Richardson, Thomas S},
  journal={Journal of the American Statistical Association},
  volume={113},
  number={522},
  pages={933--947},
  year={2018},
  publisher={Taylor \& Francis}
}

@article{krumholz2016yale,
  title={The Yale Open Data Access (YODA) project—a mechanism for data sharing},
  author={Krumholz, Harlan M and Waldstreicher, Joanne},
  journal={New England Journal of Medicine},
  volume={375},
  number={5},
  pages={403--405},
  year={2016},
  publisher={Mass Medical Soc}
}

@article{meijers2022possible,
  title={Possible alleviation of symptoms and side effects through clinicians’ nocebo information and empathy in an experimental video vignette study},
  author={Meijers, MC and Stouthard, J and Evers, AWM and Das, E and Drooger, HJ and Jansen, SJAJ and Francke, AL and Plum, N and van der Wall, E and Nestoriuc, Y and others},
  journal={Scientific reports},
  volume={12},
  number={1},
  pages={16112},
  year={2022},
  publisher={Nature Publishing Group UK London}
}

@article{cai2007non,
  title={Non-parametric bounds on treatment effects with non-compliance by covariate adjustment},
  author={Cai, Zhihong and Kuroki, Manabu and Sato, Tosiya},
  journal={Statistics in medicine},
  volume={26},
  number={16},
  pages={3188--3204},
  year={2007},
  publisher={Wiley Online Library}
}

@article{tissot2021patients,
  title={Patients with history of covid-19 had more side effects after the first dose of covid-19 vaccine},
  author={Tissot, No{\'e}mie and Brunel, Anne-Sophie and Bozon, Fabienne and Rosolen, B{\'e}atrice and Chirouze, Catherine and Bouiller, Kevin},
  journal={Vaccine},
  volume={39},
  number={36},
  pages={5087--5090},
  year={2021},
  publisher={Elsevier}
}

@article{dunkle2022efficacy,
  title={Efficacy and safety of NVX-CoV2373 in adults in the United States and Mexico},
  author={Dunkle, Lisa M and Kotloff, Karen L and Gay, Cynthia L and Anez, German and Adelglass, Jeffrey M and Barrat Hernandez, Alejandro Q and Harper, Wayne L and Duncanson, Daniel M and McArthur, Monica A and Florescu, Diana F and others},
  journal={New England Journal of Medicine},
  volume={386},
  number={6},
  pages={531--543},
  year={2022},
  publisher={Mass Medical Soc}
}

@misc{NIH_PREVENT19_2020,
  author       = {{National Institutes of Health}},
  title        = {Phase 3 trial of Novavax investigational COVID-19 vaccine opens},
  year         = {2020},
  month        = dec,
  day          = {28},
  note         = {News release},
  url          = {https://www.nih.gov/news-events/news-releases/phase-3-trial-novavax-investigational-covid-19-vaccine-opens},
  urldate      = {2025-08-29}
}

@article{Chandramohan2021NEJM,
  author  = {Chandramohan, Daniel and others},
  title   = {Seasonal Malaria Vaccination with or without Seasonal Malaria Chemoprevention},
  journal = {New England Journal of Medicine},
  year    = {2021},
  volume  = {385},
  number  = {11},
  pages   = {1005--1017},
  doi     = {10.1056/NEJMoa2026330},
  note    = {States: “All the participating children were given a long-lasting insecticide-treated bed net at the time of enrollment.”}
}

@article{gabriel2022causal,
  title={Causal bounds for outcome-dependent sampling in observational studies},
  author={Gabriel, Erin E and Sachs, Michael C and Sj{\"o}lander, Arvid},
  journal={Journal of the American Statistical Association},
  volume={117},
  number={538},
  pages={939--950},
  year={2022},
  publisher={Taylor \& Francis}
}

@article{gabriel2025elucidating,
  title={Elucidating some common biases in randomized controlled trials using directed acyclic graphs},
  author={Gabriel, Erin E and Ocampo, Alex and Sj{\"o}lander, Arvid},
  journal={European Journal of Epidemiology},
  pages={1--11},
  year={2025},
  publisher={Springer}
}

@book{ShakedShanthikumar2007,
  title={Stochastic Orders},
  author={Shaked, Moshe and Shanthikumar, J. George},
  year={2007},
  publisher={Springer},
  address={New York},
}

\newpage
\appendix
\numberwithin{table}{section}   
\numberwithin{figure}{section}   

\section*{Description of contents in the appendix}
 \begin{itemize}

\item In Appendix~\ref{app:diff_scale}, we present the LP-based and monotonicity-based bounds for the vaccine effects on the difference scale, as well as for $VE_M(a)$ for $a = 0,1$. In Appendix~\ref{app:diff.2}, we present results corresponding to Figures~\ref{fig:sub.viol.1} and~\ref{fig:sub.viol.4}--\ref{fig:DAG_with_s_violation}. In Appendix~\ref{app:diff_scale_figure_3}, we present results for Figure~\ref{fig:sub.viol.3}.

\item In Appendix~\ref{app:NPSEM_general}, we provide three proofs based on the NPSEM-IE model:
\begin{enumerate}
    \item In Appendix~\ref{app:NPSEM}, we provide a proof for Equation~\eqref{Eq:POidentBm}.
    \item In Appendix~\ref{app:indepndece.proof}, we provide a proof that $Y_i^{a,m}\perp_{\mathcal{T}_{VI}} A_i, S_i^{a,m=-1}$ under Figure~\ref{fig:sub.viol.3}.
    \item In Appendix~\ref{app:NPSEM_S}, we provide a proof that $\Pr(Y_i^{a',m}=Y_i^{a',m=-1}\mid A_i=a, B_i^{a',m=-1}=m, S_i^{a',m=-1}=s,U_i=u)=1$ under Figure~\ref{fig:violation.assump.4}.
\end{enumerate}

\item In Appendix~\ref{app:LP1}, we provide a detailed description of the linear programming (LP) method used to derive bounds for the VE parameters. This includes explicit formulations of the optimization problems, the construction of all relevant vectors and matrices, and the workflow for computing the max/min values of each parameter. 

\item In Appendix~\ref{app:non.LP.optimizer}, we provide a GitHub link for our program that calculates the LP-bounds using nonlinear optimization methods. This program maximizes and minimizes the VE expressions over all feasible values, allowing easy incorporation of additional constraints.

\item In Appendix~\ref{app:sharp.LP.bounds}, we describe how incorporating monotonicity assumptions in either $A$ or $M$ can lead to narrower LP-based bounds in Propositions~\ref{prop:LP1} and~\ref{prop:LP1_withS}. In Appendix~\ref{app:LP.sharp.assum.and_method}, we present the monotonicity assumptions required to refine the LP-based bounds and explain how these assumptions are translated into the potential response–type framework. In Appendix~\ref{app:LP.sharp.code}, we briefly explain how we implemented these assumptions in the LP procedure code. In Appendix~\ref{app:LP.sharp.2}, we present the LP-based bounds of Proposition~\ref{prop:LP1} after incorporating the additional assumptions. In Appendix~\ref{app:LP.sharp.3a}, we present the corresponding LP-based bounds of Proposition~\ref{prop:LP1_withS}.

\item In Appendix~\ref{app:eq1}, we show how the VE can be bounded without using LP methods. We present a step-by-step derivation based on the law of total probability, demonstrating that these constructive bounds are equivalent to those obtained via the LP-based approach.

\item In Appendix~\ref{app:proof of prop.bounds}, we provide a detailed derivation of the monotonicity-based bounds under Figure~\ref{fig:sub.viol.1}. In Appendix~\ref{app:proof_mon}, we prove that $p_{1.a1}\geq E(Y^{a,m=1})$. Based on this result, in Appendix~\ref{app:mon.one}, we derive the one-sided monotonicity-based bounds. 
Finally, in Appendix~\ref{app:mon.one.reversed}, we derive the monotonicity-based bounds under Assumptions~\ref{assump:pos1} and~\ref{assump:mon_in_U_general}\ref{assump:mon_in_U_general_2}.

\item In Appendix~\ref{app:proof.two.bounds}, we derive two-sided monotonicity-based bounds under Assumptions~\ref{assump:mon_in_U_general} and~\ref{assump:non.negative.m.general}. In Appendix~\ref{app:proof.two.bounds}, we prove the bounds under Assumptions~\ref{assump:mon_in_U_general}\ref{assump:mon_in_U_general_1} and~\ref{assump:non.negative.m.general}\ref{assump:non.negative.m}. 
In Appendix~\ref{app:prop2_reversed}, we derive bounds under Assumptions~\ref{assump:mon_in_U_general}\ref{assump:mon_in_U_general_2} and~\ref{assump:non.negative.m.general}\ref{assump:negative.m}.

\item In Appendix~\ref{app:LP2}, we outline the main results that lead to the LP-based bounds in Proposition~\ref{prop:LP1_withS}. We show that the structure of the response types and constraints under Figure~\ref{fig:sub.viol.3} mirrors those under Figure~\ref{fig:sub.viol.1}. 

\item In Appendix~\ref{app:proof:prop:bounds.mon.s.iv}, we provide the proof for Proposition~\ref{prop:bounds.mon.s.iv}: monotonicity-based bounds under Figure~\ref{fig:sub.viol.3}. In Appendix~\ref{app:mon.3a.one}, we derive the one-sided monotonicity-based bounds based only on Assumption~\ref{assump:U.mon.S.general}. In Appendix~\ref{app:mon.3a.two}, we derive the two-sided monotonicity-based bounds based on Assumptions~\ref{assump:U.mon.S.general} and~\ref{assump:non.negative.m.general}.

\item In Appendix~\ref{app:LP_bounds_fig4}, we provide the proof of Proposition~\ref{prop:bounds.4ab.LP}: LP-based bounds under Figures~\ref{fig:sub.viol.3}--\ref{fig:DAG_with_s_violation}. We outline the main LP method steps used to establish the bound in Proposition~\ref{prop:bounds.4ab.LP} and show that the resulting LP-based bounds under Figures~\ref{fig:sub.viol.3}--\ref{fig:DAG_with_s_violation} coincide with those obtained under Figure~\ref{fig:sub.viol.1}.

\item In Appendix~\ref{app:proof.prop.bounds2}, we provide the proof of Proposition~\ref{prop:bounds.4ab.mon}: monotonicity-based bounds under Figures~\ref{fig:sub.viol.3}--\ref{fig:DAG_with_s_violation}. We show that even though $S$ is observed, the bounds for $E(Y^{a,m})$ do not depend on $S$ and coincide with those obtained under Figure~\ref{fig:sub.viol.1}.

\item In Appendix~\ref{app:two_u}, we present Figure~\ref{fig:DAG_violations_two_U}, which includes two unmeasured common causes, as briefly described in Appendix~\ref{sec:viol.Y.S.dis}. We also show that Figure~\ref{fig:DAG_violations_two_U} is equivalent to Figures~\ref{fig:sub.viol.4}--\ref{fig:DAG_with_s_violation}, where $Y \perp_{IV} M \mid A, S, B, U_i$ for $i \in \{1,2\}$.
\item In Appendix~\ref{app:sharpening_bounds}, we present two scenarios for the covariate $L$. In Appendix \ref{sec:covariates} we provide an overview of how $L$ can be used to further narrow the LP-based and monotonicity-based bounds under suitable assumptions. In Appendix~\ref{app:sharpening_bounds_L_good}, we present Figure~\ref{fig:L_adjustment_main}, which illustrates cases in which $L$ can be used to sharpen the bounds. We then provide the formulas and a sketch of the proofs for both the LP-based and monotonicity-based bounds. We distinguish between two sub-scenarios:  
(1) when $S$ is unobserved or cannot be used to obtain narrower bounds, and  
(2) when $S$ can be used to obtain narrower bounds the bounds.  
Finally, in Appendix~\ref{app:sharpening_bounds_L_bad}, we present examples of $L$ that cannot be used to obtain narrower bounds; each example is also shown in Figure~\ref{fig:L_adjustment}.
\item In Appendix~\ref{app:sim_mon}, we provide additional results for the numerical example. In Figures~\ref{fig:app:M_mon_assump}--\ref{fig:app:SY_SU}, we present the bounds under misspecification of the underlying assumptions or the assumed DAG. In Tables~\ref{app:tab:coverage_lp}--\ref{app:tab:coverage_mon}, we show the empirical coverage of the bootstrap CIs.

\item In Appendix~\ref{app:real.data.example}, we present additional results for the real data example. In Figure~\ref{app:fig:bounds_under_figure3}, we present the results under Figure~\ref{fig:sub.viol.3}. In Tables~\ref{app:tab:bounds-figure4}--\ref{app:tab:bounds-figure3}, we show the bounds with 95\% bootstrap CIs under Figures~\ref{fig:DAG_with_s_violation} and~\ref{fig:sub.viol.3}, respectively.

\end{itemize}

\newpage
\section{The behavioral effects on the VE scale, and VE contrasts on the difference scale} \label{app:diff_scale}
This section presents all the bounds for the causal contrasts on the difference scale. The behavioral effect ($E(Y^{a,m=1})-E(Y^{a,m=0})$), the immunological effect ($E(Y^{a=1,m})-E(Y^{a=0,m})$) and the total effect ($E(Y^{a=1,m=1})-E(Y^{a=0,m=0})$). In addition we present the results for the behavioral effect on the VE scale, $VE_M(a)$. Proofs are not included here, as they follow directly from the original propositions in the main text for $VE(m)$ (for $m \in {0,1}$) and $VE_T$ on the VE scale. For example, for the proof of the LP-bound under Figure~\ref{fig:sub.viol.1}, see Proposition~\ref{prop:LP1} in Appendix Section~\ref{app:LP1}.

\subsection{Bounds for vaccine effects under Figures \ref{fig:sub.viol.1} and \ref{fig:sub.viol.4}--\ref{fig:DAG_with_s_violation}}\label{app:diff.2}

\subsubsection{LP-based bounds}
\begin{Proposition}\label{prop:app:LP_bounds_figure_2}
Consider the setting of Figures ~\ref{fig:sub.viol.1} and \ref{fig:sub.viol.4}--\ref{fig:DAG_with_s_violation}. 
The following bounds hold. 

\textbf{Behavioral effects.} For $a=1$,
$$
\begin{aligned}
  -p_{10.1} - p_{01.1}
  &\leq E(Y^{a=1,m=1}) - E(Y^{a=1,m=0})
  &\leq 1 - p_{10.1} - p_{01.1}.
\end{aligned}
$$

For $a=0$,
$$
\begin{aligned}
  -p_{10.0} - p_{01.0}
  &\leq E(Y^{a=0,m=1}) - E(Y^{a=0,m=0})
  &\leq 1 - p_{10.0} - p_{01.0}.
\end{aligned}
$$

\textbf{Immunological effects.} For $m=0$,
$$
\begin{aligned}
  -1 + p_{00.0} + p_{01.1}
  &\leq E(Y^{a=1,m=0}) - E(Y^{a=0,m=0})
  &\leq 1 - p_{00.1} - p_{01.0}.
\end{aligned}
$$

For $m=1$,
$$
\begin{aligned}
  -p_{00.1} + p_{10.0} - p_{10.1} - p_{01.1}
  &\leq E(Y^{a=1,m=1}) - E(Y^{a=0,m=1})
  &\leq p_{00.0} + p_{10.0} - p_{10.1} + p_{01.0}.
\end{aligned}
$$

\textbf{Total effect.}
$$
\begin{aligned}
  p_{00.0} - p_{00.1} - p_{10.1} - p_{01.1}
  &\leq E(Y^{a=1,m=1}) - E(Y^{a=0,m=0})
  &\leq 1 - p_{10.1} - p_{01.0}.
\end{aligned}
$$

\textbf{$\mathbf{VE_M(a)}$.} For $a \in \{0,1\}$,
$$
1 - \frac{1 - p_{10.1}}{p_{01.1}}
  \leq \mathrm{VE}_M(1)
  \leq 1 - \frac{1 - p_{00.1} - p_{10.1} - p_{01.1}}{1 - p_{00.1}},
$$
$$
1 - \frac{1 - p_{10.0}}{p_{01.0}}
  \leq \mathrm{VE}_M(0)
  \leq 1 - \frac{1 - p_{00.0} - p_{10.0} - p_{01.0}}{1 - p_{00.0}}.
$$
\end{Proposition}

\subsubsection{Monotonicity-based bounds}

\begin{Proposition}\label{prop:app:mon_bounds_figure_2}
Consider the setting of Figures~\ref{fig:sub.viol.1} and \ref{fig:sub.viol.4}--\ref{fig:DAG_with_s_violation}. 
Under Assumptions~\ref{assump:pos1},~\ref{assump:mon_in_U_general}\ref{assump:mon_in_U_general_1}, 
and~\ref{assump:non.negative.m.general}\ref{assump:non.negative.m}, 
the following bounds for the vaccine effect hold. 

\textbf{Behavioral effects.} For $a \in \{0,1\}$,
$$
\begin{aligned}
  0 
  &\leq E(Y^{a,m=1}) - E(Y^{a,m=0})
  &\leq p_{1.a1} - p_{1.a0}.
\end{aligned}
$$

\textbf{Immunological effects.} For $m=1$,
$$
\begin{aligned}
  p_{1.1} - p_{1.01} 
  &\leq E(Y^{a=1,m=1}) - E(Y^{a=0,m=1})
  &\leq p_{1.11} - p_{1.0},
\end{aligned}
$$

and for $m=0$,
$$
\begin{aligned}
  p_{1.10} - p_{1.0}
  &\leq E(Y^{a=1,m=0}) - E(Y^{a=0,m=0})
  &\leq p_{1.1} - p_{1.10}.
\end{aligned}
$$

\textbf{Total effect.}
$$
\begin{aligned}
  p_{1.1} - p_{1.0}
  &\leq E(Y^{a=1,m=1}) - E(Y^{a=0,m=0})
  &\leq p_{1.11} - p_{1.10}.
\end{aligned}
$$

\textbf{$\mathbf{VE_M(a)}$.} For $a \in \{0,1\}$,
$$
\begin{aligned}
  1 - \frac{p_{1.a1}}{p_{1.a0}}
  &\leq \mathrm{VE}_M(a)
  &\leq 0.
\end{aligned}
$$
\end{Proposition}

\subsection{Bounds under Figure \ref{fig:sub.viol.3}}\label{app:diff_scale_figure_3}

\subsubsection{LP-based bounds}
\begin{Proposition}\label{prop:app:LP_bounds_figure_3}
Consider the setting of Figure~\ref{fig:sub.viol.3}. 
The following bounds for the vaccine effect hold. 

\textbf{Behavioral effects.} For $a \in \{0,1\}$,
$$
\begin{aligned}
  \operatorname*{max}\limits_{d_1,d_2\in\{0,1\}}
  \bigl\{-p_{10.a d_1} - p_{01.a d_2}\bigr\}
  &\leq E(Y^{a,m=1}) - E(Y^{a,m=0})
  &\leq 
  \operatorname*{min}\limits_{d_1,d_2\in\{0,1\}}
  \bigl\{1 - p_{10.a d_1} - p_{01.a d_2}\bigr\}.
\end{aligned}
$$

\textbf{Immunological effects.} For $m = 0$,
$$
\begin{aligned}
  \operatorname*{max}\limits_{d_1,d_2\in\{0,1\}}
  \bigl\{-1 + p_{00.0 d_1} + p_{01.1 d_2}\bigr\}
  &\leq E(Y^{a=1,m=0}) - E(Y^{a=0,m=0})
  &\leq 
  \operatorname*{min}\limits_{d_1,d_2\in\{0,1\}}
  \bigl\{1 - p_{00.1 d_1} - p_{01.0 d_2}\bigr\},
\end{aligned}
$$

and for $m = 1$,
$$
\begin{aligned}
  \operatorname*{max}\limits_{d_1,d_2\in\{0,1\}}
  \bigl\{-1 + p_{11.1 d_1} + p_{10.0 d_2}\bigr\}
  \leq E(Y^{a=1,m=1}) - E(Y^{a=0,m=1})
  \leq 
  \operatorname*{min}\limits_{d_1,d_2\in\{0,1\}}
  \bigl\{1 - p_{10.1 d_1} - p_{11.0 d_2}\bigr\}.
\end{aligned}
$$

\textbf{Total effect.}
$$
\begin{aligned}
  \operatorname*{max}\limits_{d_1,d_2\in\{0,1\}}
  \bigl\{p_{00.0 d_1} - p_{00.1 d_2} - p_{10.1 d_2} - p_{01.1 d_2}\bigr\}
  &\leq\\ E(Y^{a=1,m=1}) - E(Y^{a=0,m=0})
  &\leq \\
  \operatorname*{min}\limits_{d_1,d_2\in\{0,1\}}&
  \bigl\{1 - p_{10.1 d_1} - p_{01.0 d_2}\bigr\}.
\end{aligned}
$$

\textbf{$\mathbf{VE_M(a)}$.} For $a \in \{0,1\}$,
$$
\begin{aligned}
  \operatorname*{max}\limits_{d_1,d_2\in\{0,1\}}
  \Bigl\{1 - \frac{1 - p_{10.a d_1}}{p_{01.a d_2}}\Bigr\}
  &\leq \mathrm{VE}_M(a)
  &\leq
  \operatorname*{min}\limits_{d_1,d_2\in\{0,1\}}
  \Bigl\{1 - \frac{1 - p_{00.a d_1} - p_{10.a d_1} - p_{01.a d_1}}{1 - p_{00.a d_2}}\Bigr\}.
\end{aligned}
$$
\end{Proposition}

\subsubsection{Monotonicity-based bounds}
 \begin{Proposition}\label{prop:app:mon_bounds_figure_3}
Consider the setting of Figure~\ref{fig:sub.viol.3}. 
Under Assumptions~\ref{assump:pos2},\ref{assump:non.negative.m.general}\ref{assump:non.negative.m} and ~\ref{assump:U.mon.S.general}\ref{assump:U.mon.S.versiona}, 
the following bounds for the vaccine effects hold. 

\textbf{Behavioral effects.} For $a \in \{0,1\}$,
$$
\begin{aligned}
  0
  &\leq E(Y^{a,m=1}) - E(Y^{a,m=0})
  &\leq 
  \operatorname*{min}\limits_{s_1,s_2\in\{0,1\}}
  \bigl\{p_{1.a1s_1} - p_{1.a0s_2}\bigr\}.
\end{aligned}
$$

\textbf{Immunological effects.} For $m = 1$,
$$
\begin{aligned}
  \operatorname*{max}\limits_{s_1\in\{0,1\}}
  \bigl\{p_{1.1} - p_{1.01s_1}\bigr\}
  &\leq E(Y^{a=1,m=1}) - E(Y^{a=0,m=1})
  &\leq
  \operatorname*{min}\limits_{s_2\in\{0,1\}}
  \bigl\{p_{1.11s_2} - p_{1.0}\bigr\},
\end{aligned}
$$

and for $m = 0$,
$$
\begin{aligned}
  \operatorname*{max}\limits_{s_1\in\{0,1\}}
  \bigl\{p_{1.10s_1} - p_{1.0}\bigr\}
  &\leq E(Y^{a=1,m=0}) - E(Y^{a=0,m=0})
  &\leq
  \operatorname*{min}\limits_{s_2\in\{0,1\}}
  \bigl\{p_{1.1} - p_{1.10s_2}\bigr\}.
\end{aligned}
$$

\textbf{Total effect.}
$$
\begin{aligned}
  p_{1.1} - p_{1.0}
  &\leq E(Y^{a=1,m=1}) - E(Y^{a=0,m=0})
  &\leq
  \operatorname*{min}\limits_{s_1,s_2\in\{0,1\}}
  \bigl\{p_{1.11s_1} - p_{1.10s_2}\bigr\}.
\end{aligned}
$$

\textbf{$\mathbf{VE_M(a)}$.} For $a \in \{0,1\}$,
$$
\begin{aligned}
  \operatorname*{max}\limits_{s_0,s_1\in\{0,1\}}
  \Bigl\{1 - \frac{p_{1.a1s_0}}{p_{1.a0s_1}}\Bigr\}
  &\leq \mathrm{VE}_M(a)
  &\leq 0.
\end{aligned}
$$
\end{Proposition}

\normalsize

\newpage

\section{Proofs using the NPSEM-IE model}\label{app:NPSEM_general}
All potential outcomes and variables are written with a subscript $i$, indicating individual-level potential outcomes and variables.

\subsection{ A proof that Assumption \ref{assump:isolation} holds under NPSEM-IE model}\label{app:NPSEM}

Under Figure~\ref{fig:sub.viol.1}, we consider the corresponding NPSEM-IE:
\begin{equation}\label{app:model.NPSEM}
\begin{aligned}
A_i &= f_A(\varepsilon_{A_i}), \\
U_i &= f_U(\varepsilon_{U_i}), \\
M_i &= f_M(\varepsilon_{M_i}),\\
B_i &= h_B(M_i,A_i,U_i,\varepsilon_{B_i})
= \Ind_{\{M_i=-1\}}f_B(A_i,U_i,\varepsilon_{B_i})+\Ind_{\{M_i\neq-1\}}M_i,\\
Y_i &= f_Y(A_i,B_i,U_i,\varepsilon_{Y_i}),
\end{aligned}
\end{equation}
for some functions $f_A, f_U, f_M, f_B,$ and $f_Y$, and where
$\varepsilon_{A_i}, \varepsilon_{U_i}, \varepsilon_{M_i}, 
\varepsilon_{B_i},$ and $\varepsilon_{Y_i}$ are mutually independent 
error terms.

It follows from the model in Equation~\eqref{app:model.NPSEM} that
\begin{align}
&\Pr\Big(Y_i^{a',m}=Y_i^{a',m=-1}\,\Big|\,A_i=a,B_i^{a',m=-1}=m,U_i=u\Big)
\notag\\
&=\Pr\Big(f_Y(a',B_i^{a',m},u,\varepsilon_{Y_i})
    =f_Y(a',B_i^{a',m=-1},u,\varepsilon_{Y_i})
    \,\Big|\,A_i=a,B_i^{a',m=-1}=m,U_i=u\Big)
    \notag\\
&=\Pr\Big(f_Y(a',m,u,\varepsilon_{Y_i})
    =f_Y(a',m,u,\varepsilon_{Y_i})
    \,\Big|\,A_i=a,B_i^{a',m=-1}=m,U_i=u\Big)
    \tag{Step 1}\label{eq:proof-step3}\\
&=\Pr\Big(f_Y(a',m,u,\varepsilon_{Y_i})
    =f_Y(a',m,u,\varepsilon_{Y_i})
    \,\Big|\,B_i^{a',m=-1}=m,U_i=u\Big)
    \tag{Step 2}\label{eq:proof-step4}\\
&=\Pr\Big(f_Y(a',m,u,\varepsilon_{Y_i})
    =f_Y(a',m,u,\varepsilon_{Y_i})
    \,\Big|\,U_i=u\Big)
    \tag{Step 3}\label{eq:proof-step5}\\
&=1,
    \tag{Step 4}\label{eq:proof-step6}
\end{align}
where: \ref{eq:proof-step3} follows from Assumption~\ref{assump:B_M}, which states that for $m\in\{0,1\}$, $B_i^{a',m}=m$, and since we are conditioning on $B_i^{a',m=-1}=m$. \ref{eq:proof-step4} follows because $A_i\!\perp\!\varepsilon_{U_i},\varepsilon_{B_i}$ and thus $A_i\!\perp\!\varepsilon_{Y_i}|B_i^{a',m=-1}=m,U_i=u$. \ref{eq:proof-step5} follows because $B_i^{a',m=-1}\!\perp\!\varepsilon_{Y_i}\,|\,U_i=u$. \ref{eq:proof-step6} holds because, for each individual $i$, the only random component $\varepsilon_{Y_i}$ is the same in both $f_Y(a',m,u,\varepsilon_{Y_i})$ and $f_Y(a',m,u,\varepsilon_{Y_i})$.


\subsection{A proof that $Y_i^{a,m}\perp_{\mathcal{T}_{VI}} A_i,S_i^{a,m=-1}$ under NPSEM-IE model and Figure \ref{fig:sub.viol.3}}\label{app:indepndece.proof}

Under Figure~\ref{fig:sub.viol.3}, we consider the corresponding NPSEM-IE:
\begin{equation}\label{app:model.NPSEM.3a}
\begin{aligned}
A_i &= f_A(\varepsilon_{A_i}), \\
U_i &= f_U(\varepsilon_{U_i}), \\
M_i &= f_M(\varepsilon_{M_i}),\\
S_i &= f_S(A_i,\varepsilon_{S_i}),\\
B_i &= h_B(M_i,S_i,U_i,\varepsilon_{B_i})
= \Ind_{\{M_i=-1\}}f_B(S_i,U_i,\varepsilon_{B_i})+\Ind_{\{M_i\neq-1\}}M_i,\\
Y_i &= f_Y(A_i,B_i,U_i,\varepsilon_{Y_i}),
\end{aligned}
\end{equation}
for some functions $f_A, f_U, f_M, f_S, f_B,$ and $f_Y$, and where 
$\varepsilon_{A_i}, \varepsilon_{U_i}, \varepsilon_{M_i}, 
\varepsilon_{S_i}, \varepsilon_{B_i},$ and $\varepsilon_{Y_i}$ are mutually independent error terms.

From the model in Equation~\eqref{app:model.NPSEM.3a},  for $m\in\{0,1\}$, it follows that 
\begin{align}
&\Pr\Big(Y_i^{a,m}=y\,\Big|\,A_i=a,S_i^{a',m=-1}\Big)
    \notag\\
&=\Pr\Big(f_Y(a,B_i^{a,m},U_i,\varepsilon_{Y_i})=y\,\Big|\,A_i,S_i^{a',m=-1}\Big)
    \notag\\
&=\Pr\Big(f_Y(a,m,U_i,\varepsilon_{Y_i})=y\,\Big|\,A_i,S_i^{a',m=-1}\Big)
    \tag{Step 1}\label{eq2:proof-step3}\\
&=\Pr\Big(f_Y(a,m,U_i,\varepsilon_{Y_i})=y\,\Big|\,S_i^{a',m=-1}\Big)
    \tag{Step 2}\label{eq2:proof-step4}\\
&=\Pr\Big(f_Y(a,m,U_i,\varepsilon_{Y_i})=y\Big)
    \tag{Step 3}\label{eq2:proof-step5}
\end{align}
where \ref{eq2:proof-step3} follows from the fact that under Assumption \ref{assump:B_M} for $m\in\{0,1\}$, $B_i^{a',m}=m$. 
\ref{eq2:proof-step4} follows since $A_i\perp U_i,\varepsilon_{Y_i}|S_i^{a',m=-1}$. 
\ref{eq2:proof-step5} follows since $S_i^{a',m=-1}=f_S(a',\varepsilon_{S_i})$, $U_i= f_U(\varepsilon_{U_i})$ and thus $\varepsilon_{S_i}\perp (\varepsilon_{Y_i},\varepsilon_{U_i})$.

\subsection{A proof that $\Pr(Y_i^{a',m}=Y_i^{a',m=-1}\mid A_i=a, B_i^{a',m=-1}=m, S_i^{a',m=-1}=s,U_i=u)=1$ under NPSEM-IE model and Figure \ref{fig:violation.assump.4}}\label{app:NPSEM_S}

Under Figure~\ref{fig:violation.assump.4}, we consider the corresponding NPSEM-IE:
\begin{equation}\label{app:model.NPSEM.3}
\begin{aligned}
A_i &= f_A(\varepsilon_{A_i}), \\
U_i &= f_U(\varepsilon_{U_i}), \\
M_i &= f_M(\varepsilon_{M_i}),\\
S_i &= h_S(A_i,U_i,\varepsilon_{S_i}),\\
B_i &= h_B(M_i,S_i,U_i,\varepsilon_{B_i})
= \Ind_{\{M_i=-1\}}f_B(S_i,U_i,\varepsilon_{B_i})+\Ind_{\{M_i\neq-1\}}M_i,\\
Y_i &= h_Y(A_i,B_i,S_i,U_i,\varepsilon_{Y_i}),
\end{aligned}
\end{equation}
for some functions $f_A, f_U, f_M, h_S, f_B,$ and $h_Y$, and where
$\varepsilon_{A_i}, \varepsilon_{U_i}, \varepsilon_{M_i}, 
\varepsilon_{S_i}, \varepsilon_{B_i},$ and $\varepsilon_{Y_i}$ are mutually independent error terms. 
The specification of $h_S$ and $h_Y$ differs across Figures~\ref{fig:sub.viol.3}--\ref{fig:DAG_with_s_violation} as follows:
\begin{itemize}
    \item Figure~\ref{fig:sub.viol.3}: $h_S(A_i,U_i,\varepsilon_{S_i})=f_S(A_i,\varepsilon_{S_i})$ and $h_Y(A_i,B_i,S_i,U_i,\varepsilon_{Y_i})=f_Y(A_i,B_i,U_i,\varepsilon_{Y_i})$.
    \item Figure~\ref{fig:sub.viol.4}: $h_S(A_i,U_i,\varepsilon_{S_i})=f_S(A_i,\varepsilon_{S_i})$ and $h_Y(A_i,B_i,S_i,U_i,\varepsilon_{Y_i})=f_Y(A_i,B_i,S_i,U_i,\varepsilon_{Y_i})$.
    \item Figure~\ref{fig:DAG_with_s_violation2}: $h_S(A_i,U_i,\varepsilon_{S_i})=f_S(A_i,U_i,\varepsilon_{S_i})$ and $h_Y(A_i,B_i,S_i,U_i,\varepsilon_{Y_i})=f_Y(A_i,B_i,U_i,\varepsilon_{Y_i})$.
    \item Figure~\ref{fig:DAG_with_s_violation}: $h_S(A_i,U_i,\varepsilon_{S_i})=f_S(A_i,U_i,\varepsilon_{S_i})$ and $h_Y(A_i,B_i,S_i,U_i,\varepsilon_{Y_i})=f_Y(A_i,B_i,S_i,U_i,\varepsilon_{Y_i})$.
\end{itemize}

Now, we show that under all Figures~\ref{fig:sub.viol.3}--\ref{fig:DAG_with_s_violation}, 
$\Pr(Y_i^{a',m}=Y_i^{a',m=-1}\mid A_i=a,B_i^{a',m=-1}=m,S_i^{a',m=-1}=s,U_i=u)=1$. 
The proof is based on the fact that under all these figures, $\Pr(S_i^{a',m=-1}=S_i^{a',m})=1$, because $f_S$ does not depend on the value of $m$. 
We present the detailed proof for Figure~\ref{fig:DAG_with_s_violation}; the proofs for the remaining figures are analogous. 
From the model in Equation~\eqref{app:model.NPSEM.3}, it follows that
\begin{align*}
&\Pr\Big(Y_i^{a',m}=Y_i^{a',m=-1}\,\Big|\,A_i=a,B_i^{a',m=-1}=m,S_i^{a',m=-1}=s,U_i=u\Big)
    \notag\\
&=\Pr\Big(f_Y(a',B_i^{a',m},S_i^{a',m},u,\varepsilon_{Y_i})
    =f_Y(a',B_i^{a',m=-1},S_i^{a',m=-1},u,\varepsilon_{Y_i})\\
&\qquad\qquad
    \Big|\,A_i=a,B_i^{a',m=-1}=m,S_i^{a',m=-1}=s,U_i=u\Big)
    \notag\\
&=\Pr\Big(f_Y(a',m,s,u,\varepsilon_{Y_i})
    =f_Y(a',m,s,u,\varepsilon_{Y_i})
    \,\Big|\,A_i=a,B_i^{a',m=-1}=m,S_i^{a',m=-1}=s,U_i=u\Big)
    \tag{Step 1}\label{eq:proof3-step3}\\
&=\Pr\Big(f_Y(a',m,s,u,\varepsilon_{Y_i})
    =f_Y(a',m,s,u,\varepsilon_{Y_i})
    \,\Big|\,B_i^{a',m=-1}=m,U_i=u\Big)
    \tag{Step 2}\label{eq:proof3-step4}\\
&=\Pr\Big(f_Y(a',m,u,\varepsilon_{Y_i})
    =f_Y(a',m,u,\varepsilon_{Y_i})
    \,\Big|\,U_i=u\Big)
    \tag{Step 3}\label{eq:proof3-step5}\\
&=1
    \tag{Step 4}\label{eq:proof3-step6}
\end{align*}
where \ref{eq:proof3-step3} follows from the fact that (I) under Assumption~\ref{assump:B_M}, for $m\in\{0,1\}$, $B_i^{a',m}=m$ and since we conditioned on $B_i^{a',m=-1}=m$, and (II) $\Pr(S_i^{a',m=-1}=S_i^{a',m})=1$, and since we conditioned on $S_i^{a',m=-1}=s$. 
\ref{eq:proof3-step4} follows since $S_i^{a',m=-1}|\{U_i=u\}= h_S(a',u,\varepsilon_{S_i})$, $A_i=f_A(\varepsilon_{A_i})$ and thus $A_i,S_i^{a',m=-1}\perp \varepsilon_{Y_i}|B_i^{a',m=-1},U_i=u$. 
\ref{eq:proof3-step5} follows since $B_i^{a',m=-1}\perp \varepsilon_{Y_i}\mid U_i=u$. 
\ref{eq:proof3-step6} follows since we assume that the only random component, $\varepsilon_{Y_i}$, is the same for each individual $i$.

\newpage
\section{Proof of Proposition \ref{prop:LP1}: LP-based bounds under Figure \ref{fig:sub.viol.1}} \label{app:LP1}

\subsection{Short description of the general LP method}
In general, the task of bounding a causal estimand can be represented as the following constrained linear program:
\begin{align*}
    \text{minimize/maximize} \; & \mathbf{c}^\top \mathbf{q} \\
    \text{subject to} \; & \mathbf{1}^\top \mathbf{q} = 1 \\
                            & \mathbf{R} \mathbf{q} = \mathbf{p} \\
                            & \mathbf{q} \geq 0
\end{align*}

Here, $\mathbf{q}$ denotes the probability vector over the possible response types, $\mathbf{c}^\top \mathbf{q}$ encodes the target causal functional, and $\mathbf{R}\mathbf{q}=\mathbf{p}$ captures the linear constraints implied by the observed data. The normalization and nonnegativity constraints ensure that $\mathbf{q}$ defines a valid probability distribution. The optimal value of this linear program yields the lower (or upper) bound for the causal effect of interest under the given assumptions. In the following section, we explicitly construct the vectors $\mathbf{q}$, $\mathbf{c}$, $\mathbf{p}$, and the matrix $\mathbf{R}$ for each setting of interest.

\subsection{List of all possible response patterns}
Let $r_{y|a,b}$ denote the potential response type of $Y^{a,m}$ given $A=a$ and $B^{a,m=-1}=b$. For clarity, when conditioning on $A=a$, we write $B^{a,m=-1}$ simply as $B$.
Without any additional assumptions, since $a\in{0,1}$ and $m\in{-1,0,1}$, there are $2^{2\times 3}=64$ possible response types. However, from Equation~\eqref{Eq:POidentBm}, we have that for each participant $i$ in the trial,
\begin{equation}
\Pr(Y_i^{a,m}=Y_i^{a,m=-1}\mid A_i=a,B_i=m,U_i=u)=1
\end{equation}
for $a,m\in{0,1}^2$.
Since this equality is required to hold at the individual level $i$ (see Section~\ref{sec:motivation}), where each individual has a fixed value of $U_i$, we can omit the condition $U_i=u$ from Equation~\eqref{Eq:POidentBm}. Consequently, the number of possible response types is reduced to 16. The following tables list all $r_{y|a,b}$ values for $a,b\in\{0,1\}^2$.
\begin{table}[!p]
  \centering
  \caption{$r_{y|a,b=0}$ response patterns}
    \begin{tabular}{|r|r|r|r|r|r|r|}
\cmidrule{2-7}    \multicolumn{1}{r|}{} & \multicolumn{6}{c|}{(a,m)} \\
    \midrule
    \multicolumn{1}{|l|}{$r_{y|a,b=0}$} & \multicolumn{1}{l}{$(0,0)$} & \multicolumn{1}{l}{$(0,1)$} & \multicolumn{1}{l}{$(1,0)$} & \multicolumn{1}{l}{$(1,1)$} & \multicolumn{1}{l}{$(0,-1)$} & \multicolumn{1}{l|}{$(1,-1)$} \\
    \midrule
    0     & 0     & 0     & 0     & 0     & 0     & 0 \\
    1     & 1     & 0     & 0     & 0     & 1     & 0 \\
    2     & 0     & 1     & 0     & 0     & 0     & 0 \\
    3     & 0     & 0     & 1     & 0     & 0     & 1 \\
    4     & 0     & 0     & 0     & 1     & 0     & 0 \\
    5     & 1     & 1     & 0     & 0     & 1     & 0 \\
    6     & 1     & 0     & 1     & 0     & 1     & 1 \\
    7     & 1     & 0     & 0     & 1     & 1     & 0 \\
    8     & 0     & 1     & 1     & 0     & 0     & 1 \\
    9     & 0     & 1     & 0     & 1     & 0     & 0 \\
    10    & 0     & 0     & 1     & 1     & 0     & 1 \\
    11    & 1     & 1     & 1     & 0     & 1     & 1 \\
    12    & 1     & 1     & 0     & 1     & 1     & 0 \\
    13    & 1     & 0     & 1     & 1     & 1     & 1 \\
    14    & 0     & 1     & 1     & 1     & 0     & 1 \\
    15    & 1     & 1     & 1     & 1     & 1     & 1 \\
    \bottomrule
    \end{tabular}%
  \label{tab:potential.responses.b0}%
\end{table}%
\begin{table}[!p]
  \centering
  \caption{$r_{y|a,b=1}$ response patterns}
    \begin{tabular}{|r|r|r|r|r|r|r|}
\cmidrule{2-7}    \multicolumn{1}{r|}{} & \multicolumn{6}{c|}{(a,m)} \\
    \midrule
    \multicolumn{1}{|l|}{$r_{y|a,b=1}$} & \multicolumn{1}{l}{$(0,0)$} & \multicolumn{1}{l}{$(0,1)$} & \multicolumn{1}{l}{$(1,0)$} & \multicolumn{1}{l}{$(1,1)$} & \multicolumn{1}{l}{$(0,-1)$} & \multicolumn{1}{l|}{$(1,-1)$} \\
    \midrule
0     & 0     & 0     & 0     & 0     & 0     & 0 \\
    1     & 1     & 0     & 0     & 0     & 0     & 0 \\
    2     & 0     & 1     & 0     & 0     & 1     & 0 \\
    3     & 0     & 0     & 1     & 0     & 0     & 0 \\
    4     & 0     & 0     & 0     & 1     & 0     & 1 \\
    5     & 1     & 1     & 0     & 0     & 1     & 0 \\
    6     & 1     & 0     & 1     & 0     & 0     & 0 \\
    7     & 1     & 0     & 0     & 1     & 0     & 1 \\
    8     & 0     & 1     & 1     & 0     & 1     & 0 \\
    9     & 0     & 1     & 0     & 1     & 1     & 1 \\
    10    & 0     & 0     & 1     & 1     & 0     & 1 \\
    11    & 1     & 1     & 1     & 0     & 1     & 0 \\
    12    & 1     & 1     & 0     & 1     & 1     & 1 \\
    13    & 1     & 0     & 1     & 1     & 0     & 1 \\
    14    & 0     & 1     & 1     & 1     & 1     & 1 \\
    15    & 1     & 1     & 1     & 1     & 1     & 1 \\
    \bottomrule
    \end{tabular}%
  \label{tab:potential.responses.b1}%
\end{table}%
We highlight the following two key characteristics of Tables~\ref{tab:potential.responses.b0} and~\ref{tab:potential.responses.b1}.
\begin{itemize}
\item From Equation~\eqref{Eq:POidentBm}, in Table~\ref{tab:potential.responses.b0}, the columns corresponding to $(a=0,m=-1)$ and $(a=1,m=-1)$ are identical to those of $(a=0,m=0)$ and $(a=1,m=0)$, respectively. Analogously, in Table~\ref{tab:potential.responses.b1}, the columns of $(a=0,m=-1)$ and $(a=1,m=-1)$ are identical to those of $(a=0,m=1)$ and $(a=1,m=1)$, respectively. This is, technically, how the number of $r_{y|a,b}$ response types is reduced from 64 to 16.
\item The value of $a$ does not affect the table rows; thus, Tables~\ref{tab:potential.responses.b0} and~\ref{tab:potential.responses.b1} hold for both $a=0$ and $a=1$.
\end{itemize}

\subsection{The target parameter VE, its building blocks and the workflow of the LP method in our context}\label{app:vE0}
Our objective is to compute bounds for vaccine efficacy (VE) parameters, such as $VE(0)$. In this section, we detail the derivation of the upper bound for $VE(0)$. The procedures for obtaining the lower bound of $VE(0)$, as well as the bounds for $VE(1)$ and $VE_T$, follow analogously.

The parameter $VE(0)$ can be written as
\begin{align*}
VE(0)
&= 1 -
  \frac{
    \pi_{0.1}\,E(Y^{a=1,m=0}\mid A=1, B=0)
    + \pi_{1.1}\,E(Y^{a=1,m=0}\mid A=1, B=1)
  }{
    \pi_{0.0}\,E(Y^{a=0,m=0}\mid A=0, B=0)
    + \pi_{1.0}\,E(Y^{a=0,m=0}\mid A=0, B=1)
  }.
\end{align*}

To compute the bounds for $VE(0)$, we proceed as follows:
\begin{enumerate}
    \item For each term in the numerator and denominator, we formulate a separate linear program (LP) to maximize or minimize the relevant expectation, subject to the constraints imposed by the observed data and structural assumptions.
    \item Specifically, we:
    \begin{enumerate}\label{eq:first.LP}
        \item Minimize $E(Y^{a=1,m=0}\mid A=1, B=0)$;
        \item Minimize $E(Y^{a=1,m=0}\mid A=1, B=1)$;
        \item Maximize $E(Y^{a=0,m=0}\mid A=0, B=0)$;
        \item Maximize $E(Y^{a=0,m=0}\mid A=0, B=1)$.
    \end{enumerate}
    Each of these LPs identifies the extreme value of its target expectation, given the observed data and model assumptions.
    \item We then substitute the extremal values obtained from these LPs into the formula for $VE(0)$. The combinations of these extremal values yield the upper and lower bounds on $VE(0)$ consistent with the data and assumptions.
\end{enumerate}

In summary, for each component of the $VE(0)$ formula, we construct a separate LP characterized by its own set of variables ($\mathbf{q}$), objective coefficients ($\mathbf{c}$), observed data ($\mathbf{p}$), and constraint matrix ($\mathbf{R}$). In the next section, we specify the construction of these elements for each LP problem.

\subsection{The target parameter in each LP problem}

For each $i\in \{0,...,15\}$, let $q_{i.ab}=\Pr(r_{y|a, b}=i)$ denote the proportion of the two-arm trial participants with potential response type $i$, among those with $A=a$ and $B=b$. Finally, let $Q_{a,b}(a,m)=\big\{r_{y|a, b}:\{Y^{a,m}|A=a,B=b\big\}=1\}$ be the set of potential response types such that with probability one, the potential outcome $Y^{a,m}$ equals to one conditionally on $A=a$ and $B=b$. 

Using this notation, the conditional expectation $E(Y^{a,m} \mid A=a, B=b)$ can be expressed as the sum of $q_{i.ab}$:
\[
E(Y^{a,m} \mid A=a, B=b) = \sum_{i \in Q_{a,b}} q_{i.ab}.
\]

For example, to compute $E(Y^{a=1,m=0} \mid A=1, B=0)$, the relevant set is 
$Q_{1,0}(1, 0) = \{3, 6, 8, 10, 11, 13, 14, 15\}$, so
\[
E(Y^{a=1,m=0} \mid A=1, B=0) = \sum_{i \in \{3, 6, 8, 10, 11, 13, 14, 15\}} q_{i.10}.
\]

This can be written in vector form as
\[
E(Y^{a=1,m=0} \mid A=1, B=0) = \mathbf{c}^\top \mathbf{q},
\]
where $\mathbf{q} = (q_{0.10}, q_{1.10}, \ldots, q_{15.10})^\top$ and 
$\mathbf{c} = (0, 0, 0, 1, 0, 0, 1, 0, 1, 0, 1, 1, 0, 1, 1, 1)$.

In each LP problem, the target parameter (i.e., the objective function) is therefore represented as a linear combination of the relevant $q_{i.ab}$, with the coefficients in $\mathbf{c}$ indicating which response types contribute to the expectation of interest.

\subsection{The constraints in each LP problem}

Using the response types, the constraints derived from the observed probabilities $p_{y.ab}$ can be written as
\[
p_{y.ab} = \sum_{i \in Q_{a,b}(a, m=-1)} q_{i.ab}.
\]
Additional constraints are implied by the fact that the $q_{i.ab}$ represent probabilities that sum to one over all potential response types. For example, in the first LP problem (for $(A=1, B=0)$) in Equation~\eqref{eq:first.LP}, the following constraints are imposed:
\begin{enumerate}
    \item $p_{0.10} = \sum_{i \in \{0,1,2,4,5,7,9,12\}} q_{i.10}$;
    \item $p_{1.10} = \sum_{i \in \{3,6,8,10,11,13,14,15\}} q_{i.10}$;
    \item $\sum_{i=0}^{15} q_{i.10} = 1$;
    \item $q_{i.10} \geq 0$ for $i = 0, \ldots, 15$.
\end{enumerate}

The corresponding constraint matrix $\mathbf{R}$ and vector $\mathbf{p}$ for the first two (probability) constraints can be written explicitly as:
\[
\mathbf{R} = 
\left(
\begin{array}{cccccccccccccccc}
1 & 1 & 1 & 0 & 1 & 1 & 0 & 1 & 0 & 1 & 0 & 0 & 1 & 0 & 0 & 0 \\
0 & 0 & 0 & 1 & 0 & 0 & 1 & 0 & 1 & 0 & 1 & 1 & 0 & 1 & 1 & 1
\end{array}
\right),
\;
\mathbf{p} =
\begin{pmatrix}
p_{0.10} \\
p_{1.10}
\end{pmatrix}.
\]

Here, $\mathbf{R}$ is a $2 \times 16$ matrix corresponding to the constraints for $Y=0$ and $Y=1$, respectively, and $\mathbf{p}$ is the vector of observed probabilities for $(A=1, B=0)$. The normalization and nonnegativity constraints are enforced separately.

The explicit construction of the vectors $\mathbf{q}$, $\mathbf{c}$, $\mathbf{p}$, and the matrix $\mathbf{R}$ for each target parameter can be found in the function \textit{compute\_bounds} in the author's GitHub repository: \url{https://github.com/xlrod1/vaccine_for_publication}.

\subsection{Solving the LP problem}
The optimal (minimum and maximum) values of each LP problem are computed using the \texttt{rcdd} package in~R, which implements exact LP via double description methods. 
All code for setting up and solving the LPs, as well as reproducing the results presented in this paper, is available in the author's GitHub repository: \url{https://github.com/xlrod1/vaccine_for_publication}.

\subsection{Finding the extremal values of VE}

Following the workflow described earlier and using the accompanying code, we obtain the following extremal values for the relevant conditional expectations:
\begin{itemize}
    \item $\min\!\left\{E(Y^{a=1,m}\mid A=1,B=m)\right\} = p_{1.10}$;
    \item $\min\!\left\{E(Y^{a=1,m=0}\mid A=1,B=1)\right\} = 0$;
    \item $\max\!\left\{E(Y^{a=0,m=0}\mid A=0,B=0)\right\} = p_{1.00}$;
    \item $\max\!\left\{E(Y^{a=0,m=0}\mid A=0,B=1)\right\} = 1$.
\end{itemize}
Thus, for example, the upper bound for $VE(0)$ is given by
\[
1 - 
\frac{
  \pi_{0.1}\,p_{1.10} + \pi_{1.1}\cdot 0
}{
  \pi_{0.0}\,p_{1.00} + \pi_{1.0}\cdot 1
},
\]
which coincides with the result stated in Proposition~\ref{prop:LP1}:
\[
VE(0) \leq 1 - \frac{p_{11.0}}{1 - p_{00.0}}.
\]

\newpage

\section{Non-linear optimization for bounding VE}\label{app:non.LP.optimizer}

In addition to the LP approach, we also computed bounds for the VE parameters using direct nonlinear optimization. Specifically, we used the function \\
\texttt{bounds\_optimized\_VE\_no\_s} (available at \url{https://github.com/xlrod1/vaccine_for_publication/tree/main/Non_linear_optimization}) to numerically maximize and minimize the VE expressions over the relevant variables. This approach allows optimization of the VE parameters directly and enables the incorporation of additional, easily specified constraints (such as monotonicity or, for example, $\mathrm{E(Y^{a=1,m=0}\mid A=1,B=0)} - \mathrm{E(Y^{a=0,m=0}\mid A=1,B=0)} \leq k$). As mentioned in Section~\ref{sec:3a.LP}, this approach is also applicable to computing $VE_M(a)$. The optimization was performed using the \texttt{nloptr} and \texttt{optim} packages in~R, ensuring robust and flexible computation of the bounds under both standard and user-defined constraints.

A running example demonstrating this workflow is available in the GitHub repository. Importantly, the results obtained via this nonlinear optimization approach exactly match those derived analytically using the LP formulas.

\newpage

\section{Obtaining narrower LP-based bounds}\label{app:sharp.LP.bounds}

In this appendix, we describe how incorporating monotonicity assumptions in either $A$ or $M$ can be used to obtain narrower LP-based bounds. First, we present the monotonicity assumptions required to refine the LP-based bounds. These assumptions resemble Assumption~\ref{assump:non.negative.m.general}, but they are formulated at the individual level and hold conditional on $A$ and $B$. Second, we explain how these assumptions are translated into the potential response-type framework and how they are implemented in the LP procedure. Finally, we present the resulted LP-based bounds corresponding to Propositions~\ref{prop:LP1} and~\ref{prop:LP1_withS}, obtained under the monotonicity assumption on $M$. For the accompanying code and illustrative examples demonstrating the resulting bounds under various constraints, see \url{https://github.com/xlrod1/vaccine_for_publication}.

\subsection{Refining the LP-based bounds in Proposition \ref{prop:LP1}}\label{app:LP.sharp.assum.and_method}

\subsubsection{\texorpdfstring{Monotonicity in \(M\)}{Monotonicity in M}}

\begin{Assumption}\label{assump:monotonic.m}
The effect of $M$ on $Y$ is monotonic. Specifically, for all $a, b, u$:
\begin{enumerate}
    \item \label{assump:monotonic.m.pos} \textbf{Non-negative effect of $M$:}
    \[
        \Pr(Y^{a,m=0} \leq Y^{a,m=1} \mid A=a, B=b, U=u) = 1;
    \]
    \item \label{assump:monotonic.m.pos.neg} \textbf{Non-positive effect of $M$:}
    \[
        \Pr(Y^{a,m=1} \leq Y^{a,m=0} \mid A=a, B=b, U=u) = 1.
    \]
\end{enumerate}
\end{Assumption}

Under Assumption~\ref{assump:monotonic.m}.\ref{assump:monotonic.m.pos}, the following response types must be assigned zero probability for all $a, b$:
\[
q_{1.ab} = q_{3.ab} = q_{6.ab} = q_{7.ab} = q_{8.ab} = q_{11.ab} = q_{13.ab} = 0,
\]
since these types violate the condition $Y^{a,m=0} \leq Y^{a,m=1}$ for all $a, b$.

Conversely, under Assumption~\ref{assump:monotonic.m}.\ref{assump:monotonic.m.pos.neg}, the following response types must be assigned zero probability for all $a, b$:
\[
q_{2.ab} = q_{4.ab} = q_{7.ab} = q_{8.ab} = q_{9.ab} = q_{12.ab} = q_{14.ab} = 0,
\]
because these types violate the condition $Y^{a,m=1} \leq Y^{a,m=0}$ for all $a, b$.

\subsubsection{\texorpdfstring{Monotonicity in \(A\) }{Monotonicity in A}}

\begin{Assumption}\label{assump:monotonic.a}
The effect of $A$ on $Y$ is monotonic. Specifically, for all $m, b, u$:
\begin{enumerate}
    \item \label{assump:monotonic.a.pos} \textbf{Non-negative effect of $A$:}
    \[
        \Pr(Y^{a=0,m} \leq Y^{a=1,m} \mid A=a, B=b, U=u) = 1;
    \]
    \item \label{assump:monotonic.a.neg} \textbf{Non-positive effect of $A$:}
    \[
        \Pr(Y^{a=1,m} \leq Y^{a=0,m} \mid A=a, B=b, U=u) = 1.
    \]
\end{enumerate}
\end{Assumption}

Under Assumption~\ref{assump:monotonic.a}.\ref{assump:monotonic.a.pos}, the following response types must be assigned zero probability for all $a, b$:
\[
q_{1.ab} = q_{2.ab} = q_{5.ab} = q_{7.ab} = q_{8.ab} = q_{11.ab} = q_{12.ab} = 0,
\]
since these types violate the condition $Y^{a=0,m} \leq Y^{a=1,m}$ for all $a, b$.

Similarly, under Assumption~\ref{assump:monotonic.a}.\ref{assump:monotonic.a.neg}, the following response types must be assigned zero probability for all $a, b$:
\[
q_{3.ab} = q_{4.ab} = q_{7.ab} = q_{8.ab} = q_{10.ab} = q_{13.ab} = q_{14.ab} = 0,
\]
because these types violate the condition $Y^{a=1,m} \leq Y^{a=0,m}$ for all $a, b$.

\subsection{Implementation in Code}\label{app:LP.sharp.code}

These zero constraints are incorporated directly into the LP formulation by setting the corresponding entries in the probability vector $\mathbf{q}$ to zero. In practice, this is implemented in the \texttt{compute\_bounds} function (see \url{https://github.com/xlrod1/vaccine_for_publication}) by identifying the indices associated with the excluded response types (as dictated by the monotonicity assumption) and adding equality constraints $\mathbf{q}_i = 0$ for those entries. The LP problem, augmented with these additional constraints, is then solved using the \texttt{rcdd} package in~R.

More generally, any inequality constraint of the form $Y^{a_1,m_1} \leq Y^{a_2,m_2}$ can be imposed in the LP framework. For greater flexibility, including inequality constraints on expectations-we also provide an implementation in~R using the \texttt{nloptr} package, which performs nonlinear optimization with user-defined constraints.

\subsection{Refined LP-based bounds under Figure~\ref{fig:sub.viol.1} and Assumption~\ref{assump:monotonic.m}.\ref{assump:monotonic.m.pos}}\label{app:LP.sharp.2}

\begin{Proposition}
\label{app:prop:LP_narrow_3a}
Consider the setting of Figure~\ref{fig:sub.viol.1}. Under Assumption~\ref{assump:monotonic.m}.\ref{assump:monotonic.m.pos}, the following bounds for $VE(0)$, $VE(1)$, and $VE_T$ hold.
\[
1 -
\frac{p_{10.1} + p_{1.11}}{1 - p_{00.0}}
\leq
VE(0)
\leq
1 - \frac{p_{10.1}}{p_{1.01} + p_{1.11}},
\]
\[
1 - \frac{p_{11.1}}{p_{10.0} + p_{11.0}}
\leq
VE(1)
\leq
1 - \frac{p_{10.1} + p_{11.1}}{p_{11.0}},
\]
\[
1 - \frac{1 - p_{01.1}}{p_{10.0}}
\leq
VE_T
\leq
1 - \frac{p_{10.1} + p_{11.1}}{p_{10.0} + p_{11.0}}.
\]
\end{Proposition}

\subsection{Refined LP-based bounds under Figure~\ref{fig:sub.viol.3}}\label{app:LP.sharp.3a}

o obtain LP-based bounds that are narrower than those in Proposition~\ref{prop:LP1}, we extend Assumption~\ref{assump:monotonic.m} to additionally condition on $S$, as follows. 

\begin{Assumption}\label{assump:monotonic.m.s}
The effect of $M$ on $Y$ is monotonic. Specifically, for all $a, b, s, u$:
\begin{enumerate}
    \item \label{assump:monotonic.m.s.pos} \textbf{Non-negative effect of $M$:}
    \[
        \Pr(Y^{a,m=0} \leq Y^{a,m=1} \mid A=a, B=b, S=s, U=u) = 1;
    \]
    \item \label{assump:monotonic.m.s.pos.neg} \textbf{Non-positive effect of $M$:}
    \[
        \Pr(Y^{a,m=1} \leq Y^{a,m=0} \mid A=a, B=b, S=s, U=u) = 1.
    \]
\end{enumerate}
\end{Assumption}

Under Assumption~\ref{assump:monotonic.m.s}, we can construct the following LP-based bounds.

\begin{Proposition}
\label{app:prop:LP_narrow}
Consider the setting of Figure~\ref{fig:sub.viol.3}. Under Assumption~\ref{assump:monotonic.m.s}.\ref{assump:monotonic.m.s.pos}, the following bounds for $VE(0)$, $VE(1)$, and $VE_T$ hold.
\[
\operatorname*{max}\limits_{s \in \{0,1\}}
\Biggl\{
1 - \frac{p_{10.1s} + p_{1.11s}}{1 - p_{00.0s}}
\Biggr\}
\leq VE(0)
\leq
\operatorname*{min}\limits_{s \in \{0,1\}}
\Biggl\{
1 - \frac{p_{10.1s}}{p_{1.01s} + p_{1.11s}}
\Biggr\},
\]
\[
\operatorname*{max}\limits_{s \in \{0,1\}}
\Biggl\{
1 - \frac{p_{11.1s}}{p_{10.0s} + p_{11.0s}}
\Biggr\}
\leq
VE(1)
\leq
\operatorname*{min}\limits_{s \in \{0,1\}}
\Biggl\{
1 - \frac{p_{10.1s} + p_{11.1s}}{p_{11.0s}}
\Biggr\},
\]
\[
\operatorname*{max}\limits_{s \in \{0,1\}}
\Biggl\{
1 - \frac{1 - p_{01.1s}}{p_{10.0s}}
\Biggr\}
\leq
VE_T
\leq
\operatorname*{min}\limits_{s \in \{0,1\}}
\Biggl\{
1 - \frac{p_{10.1s} + p_{11.1s}}{p_{10.0s} + p_{11.0s}}
\Biggr\}.
\]
\end{Proposition}

We note that, because the LP-based bounds require the monotonicity condition in Assumption~\ref{assump:monotonic.m.s} to hold for each value of $S$, the resulting bounds can be narrower than the monotonicity-based bounds under Assumption~\ref{assump:non.negative.m.general}. 
For example, the monotonicity-based upper bound for $VE_T$ is $1 - \frac{p_{1.1}}{p_{1.0}}$, whereas the LP-based upper bound is 
\[
\operatorname*{min}\limits_{s \in \{0,1\}} \Biggl\{\,1 - \frac{p_{1.1s}}{p_{1.0s}}\,\Biggr\},
\]
which is less than or equal to $1 - \frac{p_{1.1}}{p_{1.0}}$.

\newpage

\section{A non-LP approach to derive bounds} \label{app:eq1}

While the main text employed an LP-based solution, here we present an equivalent, simpler approach that is easier to explain. This method relies solely on expectations and does not require assumptions at the individual level. The central idea is to bound the components of the vaccine efficacy expressions, namely, the expectations $E(Y^{a,m})$. Each term $E(Y^{a,m})$ will be expressed as the sum of two parts: one identifiable under one extra assumption, and the other unidentifiable and thus bounded between 0 and 1. In this section, we show how $VE(0)$ can be bounded, but this procedure can similarly be applied to bound $VE(1)$ and $VE_T$.

In this section, we will assume that condition \eqref{Eq:POidentBm} holds in expectation for $a,a',m =\{0,1\}^3$:
\begin{equation}
\label{Eq:POidentBm.E}
E(Y^{a,m}|A=a,{B^{a',m=-1}=m}, U=u )=E(Y^{a,m=-1}|A=a, {B^{a,m=-1}=m}, U=u). 
\end{equation}

Now, we proceed with the derivation of the VE bounds. First, even without Assumption~\ref{assump:isolation}, under the randomization of $A$ and the law of total expectation we can rewrite $E(Y^{a,m=0})$ as:
{\footnotesize
\begin{align}
\label{eq:target.function.LP.general}
\begin{split}
E&(Y^{a,m=0}) \\
&= E(Y^{a,m=0} \mid A = a) \\
&= \Pr(B^{a,m=-1} = 0 \mid A = a) \cdot E(Y^{a,m=0} \mid A = a, B^{a,m=-1} = 0) \\
&\; + \Pr(B^{a,m=-1} = 1 \mid A = a) \cdot E(Y^{a,m=0} \mid A = a, B^{a,m=-1} = 1) \\
&= \Pr(B^{a,m=-1} = 0 \mid A = a) 
   \cdot \sum_{u} E(Y^{a,m=0} \mid A = a, B^{a,m=-1} = 0, U = u) 
   \cdot \Pr(U = u \mid A = a, B^{a,m=-1} = 0) \\
&\; + \Pr(B^{a,m=-1} = 1 \mid A = a) 
   \cdot \sum_{u} E(Y^{a,m=0} \mid A = a, B^{a,m=-1} = 1, U = u) 
   \cdot \Pr(U = u \mid A = a, B^{a,m=-1} = 1).\\
\end{split}
\end{align}
}

Now, under the condition in Equation~\eqref{Eq:POidentBm.E}, we can replace $E(\textcolor{red}{Y^{a,m=0}} \mid A = a, B^{a,m=-1} = 0, U = u)$ with $E(\textcolor{red}{Y^{a,m=-1} }\mid A = a, B^{a,m=-1}= 0, U = u)$. Thus, we obtain:
{\footnotesize
\begin{align}
\label{eq:target.function.LP.m0}
\begin{split}
&E(Y^{a,m=0}) =\\ 
&\Pr(B^{a,m=-1} = 0 \mid A = a) \cdot \sum_{u} E(Y^{a,m=-1} \mid A = a, B^{a,m=-1} = 0, U = u) \cdot \Pr(U = u \mid A = a, B^{a,m=-1}= 0) \\
&\; + \Pr(B^{a,m=-1} = 1 \mid A = a) \cdot \sum_{u} E(Y^{a,m=0} \mid A = a, B^{a,m=-1} = 1, U = u) \cdot \Pr(U = u \mid A = a, B^{a,m=-1} = 1) \\
&= \Pr(B^{a,m=-1} = 0 \mid A = a) \cdot E(Y^{a,m=-1} \mid A = a, B^{a,m=-1} = 0) \\
&\; + \Pr(B^{a,m=-1}= 1 \mid A = a) \cdot E(Y^{a,m=0} \mid A = a, B^{a,m=-1} = 1).
\end{split}
\end{align}
}
Finally, under consistency we have that
$$
E(Y^{a,m=0})=\pi_{0.a}p_{1.a0}+\pi_{1.a}E(Y^{a,m=0} \mid A = a, B = 1)
$$
That is, we showed that $E(Y^{a,m=0})$ can be expressed
as the sum of two parts: one identifiable ($p_{1.a0}$), and the other unidentifiable ($E(Y^{a,m=0} \mid A = a, B = 1)$). Therefore, by bounding the unidentifiable part between 0 and 1, we can bound $E(Y^{a,m=0})$ as:
\begin{align*}
\pi_{0.a}p_{1.a0}
\leq E(Y^{a,m=0})
\leq \pi_{0.a}p_{1.a0}+ \pi_{1.a}.
\end{align*}
Now, by substituting the lower bound for $E(Y^{a=1,m=0})$ and the upper bound for $E(Y^{a=0,m=0})$, we obtain the upper bound for $VE(0)$:
\begin{align*}
VE(0)&=1-\frac{E(Y^{a=1,m=0})}{E(Y^{a=0,m=0})}\\
&\leq1-\frac{\pi_{0.1}p_{1.10}}{\pi_{0.0}p_{1.00}+\pi_{1.0}}.
\end{align*}
By using the upper bound for $E(Y^{a=1,m=0})$ and the lower bound for $E(Y^{a=0,m=0})$, we obtain the lower bound for $VE(0)$:
\begin{align*}
VE(0)&=1-\frac{E(Y^{a=1,m=0})}{E(Y^{a=0,m=0})}\\
&\geq1-\frac{\pi_{0.1}p_{1.10}+\pi_{1.1}}{\pi_{0.0}p_{1.00}}.
\end{align*}
In total, the bounds for $VE(0)$ obtained using the method demonstrated here are equivalent to the LP-based bounds given in Proposition \ref{prop:LP1}.


\newpage
\section{Proof of the one-sided monotonicity-based bounds under Figure \ref{fig:sub.viol.1}}\label{app:proof of prop.bounds}

This section is organized as follows. In Section \ref{app:prop:one_mon} we provide the one-sided one-sided monotonicity-based bounds under Figure \ref{fig:sub.viol.1}. In Section~\ref{app.mon.lemma}, we present lemmas that will be used to prove the monotonicity-based bounds. In Section~\ref{app:proof_mon}, we prove that $E(Y \mid A=a, B=1) \geq E(Y^{a,m=1})$. Based on this result, in Section~\ref{app:mon.one} we derive the one-sided monotonicity-based bounds. 
Finally, in Section~\ref{app:mon.one.reversed}, we derive the monotonicity-based bounds under Assumptions~\ref{assump:pos1} and~\ref{assump:mon_in_U_general}\ref{assump:mon_in_U_general_2}.

\subsection{Proposition: one-sided one-sided monotonicity-based bounds under Figure \ref{fig:sub.viol.1}}\label{app:prop:one_mon}
 \begin{Proposition}\label{prop:bounds} 
 Consider the setting of Figure \ref{fig:sub.viol.1}.
Under Assumptions \ref{assump:pos1} and \ref{assump:mon_in_U_general}\ref{assump:mon_in_U_general_1}, the following bounds for $VE(0)$ and $VE_T$ hold. 
\begin{align*}
1-\frac{1}{p_{1.00}}&\leq VE(0)
\le 1-p_{1.10},\\
1-\frac{p_{1.11}}{p_{1.00}}
&\leq
VE_T \leq 1.
\end{align*}
Under Assumptions \ref{assump:pos1} and \ref{assump:mon_in_U_general} \ref{assump:mon_in_U_general_2},
 the bounds for $VE(0)$ and $VE_T$ given in Appendix \ref{app:proof of prop.bounds} hold.
\end{Proposition}

\subsection{Lemmas}\label{app.mon.lemma}

In this section, we use the following lemmas.

\begin{lemma}\citep{esary1967association}\label{lemma.sign}
Let $X=(X_1,\ldots,X_n)$ be a multivariate random variable such that each component is independent of the others. Then, for every pair of nondecreasing functions $f$ and $g$,
\begin{equation*}
    \operatorname{Cov}(f(X), g(X)) \geq 0.
\end{equation*}
\end{lemma}

\begin{lemma}\label{lemma:reversed.sign}
Let $X=(X_1,\ldots,X_n)$ be a multivariate random variable such that each component is independent of the others. Then, for every pair of nonincreasing functions $f$ and $g$,
\begin{equation*}
    \operatorname{Cov}(f(X), g(X)) \leq 0.
\end{equation*}
For discordant functions $f$ and $g$, the inequality sign is reversed.
\end{lemma}

\subsection{\texorpdfstring{Proof that \(E(Y\mid A=a,B=1)\geq E(Y^{a,m=1})\)}{Proof that E(Y|A=a,B=1) \ge E(Y a,m=1)}}\label{app:proof_mon}

For simplicity, we consider the case where $U$ is discrete, but the results also hold for continuous $U$. Under the assumption that $Y \perp\!\!\!\perp_{VI} M \mid A,B,U$ and the DAG illustrated in Figure~\ref{fig:sub.viol.1}, $E(Y^{a,m=1})$ for $a \in \{0,1\}$ is identified by 
\begin{align}\label{eq:sum_m}
\begin{split}
E(Y^{a,m=1})
&=E_{IV}\!\left(Y \mid A=a,M=1,B=1\right)\\
&=\sum_{u} E_{IV}\!\left(Y \mid A=a,M=1,B=1,U=u\right)\Pr_{IV}(U=u\mid A=a,M=1,B=1)\\
&=\sum_{u} E_{IV}\!\left(Y \mid A=a,M=1,B=1,U=u\right)\Pr_{IV}(U=u)\\
&=\sum_{u} E\!\left(Y \mid A=a,M=-1,B=1,U=u\right)\Pr(U=u)\\
&=E_{U}\!\left[E\!\left(Y \mid A=a,B=1,U\right)\right].
\end{split}
\end{align}

In addition, from the two-arm trial, $E(Y\mid A=a,B=1)$ can be written as 
\begin{align}\label{eq:sum_Minus1}
\begin{split}
E&(Y\mid A=a,B=1)\\
&=\sum_{u} E(Y\mid A=a,B=1,U=u)\Pr(U=u\mid A=a,B=1)\\
&=\sum_{u} E(Y\mid A=a,B=1,U=u)
   \frac{\Pr(B=1\mid A=a,U=u)\Pr(U=u\mid A=a)}{\Pr(B=1\mid A=a)}\\
&=\sum_{u} E(Y\mid A=a,B=1,U=u)\Pr(U=u)
   \frac{\Pr(B=1\mid A=a,U=u)}{\Pr(B=1\mid A=a)}\\
&=E_U\!\left[E(Y\mid A=a,B=1,U)\frac{\Pr(B=1\mid A=a,U)}{\Pr(B=1\mid A=a)}\right].
\end{split}
\end{align}

Let $\nu(u)=\frac{\Pr(B=1\mid A=a,U=u)}{\Pr(B=1\mid A=a)}$. Then $\nu(u)$ is nondecreasing in $u$, since $\Pr(B=1\mid A=a,U=u)$ is nondecreasing in $u$, and $E_U[\nu(U)]=1$. Thus, by Lemma~\ref{lemma.sign}, we have 
\begin{align}\label{eq:cov}
\begin{split}
E_U&\!\left[E(Y\mid A=a,B=1,U)\nu(U)\right]
 -E_U\!\left[E(Y\mid A=a,B=1,U)\right]\\
&= \operatorname{Cov}_U\!\left(E(Y\mid A=a,B=1,U),\nu(U)\right)\geq 0.
\end{split}
\end{align}

Now, by substituting Equations~\ref{eq:sum_m} and~\ref{eq:sum_Minus1} into Equation~\eqref{eq:cov}, we can show that $E(Y\mid A=a,B=1)\geq E(Y^{a,m=1})$ as follows:
\begin{align*}
E(Y\mid A=a,B=1)
&=E_U\!\left[E(Y\mid A=a,B=1,U)\nu(U)\right]\\
&\geq E_U\!\left[E(Y\mid A=a,B=1,U)\right]
=E(Y^{a,m=1}).
\end{align*}
By using Lemma~\ref{lemma:reversed.sign}, we can similarly prove the bounds for $E(Y^{a,m=0})$ and for the discordant-sign case. 

\subsection{Proof of the bounds in Proposition~\ref{prop:bounds}}\label{app:mon.one}

From the previous section, we established the following inequalities:
\begin{align*}
    &E(Y\mid A=a,B=1) \geq E(Y^{a,m=1}), \\
    &E(Y\mid A=a,B=0) \leq E(Y^{a,m=0}).
\end{align*}
Additionally, note that $E(Y^{a,m}) \leq 1$ for all $a,m$. We now derive the bounds in Proposition~\ref{prop:bounds}. Recall that $VE(0)$ and $VE_T$ are given by
$$
VE(0) = 1 - \frac{E(Y^{a=1,m=0})}{E(Y^{a=0,m=0})}, \;
VE_T = 1 - \frac{E(Y^{a=1,m=1})}{E(Y^{a=0,m=0})}.
$$

\paragraph{Upper bound for $VE(0)$:}
By substituting the lower bound for $E(Y^{a=1,m=0})$ and the upper bound for $E(Y^{a=0,m=0})$, we obtain:
\begin{align*}
    VE(0) 
    &= 1 - \frac{E(Y^{a=1,m=0})}{E(Y^{a=0,m=0})} \\
    &\leq 1 - \frac{E(Y\mid A=1,B=0)}{1} \\
    &= 1 - E(Y\mid A=1,B=0).
\end{align*}

\paragraph{Lower bound for $VE(0)$:}
Using the upper bound for $E(Y^{a=1,m=0})$ and the lower bound for $E(Y^{a=0,m=0})$, we get:
\begin{align*}
    VE(0)
    &= 1 - \frac{E(Y^{a=1,m=0})}{E(Y^{a=0,m=0})} \\
    &\geq 1 - \frac{1}{E(Y\mid A=0,B=0)}.
\end{align*}

\paragraph{Lower bound for $VE_T$:}
Similarly, for $VE_T$ we have:
\begin{align*}
    VE_T
    &= 1 - \frac{E(Y^{a=1,m=1})}{E(Y^{a=0,m=0})} \\
    &\geq 1 - \frac{1}{E(Y\mid A=0,B=0)}.
\end{align*}

\subsection{The monotonicity-based bounds under Assumptions~\ref{assump:pos1} and~\ref{assump:mon_in_U_general}\ref{assump:mon_in_U_general_2}}\label{app:mon.one.reversed}

Following the proof in Section~\ref{app:proof_mon}, it follows that under Assumption~\ref{assump:mon_in_U_general}\ref{assump:mon_in_U_general_2}, we have:
\begin{align}\label{app:reversed}
    &E(Y\mid A=a,B=1) \leq E(Y^{a,m=1}), \\
    &E(Y\mid A=a,B=0) \geq E(Y^{a,m=0}).
\end{align}

We use these inequalities to derive sharp bounds for $VE(1)$ and $VE_T$. The bounds for $VE(0)$ are the trivial bounds $-\infty < VE(0) \leq 1$.

\paragraph{Upper bound for $VE(1)$:}
By substituting the lower bound for $E(Y^{a=1,m=1})$ and the upper bound for $E(Y^{a=0,m=1})$, we obtain:
\begin{align*}
    VE(1) 
    &= 1 - \frac{E(Y^{a=1,m=1})}{E(Y^{a=0,m=1})} \\
    &\leq 1 - \frac{E(Y\mid A=1,B=1)}{1} \\
    &= 1 - E(Y\mid A=1,B=1).
\end{align*}

\paragraph{Lower bound for $VE(1)$:}
Using the upper bound for $E(Y^{a=1,m=1})$ and the lower bound for $E(Y^{a=0,m=1})$, we get:
\begin{align*}
    VE(1)
    &= 1 - \frac{E(Y^{a=1,m=1})}{E(Y^{a=0,m=1})} \\
    &\geq 1 - \frac{1}{E(Y\mid A=1,B=0)}.
\end{align*}

\paragraph{Upper bound for $VE_T$:}
Similarly, for $VE_T$ we have:
\begin{align*}
    VE_T
    &= 1 - \frac{E(Y^{a=1,m=1})}{E(Y^{a=0,m=0})} \\
    &\leq 1 - \frac{E(Y\mid A=1,B=1)}{E(Y\mid A=0,B=0)}.
\end{align*}

\newpage
\section{Proof of Proposition~\ref{prop:bounds.two.sided1}: two-sided monotonicity-based bounds under Figure~\ref{fig:sub.viol.1}}\label{app:proof.two.bounds}

Under Assumption~\ref{assump:non.negative.m.general}\ref{assump:non.negative.m} and the standard assumptions of randomization and consistency, it follows that 
\begin{equation*}
    E_{VI}(Y\mid A=a,M=0)\leq E_{VI}(Y\mid A=a,M=-1) \leq E_{VI}(Y\mid A=a,M=1).
\end{equation*}
Now, since $p_{1.a}=\sum_{m\in\{-1,0,1\}}E_{VI}(Y\mid A=a,M=m)\Pr_{VI}(M=m\mid A=a)$, we get
\begin{align*}
  p_{1.a}
  &=\sum_{m\in\{-1,0,1\}}E_{VI}(Y\mid A=a,M=m)\Pr_{VI}(M=m\mid A=a)\\
  &\leq\sum_{m\in\{-1,0,1\}}E_{VI}(Y\mid A=a,M=1)\Pr_{VI}(M=m\mid A=a)\\
  &=E_{VI}(Y\mid A=a,M=1)\\
  &=E(Y^{a,m=1}),
\end{align*}
and similarly, we get that $p_{1.a}\geq E(Y\mid A=a,M=0)=E(Y^{a,m=0})$.  
Thus, by combining these bounds with those from Section~\ref{app:proof_mon}, we obtain:
\begin{equation*}
    p_{1.a}\leq E(Y^{a,m=1})\leq p_{1.a1},
\end{equation*}
and
\begin{equation*}
    p_{1.a0}\leq E(Y^{a,m=0})\leq p_{1.a}
\end{equation*}
Using these results, we obtain the bounds presented in Proposition~\ref{prop:bounds.two.sided1}:
\[
1 - \frac{p_{1.1}}{p_{1.00}} \leq VE(0) \leq 1 - \frac{p_{1.10}}{p_{1.0}},
\]
\[
1 - \frac{p_{1.11}}{p_{1.0}} \leq VE(1) \leq 1 - \frac{p_{1.1}}{p_{1.01}},
\]
\[
1 - \frac{p_{1.11}}{p_{1.00}} \leq VE_T \leq 1 - \frac{p_{1.1}}{p_{1.0}}.
\]

\subsection{The monotonicity-based bounds under Assumptions~\ref{assump:mon_in_U_general}\ref{assump:mon_in_U_general_2} and~\ref{assump:non.negative.m.general}\ref{assump:negative.m}}\label{app:prop2_reversed}

Under Assumptions~\ref{assump:mon_in_U_general}\ref{assump:mon_in_U_general_2} and~\ref{assump:non.negative.m.general}\ref{assump:negative.m}, and by incorporating the results from Sections~\ref{app:mon.one.reversed} and~\ref{app:proof.two.bounds}, we get that
\begin{equation*}
      p_{1.a1}\leq E(Y^{a,m=1}) \leq p_{1.a},
\end{equation*}
and
\begin{equation*}
        p_{1.a}\leq E(Y^{a,m=0})\leq p_{1.a0}.
\end{equation*}
Given these inequalities, we obtain the following sharp bounds:
\[
1 - \frac{p_{1.10}}{p_{1.0}} \leq VE(0) \leq 1 - \frac{p_{1.1}}{p_{1.00}},
\]
\[
1 - \frac{p_{1.1}}{p_{1.01}} \leq VE(1) \leq 1 - \frac{p_{1.11}}{p_{1.0}},
\]
\[
1 - \frac{p_{1.1}}{p_{1.0}} \leq VE_T \leq 1 - \frac{p_{1.11}}{p_{1.00}}.
\]

\newpage

\section{Proof of Proposition~\ref{prop:LP1_withS}: LP-based bounds under Figure~\ref{fig:sub.viol.3}}
\label{app:LP2}

Since the proof is almost identical to the one presented in Section~\ref{app:LP1}, we briefly outline only the steps that differ.

The proof is based on the following result, which was proven in Appendix~\ref{app:NPSEM_S}. Analogously to Figure~\ref{fig:sub.viol.1}, for all $a, a', m, s \in \{0,1\}^4$, Assumption~\ref{assump:isolation} implies the following individual-level condition:
\begin{equation}
\Pr(Y^{a',m}=Y^{a',m=-1} \mid A=a, B^{a,m=-1}=m, S^{a,m=-1}=s, U=u)=1.
\end{equation}
Consequently, the tables of response types $r_{y \mid a, s, b}$ are identical to $r_{y \mid a, b}$ shown in Tables~\ref{tab:potential.responses.b0} and~\ref{tab:potential.responses.b1}. 

Given this result, for example, to minimize $E(Y^{a=1, m=0} \mid A=1, B=0, S=s)$, the corresponding coordinates are:
\begin{enumerate}
\item $\mathbf{q} = (q_{0.10s}, q_{1.10s}, \ldots, q_{15.10s})^T$,
\item $\mathbf{c}$ — as in Section~\ref{app:LP1},
\item $\mathbf{p} = (p_{1.00s}, p_{1.10s})$,
\item $\mathbf{R}$ — as in Section~\ref{app:LP1}.
\end{enumerate}

This algorithm is implemented in the code available at:  
\url{https://github.com/xlrod1/vaccine_for_publication/tree/main/LP_based_bounds_formulas/figure_3}.

\newpage
\section{Proof of Proposition~\ref{prop:bounds.mon.s.iv}: monotonicity-based bounds under Figure~\ref{fig:sub.viol.3}}\label{app:proof:prop:bounds.mon.s.iv}

\subsection{Proof of the one-sided monotonicity-based bounds under Figure~\ref{fig:sub.viol.3}}\label{app:mon.3a.one}
As presented in Section~\ref{sec:fig3}, under the scenario illustrated in Figure~\ref{fig:sub.viol.3}, the following two independence assumptions hold: (i) $S \perp\!\!\!\perp_{IV} Y \mid A,B,M,U$ and (ii) $S \perp\!\!\!\perp_{IV} U$. It follows that we can write $E\!\left(Y^{a,m=1}\right)$ as

\begin{align}\label{eq:sum_m_s_iv}
\begin{split}
E\!\left(Y^{a,m=1}\right)
&=E_{IV}\!\left(Y \mid A=a,M=1,B=1\right)\\
&=\sum_{u} E_{IV}\!\left(Y \mid A=a,M=1,B=1,U=u\right)\Pr_{IV}(U=u \mid A=a,M=1,B=1)\\
&=\sum_{u} E_{IV}\!\left(Y \mid A=a,M=1,B=1,U=u\right)\Pr_{IV}(U=u)\\
&=\sum_{u} E\!\left(Y \mid A=a,M=1,B=1,U=u,S=s\right)\Pr(U=u)\\
&=\sum_{u} E\!\left(Y \mid A=a,M=-1,B=1,U=u,S=s\right)\Pr(U=u) \\
&=\sum_{u} E\!\left(Y \mid A=a,B=m,U=u,S=s\right)\Pr(U=u).
\end{split}
\end{align}
Also, we can write $E(Y \mid A=a,B=1,S=s)$ as 
\small
\begin{adjustwidth}{-2.3cm}{0cm} 
\begin{align}\label{eq:sum_Minus1_s.iv}
\begin{split}
E&\!\left(Y \mid A=a,B=1,S=s\right)\\
&=\sum_{u} E\!\left(Y \mid A=a,B=1,S=s,U=u\right)\Pr(U=u \mid A=a,B=1,S=s)\\
&=\sum_{u} E\!\left(Y \mid A=a,B=1,U=u,S=s\right)
   \frac{\Pr(B=1 \mid A=a,U=u,S=s)\Pr(U=u \mid A=a,S=s)\Pr(A=a,S=s)}{\Pr(B=1 \mid A=a,S=s)\Pr(A=a,S=s)}\\
&=\sum_{u} E\!\left(Y \mid A=a,B=1,U=u,S=s\right)\Pr(U=u \mid A=a,S=s)\frac{\Pr(B=1 \mid A=a,U=u,S=s)}{\Pr(B=1 \mid A=a,S=s)}\\
&=\sum_{u} E\!\left(Y \mid A=a,B=1,U=u,S=s\right)\Pr(U=u)\frac{\Pr(B=1 \mid A=a,U=u,S=s)}{\Pr(B=1 \mid A=a,S=s)}.
\end{split}
\end{align}
\end{adjustwidth}
\normalsize
Let $\nu(u)=\frac{\Pr(B=1 \mid A=a,U=u,S=s)}{\Pr(B=1 \mid A=a,S=s)}$. Then $\nu(u)$ is nondecreasing in $u$ since $\Pr(B=1 \mid A=a,S=s,U=u)$ is nondecreasing in $u$, and since $U\perp A,S$, we have
$$
E_U[\nu(U)]=E_{U|A,S}[\nu(U)]=1.
$$
Thus, by Lemma~\ref{lemma.sign} we have 
\begin{align}
\begin{split}\label{eq:cov_s_iv}
E&\!\left(Y \mid A=a,B=1,S=s\right)-E\!\left(Y^{a,m=1}\right)\\
=&E_U\!\left[E\!\left(Y \mid A=a,B=1,U,S=s\right)\nu(U)\right]
 -E_U\!\left[E\!\left(Y \mid A=a,B=1,U,S=s\right)\right]\\
 =&E_U\!\left[E\!\left(Y \mid A=a,B=1,U,S=s\right)\nu(U)\right]
 -E_U\!\left[E\!\left(Y \mid A=a,B=1,U,S=s\right)\right]E_U[\nu(U)]\\
=&\operatorname{Cov}_U\!\left(E\!\left(Y \mid A=a,B=1,U,S=s\right),\nu(U)\right)\geq 0.
\end{split}
\end{align}

Now, by substituting Equations~\ref{eq:sum_m_s_iv} and~\ref{eq:sum_Minus1_s.iv} into Equation~\eqref{eq:cov_s_iv}, we can show that $p_{1.a1s}\geq E\!\left(Y^{a,m=1}\right)$. By using Lemma~\ref{lemma:reversed.sign}, we can similarly prove the bounds for $E\!\left(Y^{a,m=0}\right)$ and for the discordant-sign case. 

Without any further assumptions, we obtain the following inequalities:
\begin{align}
    p_{1.a1s}&\geq E\!\left(Y^{a, m=1}\right), \label{eq:lower_bound_Yam1} \\
    p_{1.a0s}&\leq E\!\left(Y^{a, m=0}\right).
\end{align}

Because these inequalities hold for each value of $s \in \{0,1\}$, it follows that
\begin{align*}
    \min_{s \in \{0,1\}} \left\{ p_{1.a1s} \right\} &\geq E\!\left(Y^{a, m=1}\right)\geq 0, \\
    \max_{s \in \{0,1\}} \left\{ p_{1.a0s} \right\} &\leq E\!\left(Y^{a, m=0}\right)\leq 1.
\end{align*}

We now derive the bounds for $VE(0)$ and the upper bound of $VE_T$. 

\paragraph{Upper bound for $VE(0)$.}
By substituting the lower bound for $E(Y^{a=1,m=0})$ and the upper bound for $E(Y^{a=0,m=0})$, we obtain:
\begin{align*}
    VE(0) 
    &= 1 - \frac{E(Y^{a=1,m=0})}{E(Y^{a=0,m=0})} \\
    &\leq 1 - \frac{\max_{s \in \{0,1\}} \left\{p_{1.a0s} \right\}}{1} \\
    &= 1 - \max_{s \in \{0,1\}} \left\{ p_{1.a0s}\right\}.
\end{align*}

\paragraph{Lower bound for $VE(0)$.}
Using the upper bound for $E(Y^{a=1,m=0})$ and the lower bound for $E(Y^{a=0,m=0})$, we get:
\begin{align*}
    VE(0)
    &= 1 - \frac{E(Y^{a=1,m=0})}{E(Y^{a=0,m=0})} \\
    &\geq 1 - \frac{1}{\max_{s \in \{0,1\}} \left\{ p_{1.00s} \right\}}.
\end{align*}

\paragraph{Lower bound for $VE_T$.}
Similarly, for $VE_T$ we have:
\begin{align*}
    VE_T
    &= 1 - \frac{E(Y^{a=1,m=1})}{E(Y^{a=0,m=0})} \\
    &\geq 1 - \frac{1}{\max_{s \in \{0,1\}} \left\{ p_{1.00s} \right\}}.
\end{align*}

\subsection{Proof of Proposition~\ref{prop:bounds.mon.s.iv}: two-sided monotonicity-based bounds under Figure~\ref{fig:sub.viol.3}}\label{app:mon.3a.two}

If we additionally assume Assumption~\ref{assump:non.negative.m.general}\ref{assump:non.negative.m}, we have
\begin{align*}
    \min_{s \in \{0,1\}} \left\{ p_{1.a1s}\right\} \geq E\!\left(Y^{a, m=1}\right) \geq p_{1.a}, \\
    \max_{s \in \{0,1\}} \left\{ p_{1.a0s} \right\} \leq E\!\left(Y^{a, m=0}\right) \leq p_{1.a}.
\end{align*}
Thus, the resulting bounds are presented in Proposition~\ref{prop:bounds.mon.s.iv}.

If, alternatively, we assume Assumptions~\ref{assump:U.mon.S.general}\ref{assump:U.mon.S.versionb} and~\ref{assump:non.negative.m.general}\ref{assump:negative.m}, then
\begin{align*}
    \max_{s \in \{0,1\}} \left\{ p_{1.a1s} \right\} \leq E\!\left(Y^{a, m=1}\right) \leq p_{1.a}, \\
    \min_{s \in \{0,1\}} \left\{ p_{1.a0s} \right\} \geq E\!\left(Y^{a, m=0}\right) \geq p_{1.a}.
\end{align*}
Accordingly, the bounds become:
\begin{align*}
\min_{s\in \{0,1\}}\!\left\{1-\frac{p_{1.1}}{p_{1.00s}}\right\}\geq
&\,VE(0) \;\ge\; \max_{s\in \{0,1\}}\!\left\{ 1-\frac{p_{1.10s}}{p_{1.0}}\right\},\\
\min_{s\in \{0,1\}}\!\left\{1-\frac{p_{1.11s}}{p_{1.0}}\right\}\ge &\,VE(1) \;\ge\; \max_{s\in \{0,1\}}\!\left\{1-\frac{p_{1.1}}{p_{1.01s}}\right\},\\
\min_{s_1,s_2\in \{0,1\}}\!\left\{1-\frac{p_{1.11s_1}}{p_{1.00s_2}}\right\}
\ge &\,VE_T
\ge 1-\frac{p_{1.1}}{p_{1.0}}.
\end{align*}

\newpage
\section{Proof of Proposition~\ref{prop:bounds.4ab.LP}: LP-based bounds under Figures~\ref{fig:sub.viol.4}--\ref{fig:DAG_with_s_violation}}\label{app:LP_bounds_fig4}

In this section, we detail the derivation of the upper bound for $VE(0)$. The procedures for obtaining the lower bound of $VE(0)$, as well as the bounds for $VE(1)$ and $VE_T$, follow analogously.

Under Figures~\ref{fig:sub.viol.3}--\ref{fig:DAG_with_s_violation} and using the result from Section~\ref{sec:LP.S.not.sharp}, the parameter $VE(0)$ can be written as
\small
\begin{align*}
\hspace*{-2cm}
&1 -
\frac{
 \sum_{s\in\{0,1\}}\gamma_{s.1}\!\left[
 \pi_{0.1s}\!\sum_{i \in Q_{1,s,0}}\!E(Y^{a=1,m=0}\mid A=1,B=0,S=s)
+\pi_{1.1s}\!\sum_{i \in Q_{1,s,1}}\!E(Y^{a=1,m=0}\mid A=1,B=1,S=s)
 \right]
}{
 \sum_{s\in\{0,1\}}\gamma_{s.0}\!\left[
 \pi_{0.0s}\!\sum_{i \in Q_{0,s,0}}\!E(Y^{a=0,m=0}\mid A=0,B=0,S=s)
+\pi_{1.0s}\!\sum_{i \in Q_{0,s,1}}\!E(Y^{a=0,m=0}\mid A=0,B=1,S=s)
 \right]
}.
\end{align*}
\normalsize

To compute the upper bound of $VE(0)$, we proceed as follows:
\begin{enumerate}
    \item For each term in the numerator and denominator, we formulate a separate linear program (LP) to maximize or minimize the relevant expectation, subject to the constraints imposed by the observed data and structural assumptions.
    \item Specifically, for each $s\in\{0,1\}$, we:
    \begin{enumerate}
        \item Minimize $E(Y^{a=1,m=0}\mid A=1, S=s,B=0)$;
        \item Minimize $E(Y^{a=1,m=0}\mid A=1, S=s,B=1)$;
        \item Maximize $E(Y^{a=0,m=0}\mid A=0,S=s,B=0)$;
        \item Maximize $E(Y^{a=0,m=0}\mid A=0,S=s,B=1)$.
    \end{enumerate}
    Each of these LPs identifies the extreme value of its target expectation given the observed data and model assumptions.
    \item We then substitute the extremal values from these LPs into the formula for $VE(0)$. The combination of these extremal values yields the upper and lower bounds on $VE(0)$ consistent with the data and assumptions.
    \item From Appendix~\ref{app:LP2}, it follows that the extrema of $E(Y^{a',m}\mid A=a,S=s,B=b)$ under Figures~\ref{fig:sub.viol.3}--\ref{fig:DAG_with_s_violation} are identical to those of $E(Y^{a',m}\mid A=a,B=b)$ under Figure~\ref{fig:sub.viol.1}, but additionally conditioned on $s$. Hence:
\begin{itemize}
    \item $\min\left\{E(Y^{a=1,m=0}\mid A=1,S=s,B=0)\right\} = p_{1.1s0}$,
    \item $\min\left\{E(Y^{a=1,m=0}\mid A=1,S=s,B=1)\right\} = 0$,
    \item $\max\left\{E(Y^{a=0,m=0}\mid A=0,S=s,B=0)\right\} = p_{1.0s0}$,
    \item $\max\left\{E(Y^{a=0,m=0}\mid A=0,S=s,B=1)\right\} = 1$.
\end{itemize}
\end{enumerate}

Thus, the upper bound for $VE(0)$ is given by 
\begin{align}
\begin{split}
VE(0)&\leq 1 - 
\frac{
 \sum_{s\in\{0,1\}}\gamma_{s.1}\Big[\pi_{0.1s}p_{1.1s0}
+\pi_{1.1s}\times 0\Big]
}{
\sum_{s\in\{0,1\}}\gamma_{s.0}\Big[\pi_{0.0s}p_{1.0s0}
+\pi_{1.0s}\times 1\Big]
}\\
&=1 - 
\frac{
 \sum_{s\in\{0,1\}}\gamma_{s.1}\Big[\pi_{0.1s}p_{1.1s0}\Big]
}{
\sum_{s\in\{0,1\}}\gamma_{s.0}\Big[\pi_{0.0s}p_{1.0s0}
+\pi_{1.0s}\Big]
}\\
&=1 - 
\frac{
 \pi_{0.1}p_{1.10}
}{
 \pi_{0.0}p_{1.00} + \pi_{1.0}
 }.
\end{split}
\end{align}

\newpage

\section{Proof of Proposition~\ref{prop:bounds.4ab.mon}: monotonicity-based bounds under Figures~\ref{fig:sub.viol.4}--\ref{fig:DAG_with_s_violation}}\label{app:proof.prop.bounds2}

In this section, we show that even though $S$ is observed, the same result as under Figure~\ref{fig:sub.viol.1} still applies; namely,
\begin{align*}
    p_{1.a1} &\geq E(Y^{a,m=1}), \\
    p_{1.a0} &\leq E(Y^{a,m=0}),
\end{align*}
with no dependency on $S$. Following this result, the one-sided monotonicity-based bounds under Figures~\ref{fig:sub.viol.4}--\ref{fig:DAG_with_s_violation} coincide with those under Figure~\ref{fig:sub.viol.1}. Consequently, adding Assumption~\ref{assump:non.negative.m.general} yields the same two-sided monotonicity-based bounds as in Figure~\ref{fig:sub.viol.1}. We will prove the result under Figure \ref{fig:DAG_with_s_violation}. Since Figures \ref{fig:sub.viol.4}--\ref{fig:DAG_with_s_violation2} are special cases of Figure \ref{fig:DAG_with_s_violation}, the result holds for them as well. In Section \ref{lemma.st1}, we recall a standard characterization of stochastic dominance; in Section \ref{lemma.st2}, we apply this result to show that $\gamma_{s.a1} \ge_{st} \gamma_{s.a}$ under a monotonicity condition; and, in Section \ref{app:proof.S.U}, we provide the proof that $E(Y \mid A=a,B=1) \ge E(Y^{a,m=1})$. Finally, in Section \ref{app:proof.prop.bounds2.reversed.bounds}, we present the assumption yielding reversed bounds. 

In this section we use the notation $ X\ge_{st} Y$ to denote that $X$ stochastically larger than $Y$.

\subsection{\citealp[Lemma (1.A.7), pg. 4]{ShakedShanthikumar2007}}\label{lemma.st1}
\begin{lemma}\label{lemma.st}
Let $X$ and $Y$ be random variables. Then
\[
X \ge_{st} Y
\]
if and only if for every increasing function $\phi$ for which the expectations exist,
\[
E[\phi(X)] \ge E[\phi(Y)].
\]
\end{lemma}

\subsection{Proof that $\gamma_{s.a1} \ge_{st} \gamma_{s.a}$}\label{lemma.st2}

\begin{lemma}\label{app:lemma.st}
  Assume that for any $a\in\{0,1\}$
 \begin{equation} \label{eq:monB}
     \Pr(B=1 \mid A=a,S=1) \;\ge\; \Pr(B=1 \mid A=a,S=0).
 \end{equation}
Then, the distribution of $S$ given $(A=a,B=1)$ stochastically dominates the distribution of $S$ given $A=a$, i.e.
\[
\gamma_{s.a1} \ge_{st} \gamma_{s.a}.
\]
\end{lemma}

\paragraph{Proof.}
Since $S\in\{0,1\}$, first-order stochastic dominance reduces to verifying
\begin{equation}\label{eq:FOSD-binary}
    \Pr(S=1 \mid A=a,B=1) \;\ge\; \Pr(S=1 \mid A=a). 
\end{equation}

Let $p=\Pr(S=1\mid A=a)$ and write $w_s=\Pr(B=1\mid A=a,S=s)$ for $s=0,1$.
By Bayes' rule,
$$
\Pr(S=1\mid A=a,B=1)
= \frac{w_1 p}{\Pr(B=1\mid A=a)}.
$$
The law of total probability gives
$$
\Pr(B=1\mid A=a)=w_1 p + w_0(1-p).
$$
Substituting yields
$$
\Pr(S=1\mid A=a,B=1)
= \frac{w_1 p}{w_1 p + w_0(1-p)}.
$$

To show~\eqref{eq:FOSD-binary}, note that
$$
\frac{w_1 p}{w_1 p + w_0(1-p)} \ge p
\quad\Longleftrightarrow\quad
p(1-p)(w_1 - w_0) \ge 0,
$$
which holds because $0\le p\le 1$ and $w_1\ge w_0$ by assumption~\eqref{eq:monB}.
Thus $\gamma_{s.a1}\ge_{st}\gamma_{s.a}$.

\subsection{Proof that $E(Y \mid A=a,B=1)\ge E(Y^{a,m=1})$}\label{app:proof.S.U}
Let $\gamma^{IV}_{s.a}=Pr_{IV}(S=s \mid A=a)$. Under Figure \ref{fig:DAG_with_s_violation} and the assumptions (1) $Y \perp M \mid_{VI} A,B,S,U$, (2) $\gamma^{IV}_{s.a}=\gamma_{s.a}$, and (3) $\Pr_{IV}(U=u|S=s,A=a)=\Pr(U=u|S=s,A=a)$, it follows that $E(Y^{a,m=1})$ for $a \in \{0,1\}$ is identified by
\begin{align}
\begin{split}
E&(Y^{a,m=1}) \\
&= \sum_s E_{IV}\!\left(Y \mid A=a,M=1,B=1,S=s\right)\gamma^{IV}_{s.a}\\
&= \sum_{s,u} E_{IV}\!\left(Y \mid A=a,M=1,B=1,S=s,U=u\right)\Pr_{IV}(U=u \mid A=a,M=1,B=1,S=s)\gamma^{IV}_{s.a}\\
&= \sum_{s,u} E_{IV}\!\left(Y \mid A=a,M=1,B=1,S=s,U=u\right)\Pr_{IV}(U=u|{S=s,A=a})\gamma^{IV}_{s.a}\\
&= \sum_{s,u} E\!\left(Y \mid A=a,M=-1,B=1,S=s,U=u\right)\Pr(U=u|{S=s,A=a})\gamma^{IV}_{s.a}\\
&= \sum_{s} E_{U}\!\left[E\!\left(Y \mid A=a,B=1,S=s,U\right)\gamma_{s.a}|{S=s,A=a}\right].
\end{split}
\end{align}
Similarly, $E(Y \mid A=a,B=1)$ can be written as 
\small
\begin{align}
\begin{split}
E&(Y \mid A=a,B=1)\\
&= \sum_{s} E\!\left(Y \mid A=a,B=1,S=s\right)\gamma_{s.a{b}}\\
&= \sum_{s,u} E\!\left(Y \mid A=a,B=1,S=s,U=u\right)\gamma_{s.a{b}}\Pr(U=u \mid A=a,B=1,S=s)\\
&= \sum_{s,u} E\!\left(Y \mid A=a,B=1,S=s,U=u\right)\gamma_{s.a{b}}
    \frac{\Pr(B=1 \mid A=a,U=u,S=s)\Pr(U=u \mid A=a,S=s)}{\Pr(B=1 \mid A=a,S=s)}\\
&= \sum_{s,u} E\!\left(Y \mid A=a,B=1,S=s,U=u\right)\gamma_{s.a{b}}\Pr(U=u|{A=a,S=s})
    \frac{\Pr(B=1 \mid A=a,U=u,S=s)}{\Pr(B=1 \mid A=a,S=s)}\\
&= \sum_{s} E_{U}\!\left[E\!\left(Y \mid A=a,B=1,S=s,U\right)\gamma_{s.a{b}}
    \frac{\Pr(B=1 \mid A=a,S=s,U)}{\Pr(B=1 \mid A=a,S=s)}\bigg|{A=a,S=s}\right].
\end{split}
\end{align}
\normalsize

Now, we are going to prove that $E(Y \mid A=a,B=1)-E(Y^{a,m=1})\ge 0$. Let $\nu_s(u) = \frac{\Pr(B=1 \mid A=a,S=s,U=u)}{\Pr(B=1 \mid A=a,S=s)}$ and $g_s(U)=E\!\left(Y \mid A=a,B=1,S=s,U\right)$. Assume that (1) $\Pr(B=1 \mid A=a,S=s,U=u)$ and $E\!\left(Y \mid A=a,B=1,S=s,U\right)$ are both nondecreasing in $u$ for all $a$ and $s$, (2) $Pr(B=1|A=a,S=1)\geq Pr(B=1|A=a,S=0)$ and (3) $E_{U}(g_s(U)|A=a,S=s)=E(Y^{a,m}|A=a,S=s)$ is nondecreasing in $s$. Then, $E(Y \mid A=a,B=1)-E(Y^{a,m=1})$ can be written as
\begin{align}
\begin{split}
E&(Y \mid A=a,B=1)-E(Y^{a,m=1})\\
&\overset{\mathrm{(1)}}{=} \sum_{s} E_{U}\!\left[{\gamma_{s.a1}}g_s(U)\nu_s(U)|A=a,S=s\right]- \sum_{s}E_{U}\!\left[{\gamma_{s.a}}g_s(U)|A=a,S=s\right]  \\
&\overset{\mathrm{}}{=}\sum_{s} \bigg[[E_{U}\!\left[{\gamma_{s.a1}}g_s(U)\nu_s(U)|A=a,S=s\right]-E_{U}\!\left[{\gamma_{s.a}}g_s(U)|A=a,S=s\right]\bigg]  \\
&\overset{\mathrm{(2)}}{=} \sum_{s} \bigg[{\gamma_{s.a1}}\operatorname{Cov}_{U}\!\left(g_s(U),\, \nu_s(U)|A=a,S=s\right) +(\gamma_{s.a1}-\gamma_{s.a})E_{U}(g_s(U)|A=a,S=s)\bigg]\\
&\overset{\mathrm{}}{=}E_{s.a1}\bigg[\operatorname{Cov}_{U}\!\left(g_s(U),\, \nu_s(U)|A=a,S=s\right)\bigg] \\&+\big( E_{\gamma_{s.a1}}E_{U}(g_s(U)|A=a,S=s)-E_{\gamma_{s.a}}E_{U}(g_s(U)|A=a,S=s)\big)\\
&\overset{\mathrm{(3)}}{\ge}\ 0
\end{split}
\end{align}
Step (1) follows from the derivations above and the definitions of $\nu_s(u)$ and $g_s(U)$. Step (2) follows since $E_{U}[\nu_s(U)|A=a,S=s] = 1$ and since it is easy to show that for any random variables $X,Y$, and constants $a,b$, $E(aXY|C)-E(bX|C)E(Y|C)=aCov(X,Y|C)+(a-b)E(X|C)E(Y|C)$ where $C=\{A=a,S=s\}$. Step (3) follows from the following two points
\begin{enumerate}
    \item We assume that $g_s(U)$ is nondecreasing in $u$ and that $\Pr(B=1 \mid A=a,S=s,U=u)$ is nondecreasing in $u$ and thus $\nu_s(u)$ is nondecreasing in $u$. From Lemma \ref{lemma.sign} it follows that $\operatorname{Cov}_{U}\!\left(g_s(U),\, \nu_s(U)\right|A=a,S=s)\ge 0$.
    \item We assume that $E_{U}(g_s(U)|A=a,S=s)$ is nondecreasing in $s$. By Lemma \ref{app:lemma.st} we have that $\gamma_{s.a1}\geq_{st}\gamma_{s.a}$. Thus, by Lemma \ref{lemma.st} we have that $\big( E_{\gamma_{s.a1}}E_{U}(g_s(U)|A=a,S=s)-E_{\gamma_{s.a}}E_{U}(g_s(U)|A=a,S=s)\big)\ge 0$.
\end{enumerate}

\subsection{Assumption guaranteeing reversed bounds}\label{app:proof.prop.bounds2.reversed.bounds}

\begin{Assumption}\label{assump:U.mon.S.general.SU.reveresed}($U|S$ monotonicity)
    Consider the setting of Figures \ref{fig:sub.viol.4}--\ref{fig:DAG_with_s_violation}. Suppose that $Y\perp_{\mathcal{T}_{VI}} M|A,B,S,U$ the following conditions hold:
\begin{enumerate}[label=(\roman*)]

\item \label{assump:U.mon.S.versionb.SU}
\item For all $a$ and $s$, exactly one of 
$\Pr\!\left(Y=1 \mid A=a,B=1,S=s,U=u\right)$ and 
$\Pr(B=1 \mid A=a,S=s,U=u)$
is nondecreasing in $u$, while the other is nonincreasing in $u$.
\item For all $a$, $\Pr(B=1 \mid A=a,S=s)$ is nondecreasing in $s$.
\item For all $a$, $E(Y^{a,m}\mid A=a,S=s)$ is nonincreasing in $s$.
\end{enumerate} 
\end{Assumption}

\newpage
\section{Discussion of scenarios with two unmeasured common causes: $U_1$ and $U_2$}\label{app:two_u}

\begin{figure}[ht]
\centering
\begin{subfigure}[b]{0.45\textwidth}
\centering
\resizebox{\linewidth}{!}{
\begin{tikzpicture}
\tikzset{line width=1.5pt, outer sep=0pt, ell/.style={draw,fill=white, inner sep=2pt,
line width=1.5pt},  swig vsplit={gap=5pt, inner line width right=0.5pt}};
\node[name=A,ell,  shape=ellipse] at (-9,-1) {$A$};
\node[name=S,ell,  shape=ellipse] at (-6,0) {$S$};
\node[name=B,ell,  shape=ellipse] at (-3,0) {$B$};
\node[name=E,ell,  shape=ellipse] at (0,0) {$E$};
\node[name=U_1,ell,  shape=ellipse] at (0,2) {$U_1$};
\node[name=U_2,ell,  shape=ellipse] at (-3,-1) {$U_2$};
\node[name=Y,ell,  shape=ellipse] at (3,0) {$Y$};
\node[name=M,ell,  shape=ellipse] at (-9,1) {$M$};
\begin{scope}[>={Stealth[black]},
every edge/.style={draw=black,very thick}]
\path [->] (E) edge (Y);
\path [->] (A) edge (S);
\path [->] (S) edge (B);
\path [->] (M) edge (B);
\path [->] (B) edge (E);
\path [->] (U_2) edge[dashed] (S);
\path [->] (U_2) edge[dashed] (E);
\path [->] (U_2) edge[dashed] (Y);
\path [->] (U_1) edge[dashed] (B);
\path [->] (U_1) edge[dashed] (E);
\path [->] (E) edge (Y);  
\path [->] (A) edge[bend left=-30] (Y);    
\end{scope}
\end{tikzpicture}}
\subcaption{Assumption~\ref{assump:y.s.dis} is violated due to the dashed arrows.}
\label{fig:DAG_two_Ua}
\end{subfigure}
\hfill
\begin{subfigure}[b]{0.45\textwidth}
\centering
\resizebox{\linewidth}{!}{
\begin{tikzpicture}
\tikzset{line width=1.5pt, outer sep=0pt, ell/.style={draw,fill=white, inner sep=2pt,
line width=1.5pt},swig vsplit={gap=5pt, inner line width right=0.5pt}};
\node[name=A,ell,  shape=ellipse] at (-9,-1) {$A$};
\node[name=S,ell,  shape=ellipse] at (-6,0) {$S$};
\node[name=B,ell,  shape=ellipse] at (-3,0) {$B$};
\node[name=E,ell,  shape=ellipse] at (0,0) {$E$};
\node[name=U_1,ell,  shape=ellipse] at (-3,2) {$U_1$};
\node[name=U_2,ell,  shape=ellipse] at (-3,-1) {$U_2$};
\node[name=Y,ell,  shape=ellipse] at (3,0) {$Y$};
\node[name=M,ell,  shape=ellipse] at (-9,1) {$M$};
\begin{scope}[>={Stealth[black]},
every edge/.style={draw=black,very thick}]
\path [->] (E) edge (Y);
\path [->] (A) edge (S);
\path [->] (S) edge (B);
\path [->] (S) edge[bend left=-35] (Y);    
\path [->] (M) edge (B);
\path [->] (B) edge (E);
\path [->] (U_2) edge[dashed] (S);
\path [->] (U_2) edge[dashed] (E);
\path [->] (U_2) edge[dashed] (Y);
\path [->] (U_1) edge[dashed] (B);
\path [->] (U_1) edge[dashed] (S);
\path [->] (E) edge (Y);  
\path [->] (A) edge[bend left=-30] (Y);    
\end{scope}
\end{tikzpicture}
}
\subcaption{Assumption~\ref{assump:y.s.dis} is violated due to the dashed arrows.}
\label{fig:DAG_two_Ub}
\end{subfigure}
\caption{DAGs describing hypothetical six-arm trials $\mathcal{T}_{VI}$. In these trials, the VEs are not identifiable because Assumption~\ref{assump:y.s.dis} is violated.}
\label{fig:DAG_violations_two_U}
\end{figure}
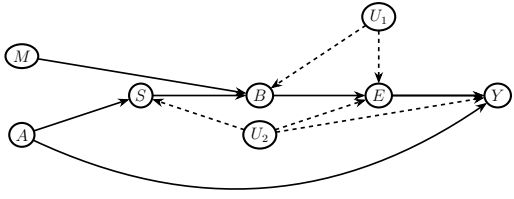
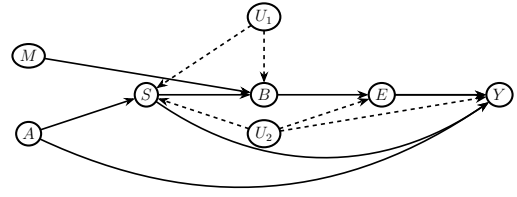

In this Appendix, we show that although Figure~\ref{fig:DAG_violations_two_U} includes two common causes, it is still equivalent to Figures~\ref{fig:sub.viol.4}--\ref{fig:DAG_with_s_violation} in the sense that it suffices to condition on only one of the $U$s, such that $Y \perp_{IV} M \mid A, S, B, U_i$ for $i \in \{1,2\}$. Therefore, the bounds in Propositions~\ref{prop:bounds.4ab.LP} and~\ref{prop:bounds.4ab.mon} remain valid under Figure~\ref{fig:DAG_violations_two_U}.

\paragraph{Figure~\ref{fig:DAG_two_Ua}.} In Figure~\ref{fig:DAG_two_Ua}, the paths from $M$ to $Y$ are:
\begin{enumerate}
    \item $M \rightarrow B \rightarrow E \rightarrow Y$,
    \item $M \rightarrow B \leftarrow S \leftarrow U_2 \rightarrow Y$,
    \item $M \rightarrow B \leftarrow S \leftarrow A \rightarrow Y$,
    \item $M \rightarrow B \leftarrow U_1 \rightarrow E \rightarrow Y$.
\end{enumerate}

The four paths from $M$ to $Y$ can each be blocked as follows:  
(1) $M \rightarrow B \rightarrow E \rightarrow Y$ is blocked by $\Sq B$;  
(2) $M \rightarrow B \leftarrow S \leftarrow U_2 \rightarrow Y$ is a collider path, closed by default, and remains blocked when both $\Sq B$ and $\Sq S$ are applied;  
(3) $M \rightarrow B \leftarrow S \leftarrow A \rightarrow Y$ is similarly closed by default and, once opened by $\Sq B$, becomes blocked again by $\Sq S$ or $\Sq A$; and  
(4) $M \rightarrow B \leftarrow U_1 \rightarrow E \rightarrow Y$ is blocked by $\Sq B$ together with $\Sq U_1$.  
Therefore, conditioning on $A, S, B, U_i$ for either $i \in \{1,2\}$ blocks all four paths, implying $Y \perp_{IV} M \mid A, S, B, U_i$.

\paragraph{Figure~\ref{fig:DAG_two_Ub}.} In Figure~\ref{fig:DAG_two_Ub}, the paths from $M$ to $Y$ are:
\begin{enumerate}
    \item $M \rightarrow B \rightarrow E \rightarrow Y$,
    \item $M \rightarrow B \leftarrow S \leftarrow U_2 \rightarrow Y$,
    \item $M \rightarrow B \leftarrow S \rightarrow Y$,
    \item $M \rightarrow B \leftarrow S \leftarrow A \rightarrow Y$,
    \item $M \rightarrow B \leftarrow U_1 \rightarrow S \leftarrow U_2 \rightarrow Y$,
    \item $M \rightarrow B \leftarrow U_1 \rightarrow S \leftarrow A \rightarrow Y$,
    \item $M \rightarrow B \leftarrow U_1 \rightarrow S \rightarrow Y$.
\end{enumerate}

The seven paths from $M$ to $Y$ can each be blocked as follows:  
(1) $M \rightarrow B \rightarrow E \rightarrow Y$ is blocked by $\Sq B$;  
(2) $M \rightarrow B \leftarrow S \leftarrow U_2 \rightarrow Y$ is a collider path, closed by default, and remains blocked when both $\Sq B$ and $\Sq S$ are applied;  
(3) $M \rightarrow B \leftarrow S \rightarrow Y$ and (4) $M \rightarrow B \leftarrow S \leftarrow A \rightarrow Y$ are also closed by default and, once opened by $\Sq B$, are blocked again by $\Sq S$ or $\Sq A$;  
(5) $M \rightarrow B \leftarrow U_1 \rightarrow S \leftarrow U_2 \rightarrow Y$ is blocked by $\Sq B$ together with either $\Sq U_1$ or $\Sq U_2$;  
(6) $M \rightarrow B \leftarrow U_1 \rightarrow S \leftarrow A \rightarrow Y$ is blocked by $\Sq B$ and $\Sq U_1$ or $\Sq A$; and  
(7) $M \rightarrow B \leftarrow U_1 \rightarrow S \rightarrow Y$ is blocked by $\Sq B$ and $\Sq U_1$ or $\Sq S$.  
Therefore, conditioning on $A, S, B, U_i$ for either $i \in \{1,2\}$ blocks all seven paths, implying $Y \perp_{IV} M \mid A, S, B, U_i$.

\newpage
\section{Narrowing bounds with additional measured covariates}\label{app:sharpening_bounds}

In this section, we use the notation
$p^{\,l}_{yb.al}=\Pr(Y=y,B=b\mid A=a,L=l)$,
$p^{\,l}_{y.abl}=\Pr(Y=y\mid A=a,B=b,L=l)$,
$p^{\,l}_{y.abls}=\Pr(Y=y\mid A=a,B=b,L=l,S=s)$,
$\pi^{\,l}_{b.al}=\Pr(B=b\mid A=a,L=l)$, and
$\pi^{\,l}_{b.als}=\Pr(B=b\mid A=a,L=l,S=s)$.


\subsection{Narrowing bounds with additional measured covariates}\label{sec:covariates}

In practice, vaccine trials typically collect a set of baseline covariates (other than $S$), denoted by $L$. We begin with the case where certain covariates $L$ can be used to obtain narrower bounds under the conditions described below. When $S$ is unobserved (Figure \ref{fig:sub.viol.1}) or cannot be exploited to tighten the bounds (as in Figures \ref{fig:sub.viol.4}--\ref{fig:DAG_with_s_violation}), such covariates $L$ may serve this purpose. When conditioning on $S$ leads to narrower bounds (as in Figure~\ref{fig:sub.viol.3}), replacing stratification by $S$ alone with joint stratification by $(S,L)$ may further narrow them. For illustration, we present only the LP-based bounds for $VE(0)$ in the main text. The remaining LP-based bounds, the monotonicity-based bounds, and the full proofs are deferred to the rest of Appendix~\ref{app:sharpening_bounds}.

From Section~\ref{sec:3c-d.LP}, we know that $S$ can be used to obtain narrower bounds if it satisfies $S \perp_{\mathcal{T}_{VI}} U$ and $Y \perp_{\mathcal{T}_{VI}} S \mid A,B,M,U$. The analogous result holds for $L$, as summarized in the following proposition.  

\begin{Proposition}\label{prop:L_sharp}
Any covariate $L$ that satisfies $L \perp_{\mathcal{T}_{VI}} (A,M,S,U)$ and 
$Y \perp_{\mathcal{T}_{VI}} L \mid A,S,B,M,U$ can be used to obtain narrower bounds under both the LP-based and the monotonicity-based approaches.
\end{Proposition}

An example of an $L$ that might satisfy the above conditions is participants’ awareness of the randomization ratio. Participants’ awareness of the ratio prior to enrollment could affect their belief about treatment assignment without directly influencing the risk of infection. For instance, in the Novavax COVID-19 vaccine trial, the randomization ratio was 2:1 (vaccine:placebo), and participants were informed of this ratio before enrollment \citep{dunkle2022efficacy,NIH_PREVENT19_2020}. In practice, however, not all participants may have been equally attentive to or aware of this information. If only some participants are aware of the randomization ratio (which could be measured by a survey), conditioning on awareness can be used to obtain narrower bounds. An example of an $L$ that fails to meet the conditions above is when $L$ represents the observed part of $U$, that is, a measured common cause of both $B$ and $Y$. In this case, as well as in the other cases illustrated in Figure~\ref{fig:L_adjustment}, $L$ cannot be used to obtain narrower bounds. Further discussion with real data examples appears in Appendix \ref{app:sharpening_bounds}. In practice, most baseline covariates in vaccine trials are connected to $U$, $Y$, or both, and therefore cannot be used to obtain narrower bounds. 

When $S$ is unobserved (Figure \ref{fig:sub.viol.1}) or does not lead to narrower bounds (Figures \ref{fig:sub.viol.4}--\ref{fig:DAG_with_s_violation}), a covariate $L$ satisfying the conditions of Proposition \ref{prop:L_sharp} may be used instead. We can apply the results for $S$ in Section~\ref{sec:fig3} and replace $S$ with $L$ in the LP-based bounds of Proposition~\ref{prop:LP1_withS} 
and in the monotonicity-based bounds of Proposition~\ref{prop:bounds.mon.s.iv}. 
For example, under Figure \ref{fig:DAG_with_s_violation2_with_L} the LP-based bounds for $VE(0)$ now take the form
$$
    \max_{l_1,l_2 \in \{0,1\}^2}
    \left\{
      1-\frac{1-p^{l}_{00.1l_1}}{p^{l}_{10.0l_2}}
    \right\}
   \leq
   VE(0)
   \leq
   \min_{l_1,l_2 \in \{0,1\}^2}
    \left\{
      1-\frac{p^{l}_{10.1l_1}}{1-p^{l}_{00.0l_2}}
    \right\},
$$
where $p^{l}_{yb.al} = \Pr(Y=y,B=b \mid A=a,L=l)$.

By contrast, when $S$ can be used to obtain narrower bounds (Figure~\ref{fig:sub.viol.3}), 
these results can be further refined when $L$ also satisfies Proposition \ref{prop:L_sharp} conditions, 
by replacing stratification by $S$ alone with joint stratification by $(S,L)$. 
For example, under Figure \ref{fig:sub.viol.3.with.L} the LP-based bounds for $VE(0)$ now become
$$
    \max_{s_1,s_2,l_1,l_2 \in \{0,1\}^4}
    \left\{
      1 - \frac{1 - p^{l}_{00.1\,s_1 l_1}}{p^{l}_{10.0\,s_2 l_2}}
    \right\}
    \;\le\;
    VE(0)
    \;\le\;
    \min_{s_1,s_2,l_1,l_2 \in \{0,1\}^4}
    \left\{
      1 - \frac{p^{l}_{10.1\,s_1 l_1}}{1 - p^{l}_{00.0\,s_2 l_2}}
    \right\},
$$
where $p^{l}_{yb.asl} = \Pr(Y=y, B=b \mid A=a, S=s, L=l)$.

\subsection{Case 1: when $L$ can be used to obtain narrower bounds}\label{app:sharpening_bounds_L_good}

As described in Section~\ref{sec:covariates}, when $L$ satisfies the conditions in Proposition~\ref{prop:L_sharp}, it can be used to narrow the LP-based bounds in Proposition~\ref{prop:LP1} and the monotonicity-based bounds in Proposition~\ref{prop:bounds.two.sided1}. Examples of such variables $L$ are presented in Figure~\ref{fig:L_adjustment_main}. In this section, we will proof the bounds for three scenarios.
\begin{enumerate}
    \item \textbf{Case 1a.} When $L$ can be used to obtain narrower bounds but $S$ cannot be used to obtain narrower bounds (Figure \ref{fig:DAG_with_s_violation2_with_L}).
    \item \textbf{Case 2b.} When both $L$ and $S$ can be used to obtain narrower bounds (Figure \ref{fig:sub.viol.3.with.L}). 
    \item \textbf{Case 3c.} When $L$ can be used to obtain narrower bounds and $S$ is unobserved (Figure \ref{fig:L_adjustment_S_not_observed}). 
\end{enumerate}
Although other DAGs besides the one in Figure~\ref{fig:DAG_with_s_violation2_with_L} could represent the first scenario, we focus on this DAG and prove the bounds under Figure~\ref{fig:DAG_with_s_violation2_with_L}, since it is the most ``complicated'' in the sense that both paths $S \rightarrow Y$ and $S \leftarrow U$ are present.

\begin{figure}[h]
\centering
\begin{subfigure}[b]{0.45\textwidth}
\centering
\resizebox{\linewidth}{!}{
\begin{tikzpicture}
\tikzset{line width=1.5pt, outer sep=0pt, ell/.style={draw,fill=white, inner sep=2pt,
line width=1.5pt},swig vsplit={gap=5pt, inner line width right=0.5pt}};
\node[name=A,ell,ellipse] at (-9,-1) {$A$};
\node[name=S,ell,ellipse] at (-6,0) {$S$};
\node[name=B,ell,ellipse] at (-3,0) {$B$};
\node[name=E,ell,ellipse] at (0,0) {$E$};
\node[name=U,ell,ellipse] at (0,2) {$U$};
\node[name=Y,ell,ellipse] at (3,0) {$Y$};
\node[name=M,ell,ellipse] at (-9,1) {$M$};
\node[name=L,ell,ellipse] at (-6,2) {$L$};
\begin{scope}[>={Stealth[black]},every edge/.style={draw=black,very thick}]
\path [->] (E) edge (Y);
\path [->] (A) edge (S);
\path [->] (S) edge (B);
\path [->] (S) edge [bend left=-30] (Y);
\path [->] (M) edge (B);
\path [->] (L) edge (B);
\path [->] (B) edge (E);
\path [->] (U) edge (E);
\path [->] (U) edge[dashed] (B);
\path [->] (U) edge (S);
\path [->] (U) edge (Y);
\path [->] (A) edge[bend left=-30] (Y);
\end{scope}
\end{tikzpicture}}
\subcaption{\textbf{Case 1a.} Similar to Figure~\ref{fig:DAG_with_s_violation} but with $L$ connected only to $B$.}
\label{fig:DAG_with_s_violation2_with_L}
\end{subfigure}
\hfill
\begin{subfigure}[b]{0.45\textwidth}
\centering
\resizebox{\linewidth}{!}{
\begin{tikzpicture}
\tikzset{line width=1.5pt, outer sep=0pt, ell/.style={draw,fill=white, inner sep=2pt,
line width=1.5pt},swig vsplit={gap=5pt, inner line width right=0.5pt}};
\node[name=A,ell,ellipse] at (-9,-1) {$A$};
\node[name=S,ell,ellipse] at (-6,0) {$S$};
\node[name=B,ell,ellipse] at (-3,0) {$B$};
\node[name=E,ell,ellipse] at (0,0) {$E$};
\node[name=L,ell,ellipse] at (-6,2) {$L$};
\node[name=U,ell,ellipse] at (0,2) {$U$};
\node[name=Y,ell,ellipse] at (3,0) {$Y$};
\node[name=M,ell,ellipse] at (-9,1) {$M$};
\begin{scope}[>={Stealth[black]},every edge/.style={draw=black,very thick}]
\path [->] (E) edge (Y);
\path [->] (A) edge (S);
\path [->] (S) edge (B);
\path [->] (L) edge (B);
\path [->] (M) edge (B);
\path [->] (B) edge (E);
\path [->] (U) edge (E);
\path [->] (U) edge[dashed] (B);
\path [->] (U) edge (Y);
\path [->] (A) edge[bend left=-30] (Y);
\end{scope}
\end{tikzpicture}}
\subcaption{\textbf{Case 1b.} Similar to Figure~\ref{fig:sub.viol.3} but with $L$ connected only to $B$.}
\label{fig:sub.viol.3.with.L}
\end{subfigure}
\begin{subfigure}[b]{0.45\textwidth}
\centering
\resizebox{\linewidth}{!}{
\begin{tikzpicture}
\tikzset{line width=1.5pt, outer sep=0pt, ell/.style={draw,fill=white, inner sep=2pt,
line width=1.5pt},swig vsplit={gap=5pt, inner line width right=0.5pt}};
\node[name=A,ell,ellipse] at (-9,-1) {$A$};
\node[name=B,ell,ellipse] at (-3,0) {$B$};
\node[name=E,ell,ellipse] at (0,0) {$E$};
\node[name=U,ell,ellipse] at (0,2) {$U$};
\node[name=Y,ell,ellipse] at (3,0) {$Y$};
\node[name=M,ell,ellipse] at (-9,1) {$M$};
\node[name=L,ell,ellipse] at (-6,2) {$L$};
\begin{scope}[>={Stealth[black]},every edge/.style={draw=black,very thick}]
\path [->] (E) edge (Y);
\path [->] (A) edge (B);
\path [->] (M) edge (B);
\path [->] (L) edge (B);
\path [->] (B) edge (E);
\path [->] (U) edge (E);
\path [->] (U) edge[dashed] (B);
\path [->] (U) edge (Y);
\path [->] (A) edge[bend left=-30] (Y);
\end{scope}
\end{tikzpicture}}
\subcaption{\textbf{Case 1c.} Similar to Figure~\ref{fig:sub.viol.1} but with $L$ connected only to $B$.}
\label{fig:L_adjustment_S_not_observed}
\end{subfigure}
\caption{DAGs describing scenarios with observed covariate $L$ that can be used to obtain narower bounds.}
\label{fig:L_adjustment_main}
\end{figure}

\subsubsection{A proof that under Proposition~\ref{prop:L_sharp}, $Y_i^{a,m}\perp_{\mathcal{T}_{VI}}(A_i,L_i)$ holds}\label{app:proof.L.indep}
For the LP-based bounds, we need to show that if $L$ satisfies the conditions in Proposition~\ref{prop:L_sharp}, then
$Y_i^{a,m}\perp_{\mathcal{T}_{VI}}(A_i,L_i)$ also holds under the NPSEM-IE model. We prove this independence under Figure~\ref{fig:DAG_with_s_violation2_with_L}; similar arguments apply to
Figures~\ref{fig:sub.viol.3.with.L} and~\ref{fig:L_adjustment_S_not_observed}.

Under Figure~\ref{fig:DAG_with_s_violation2_with_L}, we consider the corresponding NPSEM-IE:
\begin{equation}\label{app:model.NPSEM.L}
\begin{aligned}
A_i &= f_A(\varepsilon_{A_i}), \\
L_i &= f_L(\varepsilon_{L_i}), \\
U_i &= f_U(\varepsilon_{U_i}), \\
M_i &= f_M(\varepsilon_{M_i}),\\
S_i &= f_S(A_i,U_i,\varepsilon_{S_i}),\\
B_i &= h_B(M_i,S_i,L_i,U_i,\varepsilon_{B_i})
= \Ind_{\{M_i=-1\}}f_B(S_i,L_i,U_i,\varepsilon_{B_i})+\Ind_{\{M_i\neq-1\}}M_i,\\
Y_i &= f_Y(A_i,B_i,S_i,U_i,\varepsilon_{Y_i}),
\end{aligned}
\end{equation}
for some functions $f_A,  f_L,f_U, f_M, f_S, f_B,$ and $f_Y$, and where 
$\varepsilon_{A_i}, \varepsilon_{L_i},\varepsilon_{U_i}, \varepsilon_{M_i}, 
\varepsilon_{S_i}, \varepsilon_{B_i},$ and $\varepsilon_{Y_i}$ are mutually independent error terms.

From the model in Equation~\eqref{app:model.NPSEM.L},  for $m\in\{0,1\}$, it follows that 
\begin{align}
&Pr_{II}\Big(Y_i^{a,m}=y\,\Big|\,A_i,L_i\Big)\\
&=Pr_{IV}\Big(Y_i^{a,m}=y\,\Big|\,A_i,L_i\Big)
    \notag\\
&=Pr_{IV}\Big(f_Y(a,B_i^{a,m},S^a_i,U_i,\varepsilon_{Y_i})=y\,\Big|\,A_i,L_i\Big)
    \notag\\
&=Pr_{IV}\Big(f_Y(a,m,f_S(a,f_U(\varepsilon_{U_i}),\varepsilon_{S_i}),f_U(\varepsilon_{U_i}),\varepsilon_{Y_i})=y\,\Big|\,f_A(\varepsilon_{A_i}),f_L(\varepsilon_{L_i})\Big)
    \tag{Step 1}\label{eq4:proof-step3}\\
&=Pr_{IV}\Big(f_Y(a,m,f_S(a,f_U(\varepsilon_{U_i}),\varepsilon_{S_i}),f_U(\varepsilon_{U_i}),\varepsilon_{Y_i})=y\Big)
    \tag{Step 2}\label{eq4:proof-step5}
\end{align}
where \ref{eq4:proof-step3} follows from the fact that under Assumption \ref{assump:B_M} for $m\in\{0,1\}$, $B_i^{a,m}=m$. \ref{eq4:proof-step5} follows since $(\varepsilon_{A_i},\varepsilon_{L_i})\perp (\varepsilon_{U_i},\varepsilon_{S_i})$.

\subsubsection{Case 1a. LP-based bounds }
\begin{Proposition}
  Consider the setting of Figure~\ref{fig:DAG_with_s_violation2_with_L}. The LP-based bounds are
\[
\max_{l_1,l_2\in \{0,1\}}\!\left\{1-\frac{1-p^{\,l}_{00.1l_1}}{p^{\,l}_{10.0l_2}}\right\}
\leq VE(0) \leq
\min_{l_1,l_2\in \{0,1\}}\!\left\{1-\frac{p^{\,l}_{10.1l_1}}{1-p^{\,l}_{00.0l_2}}\right\},
\]
\[
\max_{l_1,l_2\in \{0,1\}}\!\left\{1-\frac{1-p^{\,l}_{01.1l_1}}{p^{\,l}_{11.0l_2}}\right\}
\leq VE(1) \leq
\min_{l_1,l_2\in \{0,1\}}\!\left\{1-\frac{p^{\,l}_{11.1l_1}}{1-p^{\,l}_{01.0l_2}}\right\},
\]
\[
\max_{l_1,l_2\in \{0,1\}}\!\left\{1-\frac{1-p^{\,l}_{01.1l_1}}{p^{\,l}_{10.0l_2}}\right\}
\leq VE_T \leq
\min_{l_1,l_2\in \{0,1\}}\!\left\{1-\frac{p^{\,l}_{11.1l_1}}{1-p^{\,l}_{00.0l_2}}\right\}.
\]
\end{Proposition}

\textbf{Proof}. 
We begin by outlining the main steps of the proof for deriving the LP-based bounds. Let $r_{y\mid a,l,b}$ denote the potential response of $Y^{a,m}$ given $A=a$, $L=l$, and $B^{a,m=-1}=b$, for all $a,m\in\{0,1\}^2$ in the two-arm trial. As in Section~\ref{sec:3a.LP}, and for clarity, when conditioning on $A=a$ we write $B^{a,m=-1}$ simply as $B$. Furthermore, let $q_{i.alb}=\Pr(r_{y\mid a,l,b}=i)$ and define
\[
Q_{a,l,b}(a,m)=\{r_{y\mid a,l,b}:\, Y^{a,m}=1\ \text{given } A=a,L=l,B=b\}.
\]
Then, similarly to Section~\ref{sec:3a.LP}, and using the fact that $Y^{a,m}\perp_{\mathcal{T}_{VI}}(A,L)$, we can express the target estimand $E(Y^{a,m})$ as
\begin{align}
\begin{split}
E(Y^{a,m})
&=E(Y^{a,m}\mid A=a,L=l)\\
&=\pi^{\,l}_{0.al}E(Y^{a,m}\mid A=a,L=l,B=0)+\pi^{\,l}_{1.al}E(Y^{a,m}\mid A=a,L=l,B=1)\\
&=\pi^{\,l}_{0.al}\sum_{i \in Q_{a,l,0}} q^{\,l}_{i.al0}
+\pi^{\,l}_{1.al}\sum_{i \in Q_{a,l,1}} q^{\,l}_{i.al1}.
\end{split}
\end{align}
The constraints and the remaining steps of the proof are identical to those described in Section~\ref{sec:3a.LP}, except that $S$ is replaced by $L$.

\subsubsection{Case 1a. Monotonicity-based bounds}
Consider the following assumption.
\begin{Assumption}\label{assump:U.mon.S.general.SUL}($U|S,L$ monotonicity)
    Consider the setting of Figures \ref{fig:DAG_with_s_violation2_with_L}. Suppose that $Y\perp_{\mathcal{T}_{VI}} M|A,B,S,U,L$ and the following conditions hold:
\begin{enumerate}
    \item  $\Pr\left(Y=1|A=a,B=1,S=s,L=l,U=u\right)$ and $\Pr(B=1|A=a,S=s,L=l,U=u)$ are both nondecreasing or both nonincreasing in $u$, for all $a$, $s$ and $l$.
    \item \label{assump:a} $\Pr(B=1|A=a,S=s,L=l)$ is nondecreasing in $s$, for all $a$ and $l$.
    \item \label{assump:b} $E\!\left(Y \mid A=a,B=1,S=s,L=l,U\right)$ is nondecreasing in $s$, for all $a$ and $l$.
\end{enumerate} 
\end{Assumption}

\begin{Proposition}\label{app:prop:L_S_bad}
    Consider the setting of Figure~\ref{fig:DAG_with_s_violation2_with_L}. Under Assumptions~\ref{assump:non.negative.m.general} and \ref{assump:U.mon.S.general.SUL}, the monotonicity-based bounds are
\begin{alignat*}{3}
\max_{l\in \{0,1\}}\!\left\{1-\frac{p_{1.1}}{p^{\,l}_{1.00l}}\right\}
&\;\le\;& VE(0) &\;\le\;
\min_{l\in \{0,1\}}\!\left\{1-\frac{p^{\,l}_{1.10l}}{p_{1.0}}\right\}, \\[1ex]
\max_{l\in \{0,1\}}\!\left\{1-\frac{p^{\,l}_{1.11l}}{p_{1.0}}\right\}
&\;\le\;& VE(1) &\;\le\;
\min_{l\in \{0,1\}}\!\left\{1-\frac{p_{1.1}}{p^{\,l}_{1.01l}}\right\}, \\[1ex]
\max_{l_1,l_2\in \{0,1\}}\!\left\{1-\frac{p^{\,l}_{1.11l_1}}{p^{\,l}_{1.00l_2}}\right\}
&\;\le\;& VE_T &\;\le\;
1-\frac{p_{1.1}}{p_{1.0}}.
\end{alignat*}
\end{Proposition}

\textbf{Proof}. 
In Appendix \ref{app:proof.S.U} we proved that 
$$
E(Y^{a,m=1})=\sum_{s} E_{U}\!\left[E\!\left(Y \mid A=a,B=1,S=s,U\right)\gamma_{s.a}|S=s,A=a\right].
$$
Since under Assumption \ref{assump:U.mon.S.general.SUL} we assume that $Y \perp_{\mathcal{T}_{VI}} L \mid A,S,B,M,U$, so we can also write that 
$$
E(Y^{a,m=1})=\sum_{s} E_{U}\!\left[E\!\left(Y \mid A=a,B=1,S=s,L=l,U\right)\gamma_{s.a}|S=s,A=a\right].
$$
From the other side. Under Figure \ref{fig:DAG_with_s_violation2_with_L}, we have that $U\perp L|A,S$, thus $E(Y \mid A=a,B=1,L=l)$ can be written as 
\begin{adjustwidth}{-1.5cm}{0cm}
\footnotesize
\begin{align}
\begin{split}
E&(Y \mid A=a,B=1,L=l)\\
&= \sum_{s} E\!\left(Y \mid A=a,B=1,L=l,S=s\right)\gamma_{s.abl}\\
&= \sum_{s,u} E\!\left(Y \mid A=a,B=1,S=s,L=l,U=u\right)\gamma_{s.abl}\Pr(U=u \mid A=a,B=1,S=s,L=l)\\
&= \sum_{s,u} E\!\left(Y \mid A=a,B=1,S=s,L=l,U=u\right)\gamma_{s.abl}
    \frac{\Pr(B=1 \mid A=a,U=u,S=s,L=l)\Pr(U=u \mid A=a,S=s,L=l)}{\Pr(B=1 \mid A=a,S=s,l=l)}\\
&= \sum_{s,u} E\!\left(Y \mid A=a,B=1,S=s,L=l,U=u\right)\gamma_{s.abl}\Pr(U=u|A=a,S=s)
    \frac{\Pr(B=1 \mid A=a,U=u,S=s,L=l)}{\Pr(B=1 \mid A=a,S=s,L=l)}\\
&= \sum_{s} E_{U}\!\left[E\!\left(Y \mid A=a,B=1,S=s,L=l,U\right)\gamma_{s.abl}
    \frac{\Pr(B=1 \mid A=a,S=s,L=l,U)}{\Pr(B=1 \mid A=a,S=s,L=l)}\bigg|{A=a,S=s}\right].
\end{split}
\end{align}
\end{adjustwidth}
\normalsize
Now, we prove that 
$E(Y \mid A=a,B=1,L=l) - E(Y^{a,m=1}) \ge 0$.
Let $\nu_{sl}(u)
= \frac{\Pr(B=1 \mid A=a,S=s,L=l,U=u)}
       {\Pr(B=1 \mid A=a,S=s,L=l)}$
and
$g_{sl}(u)
= E\!\left(Y \mid A=a,B=1,S=s,L=l,U=u\right)$.
Under Assumption~\ref{assump:U.mon.S.general.SUL}, and following the steps in
Appendix~\ref{app:proof.S.U}, we obtain that
$E(Y \mid A=a,B=1,L=l) - E(Y^{a,m=1})$ can be written as
\begin{align}
\begin{split}
&E_{s.a1l}\bigg[\operatorname{Cov}_{U}\!\left(g_{sl}(U),\, \nu_{sl}(U)|A=a,S=s\right)\bigg] \\&+\big( E_{\gamma_{s.a1l}}E_{U}(g_{sl}(U)|A=a,S=s)-E_{\gamma_{s.a}}E_{U}(g_{sl}(U)|A=a,S=s)\big)\\
\end{split}
\end{align}
Which is greater or equal to zero since:
\begin{enumerate}
    \item We assume that $g_{sl}(U)$ is nondecreasing in $u$ and that $\Pr(B=1 \mid A=a,S=s,L=l,U=u)$ is nondecreasing in $u$ and thus $\nu_{sl}(u)$ is nondecreasing in $u$. From Lemma \ref{lemma.sign} it follows that $\operatorname{Cov}_{U}\!\left(g_s(U),\, \nu_s(U)\right|A=a,S=s)\ge 0$.
    \item We assume that $E_{U}(g_{sl}(U)|A=a,S=s)$ is nondecreasing in $s$. By Lemma \ref{app:lemma.st} we have that $\gamma_{s.a1l}\geq_{st}\gamma_{s.al}$. Thus, by Lemma \ref{lemma.st} we have that $\big( E_{\gamma_{s.a1l}}E_{U}(g_{sl}(U)|A=a,S=s)-E_{\gamma_{s.a}}E_{U}(g_{sl}(U)|A=a,S=s)\big)\ge 0$.
\end{enumerate}

\subsubsection{Case 1b. LP-based bounds}
\begin{Proposition}
    Consider the setting of Figure~\ref{fig:sub.viol.3.with.L}. The LP-based bounds are
\[
\max_{s_1,s_2,l_1,l_2\in \{0,1\}^4}\!\left\{1-\frac{1-p^{\,l}_{00.1l_1s_1}}{p^{\,l}_{10.0l_2s_2}}\right\}
\leq VE(0) \leq
\min_{s_1,s_2,l_1,l_2\in \{0,1\}^4}\!\left\{1-\frac{p^{\,l}_{10.1l_1s_1}}{1-p^{\,l}_{00.0l_2s_2}}\right\},
\]
\[
\max_{s_1,s_2,l_1,l_2\in \{0,1\}^4}\!\left\{1-\frac{1-p^{\,l}_{01.1l_1s_1}}{p^{\,l}_{11.0l_2s_2}}\right\}
\leq VE(1) \leq
\min_{s_1,s_2,l_1,l_2\in \{0,1\}^4}\!\left\{1-\frac{p^{\,l}_{11.1l_1}}{1-p^{\,l}_{01.0l_2}}\right\},
\]
\[
\max_{s_1,s_2,l_1,l_2\in \{0,1\}^4}\!\left\{1-\frac{1-p^{\,l}_{01.1l_1s_1}}{p^{\,l}_{10.0l_2s_2}}\right\}
\leq VE_T \leq
\min_{s_1,s_2,l_1,l_2\in \{0,1\}^4}\!\left\{1-\frac{p^{\,l}_{11.1l_1s_1}}{1-p^{\,l}_{00.0l_2s_2}}\right\}.
\]
\end{Proposition}

\textbf{Proof. }We again outline the main steps of the proof for deriving the LP-based bounds. Let $r_{y\mid a,l,s,b}$ denote the potential response of $Y^{a,m}$ given $A=a$, $S^{a,m=-1}=s$, $L=l$, and $B^{a,m=-1}=b$, for all $a,m\in\{0,1\}^2$ in the two-arm trial. As in Section~\ref{sec:3a.LP}, and for clarity, when conditioning on $A=a$ we write $B^{a,m=-1}$ simply as $B$ and $S^{a,m=-1}$ as $S$. Furthermore, let $q_{i.alsb}=\Pr(r_{y\mid a,l,s,b}=i)$ and define
\[
Q_{a,l,s,b}(a,m)=\{r_{y\mid a,l,s,b}:\, Y^{a,m}=1\ \text{given } A=a,L=l,S=s,B=b\}.
\]
Then, similarly to Section~\ref{sec:3a.LP}, and using the fact that $Y^{a,m}\perp_{\mathcal{T}_{VI}}(A,L,S)$, we can express the target estimand $E(Y^{a,m})$ as
\begin{align}
\begin{split}
E(Y^{a,m})
&=E(Y^{a,m}\mid A=a,L=l,S=s)\\
&=\pi^{\,l}_{0.als}E(Y^{a,m}\mid A=a,L=l,S=s,B=0)+\pi^{\,l}_{1.als}E(Y^{a,m}\mid A=a,L=l,S=s,B=1)\\
&=\pi^{\,l}_{0.als}\sum_{i \in Q_{a,l,s,0}} q^{\,l}_{i.als0}
+\pi^{\,l}_{1.als}\sum_{i \in Q_{a,l,s,1}} q^{\,l}_{i.als1}.
\end{split}
\end{align}
The constraints and the remaining steps of the proof are identical to those described in Section~\ref{sec:3a.LP}, except that we condition, in addition to $S$, also on $L$.

\subsubsection{Case 1b. Monotonicity-based bounds}

Consider the following assumption.
\begin{Assumption}\label{assump:U.mon.L.general}($U,L,S$ monotonicity)
Consider the setting of Figure~\ref{fig:sub.viol.3.with.L}. Suppose that $Y\perp_{\mathcal{T}_{VI}} M\mid A,B,L,S,U$ and one of the following conditions holds:
\begin{enumerate}
\item $\Pr(Y=1\mid A=a,B=1,L=l,S=s,U=u)$ and $\Pr(B=1\mid A=a,L=l,S=s,U=u)$ are both nondecreasing or both nonincreasing in $u$, for all $a$, $l$, and $s$.
\item One of $\Pr(Y=1\mid A=a,B=1,L=l,S=s,U=u)$ or $\Pr(B=1\mid A=a,L=l,S=s,U=u)$ is nondecreasing in $u$ and the other is nonincreasing in $u$, for all $a$, $l$, and $s$.
\end{enumerate}
\end{Assumption}

\begin{Proposition}
  Consider the setting of Figure~\ref{fig:sub.viol.3.with.L}.  Under Assumptions~\ref{assump:non.negative.m.general} and~\ref{assump:U.mon.L.general}, the monotonicity-based bounds are
\begin{alignat*}{3}
\max_{l,s\in \{0,1\}^2}\!\left\{1-\frac{p_{1.1}}{p^{\,l}_{1.00ls}}\right\}
&\;\le\;& VE(0) &\;\le\;
\min_{l,s\in \{0,1\}^2}\!\left\{1-\frac{p^{\,l}_{1.10ls}}{p_{1.0}}\right\}, \\[1ex]
\max_{l,s\in \{0,1\}^2}\!\left\{1-\frac{p^{\,l}_{1.11ls}}{p_{1.0}}\right\}
&\;\le\;& VE(1) &\;\le\;
\min_{l,s\in \{0,1\}^2}\!\left\{1-\frac{p_{1.1}}{p^{\,l}_{1.01ls}}\right\}, \\[1ex]
\max_{l_1,l_2,s_1,s_2\in \{0,1\}^4}\!\left\{1-\frac{p^{\,l}_{1.11l_1s_1}}{p^{\,l}_{1.00l_2s_2}}\right\}
&\;\le\;& VE_T &\;\le\;
1-\frac{p_{1.1}}{p_{1.0}}.
\end{alignat*}
\end{Proposition}. 

\paragraph{\textbf{Proof.}}
As a result from Appendix \ref{app:proof:prop:bounds.mon.s.iv}, under  Figure~\ref{fig:sub.viol.3.with.L}, $E\!\left(Y^{a,m=1}\right)$ can be written as
\begin{align}\label{eq:sum_m_s_iv2}
\begin{split}
E_U\big[ E\!\left(Y \mid A=a,B=m,U=u,S=s\right)\big].
\end{split}
\end{align}
Since under Assumption \ref{assump:U.mon.L.general}, $Y \perp_{\mathcal{T}_{VI}} L \mid A,S,B,M,U$, we can also write that 
$$
E(Y^{a,m=1})=E_U\big[E\!\left(Y \mid A=a,B=m,U=u,S=s,L=l\right)\big]
$$
Also, from Appendix \ref{app:proof:prop:bounds.mon.s.iv}, $E(Y \mid A=a,B=1,S=s,L=l)$ can be written as 
\small
\begin{align}\label{eq:sum_Minus1_s.iv2}
\begin{split}
E_U\bigg[ E\!\left(Y \mid A=a,B=1,U=u,S=s,L=l\right)\frac{\Pr(B=1 \mid A=a,U=u,S=s,L=l)}{\Pr(B=1 \mid A=a,S=s,L=l)}\bigg].
\end{split}
\end{align}
\normalsize
Let $\nu(u)=\frac{\Pr(B=1 \mid A=a,U=u,S=s,L=l)}{\Pr(B=1 \mid A=a,S=s,L=l)}$. Then $\nu(u)$ is nondecreasing in $u$ since $\Pr(B=1 \mid A=a,S=s,U=u,L=l)$ is nondecreasing in $u$, and since under Figure \ref{fig:sub.viol.3.with.L}, $U\perp A,S,L$, we have
$$
E_U[\nu(U)]=E_{U|A,S,L}[\nu(U)]=1.
$$
Thus, by Lemma~\ref{lemma.sign} we have 
\begin{align}
\begin{split}\label{eq:cov_s_iv2}
E&\!\left(Y \mid A=a,B=1,S=s\right)-E\!\left(Y^{a,m=1}\right)\\
=&\operatorname{Cov}_U\!\left(E\!\left(Y \mid A=a,B=1,U,S=s,L=l\right),\nu(U)\right)\geq 0.
\end{split}
\end{align}
The rest of the proof is similar to the proof in Appendix \ref{app:proof:prop:bounds.mon.s.iv}.

\subsubsection{Case 1c. LP-based and monotonicity-based bounds}
When $S$ is unobserved, as in Figure~\ref{fig:L_adjustment_S_not_observed}, we can apply Propositions~\ref{prop:LP1_withS}--\ref{prop:bounds.mon.s.iv} and their underlying assumptions by replacing $S$ with $L$, which yields the bounds stated in Proposition~\ref{app:prop:L_S_bad}.

\subsection{Case 2: when $L$ does not lead to narrower bounds}\label{app:sharpening_bounds_L_bad}

In this section, we present examples of covariates $L$ that do not satisfy the conditions of Proposition~\ref{prop:L_sharp} and therefore cannot be used to obtain narrower bounds. Inspired by \citet{cai2007non}, we classify the observed covariate $L$ into five scenarios, illustrated in Figure~\ref{fig:L_adjustment}.

Figures~\ref{fig:L_confounder}--\ref{fig:L_confounder_U} depict scenarios in which $L$ is a confounder affecting both $B$ and $Y$. In Figure~\ref{fig:L_confounder_U}, $L$ is also influenced by $U$. An example corresponding to Figure~\ref{fig:L_confounder} is age. In many clinical trials, including the ENSEMBLE2 trial, AEs are more common among younger participants \citep{hardt2022efficacy}. As previously argued, an AE may affect participants' beliefs. Moreover, age can influence exposure to the pathogen, as younger individuals tend to be more socially active. An example of $L$ that is a confounder and is also affected by $U$ is prior infection with the pathogen, which was measured, for instance, in the Pfizer COVID-19 vaccine trial \citep{polack2020safety}. \citet{tissot2021patients} found that prior COVID-19 infection is associated with increased risk of AEs (which may influence belief) and with reduced risk of reinfection.

Figure~\ref{fig:L_prognostic} illustrates a scenario in which $L$ is a prognostic factor for $Y$. Here, $L$ may represent baseline comorbidities that affect infection risk. For example, in the ENSEMBLE2 trial, comorbidities such as diabetes and high blood pressure were reported to influence the outcome \citep{hardt2022efficacy}.

Figures~\ref{fig:L_intermidiate}--\ref{fig:L_intermidiate_U} represent cases where $L$ is an intermediate variable between $B$ and $Y$. A simple example is when exposure $E$ is observed. In malaria vaccine trials, for instance, the use of bed nets was recorded; this behavior may be influenced by participants' beliefs and, in turn, may affect infection risk \citep{Chandramohan2021NEJM}.

\begin{figure}[h]
\centering
\begin{subfigure}[b]{0.45\textwidth}
\centering
\resizebox{\linewidth}{!}{
\begin{tikzpicture}
\tikzset{line width=1.5pt, outer sep=0pt, ell/.style={draw,fill=white, inner sep=2pt,
line width=1.5pt},  swig vsplit={gap=5pt, inner line width right=0.5pt}};
\node[name=A,ell,ellipse] at (-9,-1) {$A$};
\node[name=L,ell,ellipse] at (0,-1) {$L$};
\node[name=B,ell,ellipse] at (-3,0) {$B$};
\node[name=U,ell,ellipse] at (0,2) {$U$};
\node[name=Y,ell,ellipse] at (3,0) {$Y$};
\node[name=M,ell,ellipse] at (-9,1) {$M$};
\begin{scope}[>={Stealth[black]},every edge/.style={draw=black,very thick}]
\path [->] (A) edge (B);
\path [->] (B) edge (Y);
\path [->] (L) edge (B);
\path [->] (M) edge (B);
\path [->] (U) edge[dashed] (B);
\path [->] (U) edge[dashed] (Y);
\path [->] (A) edge[bend left=-30] (Y);
\path [->] (L) edge (Y);
\end{scope}
\end{tikzpicture}}
\subcaption{}
\label{fig:L_confounder}
\end{subfigure}
\hfill
\begin{subfigure}[b]{0.45\textwidth}
\centering
\resizebox{\linewidth}{!}{
\begin{tikzpicture}
\tikzset{line width=1.5pt, outer sep=0pt, ell/.style={draw,fill=white, inner sep=2pt,
line width=1.5pt},swig vsplit={gap=5pt, inner line width right=0.5pt}};
\node[name=A,ell,ellipse] at (-9,-1) {$A$};
\node[name=L,ell,ellipse] at (0,-1) {$L$};
\node[name=B,ell,ellipse] at (-3,0) {$B$};
\node[name=U,ell,ellipse] at (0,2) {$U$};
\node[name=Y,ell,ellipse] at (3,0) {$Y$};
\node[name=M,ell,ellipse] at (-9,1) {$M$};
\begin{scope}[>={Stealth[black]},every edge/.style={draw=black,very thick}]
\path [->] (A) edge (B);
\path [->] (B) edge (Y);
\path [->] (L) edge (B);
\path [->] (M) edge (B);
\path [->] (U) edge[dashed] (B);
\path [->] (U) edge[dashed] (Y);
\path [->] (U) edge[dashed] (L);
\path [->] (A) edge[bend left=-30] (Y);
\path [->] (L) edge (Y);
\end{scope}
\end{tikzpicture}}
\subcaption{}
\label{fig:L_confounder_U}
\end{subfigure}

\hfill
\begin{subfigure}[b]{0.45\textwidth}
\centering
\resizebox{\linewidth}{!}{
\begin{tikzpicture}
\tikzset{line width=1.5pt, outer sep=0pt, ell/.style={draw,fill=white, inner sep=2pt,
line width=1.5pt},swig vsplit={gap=5pt, inner line width right=0.5pt}};
\node[name=A,ell,ellipse] at (-9,-1) {$A$};
\node[name=L,ell,ellipse] at (3,2) {$L$};
\node[name=B,ell,ellipse] at (-3,0) {$B$};
\node[name=U,ell,ellipse] at (0,2) {$U$};
\node[name=Y,ell,ellipse] at (3,0) {$Y$};
\node[name=M,ell,ellipse] at (-9,1) {$M$};
\begin{scope}[>={Stealth[black]},every edge/.style={draw=black,very thick}]
\path [->] (A) edge (B);
\path [->] (B) edge (Y);
\path [->] (L) edge (Y);
\path [->] (M) edge (B);
\path [->] (U) edge[dashed] (B);
\path [->] (U) edge[dashed] (Y);
\path [->] (A) edge[bend left=-30] (Y);
\end{scope}
\end{tikzpicture}}
\subcaption{}
\label{fig:L_prognostic}
\end{subfigure}
\begin{subfigure}[b]{0.45\textwidth}
\centering
\resizebox{\linewidth}{!}{
\begin{tikzpicture}
\tikzset{line width=1.5pt, outer sep=0pt, ell/.style={draw,fill=white, inner sep=2pt,
line width=1.5pt},swig vsplit={gap=5pt, inner line width right=0.5pt}};
\node[name=A,ell,ellipse] at (-9,-1) {$A$};
\node[name=L,ell,ellipse] at (0,0) {$L$};
\node[name=B,ell,ellipse] at (-3,0) {$B$};
\node[name=U,ell,ellipse] at (0,2) {$U$};
\node[name=Y,ell,ellipse] at (3,0) {$Y$};
\node[name=M,ell,ellipse] at (-9,1) {$M$};
\begin{scope}[>={Stealth[black]},every edge/.style={draw=black,very thick}]
\path [->] (A) edge (B);
\path [->] (B) edge (L);
\path [->] (L) edge (Y);
\path [->] (M) edge (B);
\path [->] (U) edge[dashed] (B);
\path [->] (U) edge[dashed] (Y);
\path [->] (A) edge[bend left=-30] (Y);
\end{scope}
\end{tikzpicture}}
\subcaption{}
\label{fig:L_intermidiate}
\end{subfigure}
\begin{subfigure}[b]{0.45\textwidth}
\centering
\resizebox{\linewidth}{!}{
\begin{tikzpicture}
\tikzset{line width=1.5pt, outer sep=0pt, ell/.style={draw,fill=white, inner sep=2pt,
line width=1.5pt},swig vsplit={gap=5pt, inner line width right=0.5pt}};
\node[name=A,ell,ellipse] at (-9,-1) {$A$};
\node[name=L,ell,ellipse] at (0,0) {$L$};
\node[name=B,ell,ellipse] at (-3,0) {$B$};
\node[name=U,ell,ellipse] at (0,2) {$U$};
\node[name=Y,ell,ellipse] at (3,0) {$Y$};
\node[name=M,ell,ellipse] at (-9,1) {$M$};
\begin{scope}[>={Stealth[black]},every edge/.style={draw=black,very thick}]
\path [->] (A) edge (B);
\path [->] (B) edge (L);
\path [->] (L) edge (Y);
\path [->] (M) edge (B);
\path [->] (U) edge[dashed] (B);
\path [->] (U) edge[dashed] (L);
\path [->] (U) edge[dashed] (Y);
\path [->] (A) edge[bend left=-30] (Y);
\end{scope}
\end{tikzpicture}}
\subcaption{}
\label{fig:L_intermidiate_U}
\end{subfigure}
\caption{DAGs illustrating five scenarios where the observed covariate $L$ cannot be used to obtain narrower bounds. For clarity, $S$ and $E$ are omitted.}
\label{fig:L_adjustment}
\end{figure}

\clearpage
\newpage
\section{Additional results for the numerical example}\label{app:sim_mon}

\begin{figure}[!p]
    \centering
    \includegraphics[width=1\linewidth]{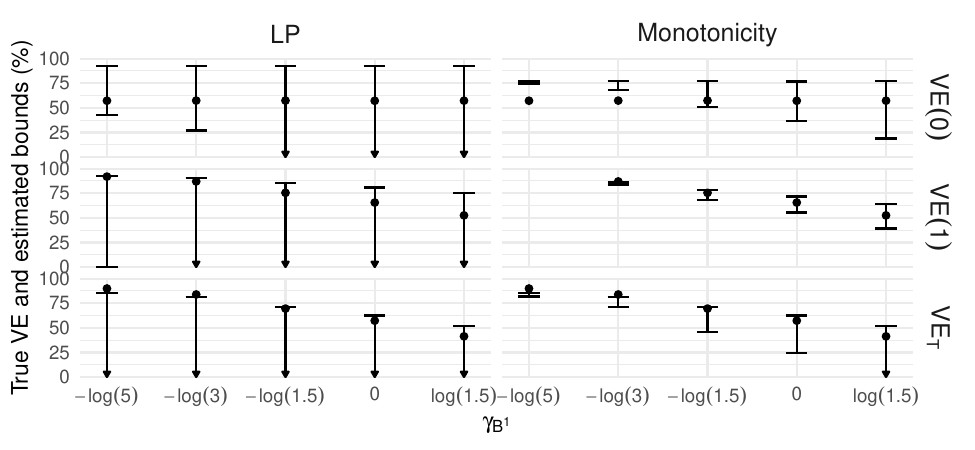}
    \caption{The relationship between the magnitude of $\gamma_{B^1}$ and the width of the bounds for $\mathrm{VE}(0)$, $\mathrm{VE}(1)$, and $\mathrm{VE}_T$ under the setting of Figure~\ref{fig:DAG_with_s_violation}. Assumption~\ref{assump:non.negative.m.general}\ref{assump:non.negative.m} is preserved when $\gamma_{B^1}>0$ and increasingly violated as $\gamma_{B^1}$ becomes negative, with $\gamma_{B^0}>0$ throughout. The parameters $\gamma_{B^0}$ and $\gamma_{B^1}$ control whether this assumption holds: when both were positive, the assumption is preserved, whereas more negative values of $\gamma_{B^1}$ correspond to stronger violations. Circles represent the true values. LP: bounds constructed via Proposition~\ref{prop:bounds.4ab.LP} and the results in Appendix~\ref{app:sharp.LP.bounds}; Monotonicity: bounds constructed via Proposition~\ref{prop:bounds.4ab.mon}. Triangles at the left boundary indicate bounds that extended below 0\% but were truncated for display. No bounds are displayed when the estimated lower bound exceeds the upper bound, reflecting violation of the assumed monotonicity assumption.
}
    \label{fig:app:M_mon_assump}
\end{figure}
\begin{figure}[!p]
    \centering
    \includegraphics[width=1\linewidth]{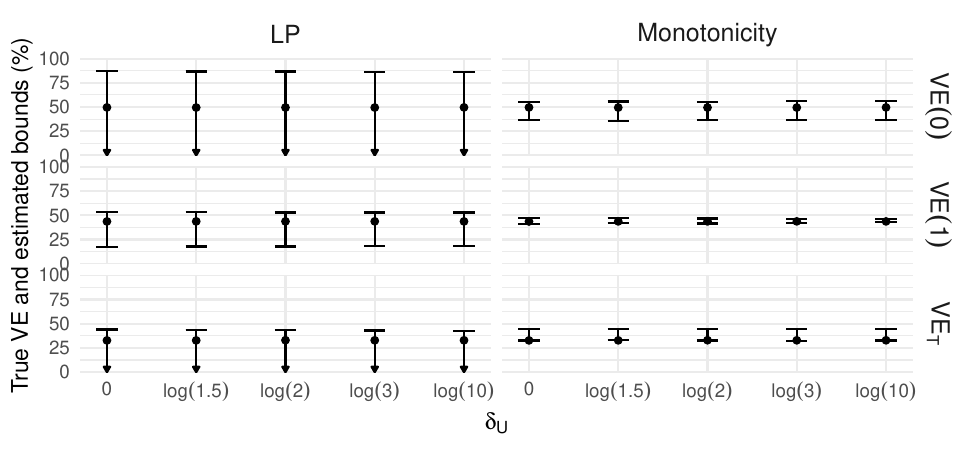}
    \caption{The relationship between the magnitude of $\delta_U$ and the width of the bounds for $\mathrm{VE}(0)$, $\mathrm{VE}(1)$, and $\mathrm{VE}_T$ under the setting of Figure \ref{fig:DAG_with_s_violation2} ($\beta_S=0$). The case $\delta_U=0$ corresponded to the situation where our assumed DAG matched the true data-generating DAG. As $\delta_U$ increased, however, the true data moved further away from this assumed structure, since the effect of $U$ on $S$ became stronger (Figures \ref{fig:sub.viol.3} and \ref{fig:DAG_with_s_violation2}).Circles represent the true values. LP-based bounds constructed via Proposition~\ref{prop:LP1_withS} and the results in Appendix~\ref{app:sharp.LP.bounds}. Triangles at the left boundary indicate bounds that extended below 0\% but were truncated for display. 
}
    \label{fig:app:SY}
\end{figure}

\begin{figure}[!p]
    \centering
    \includegraphics[width=1\linewidth]{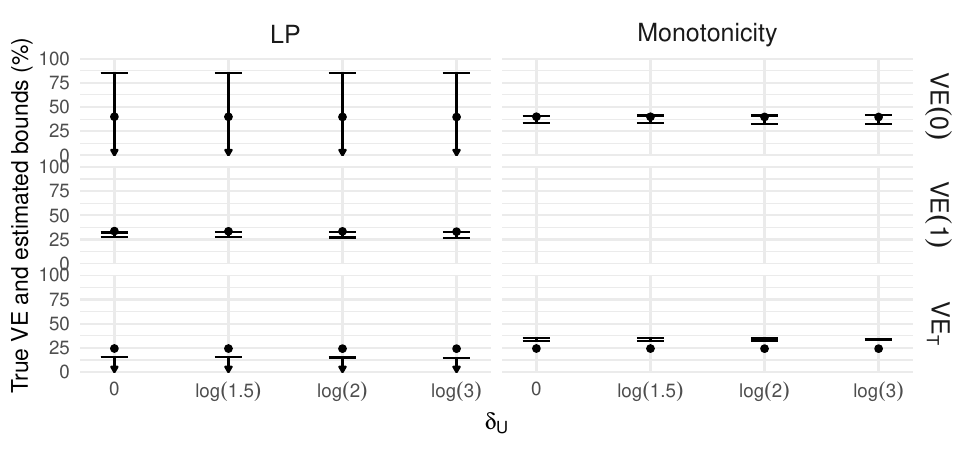}
    \caption{The relationship between the magnitude of $\delta_U$ and the width of the bounds for $\mathrm{VE}(0)$, $\mathrm{VE}(1)$, and $\mathrm{VE}_T$ under the setting of Figure~\ref{fig:DAG_with_s_violation}. Here $\beta_S=log(2)$, so the data were generated with both $U$ and $S$ affecting $Y$, while the assumed DAG omitted these effects (Figure~\ref{fig:sub.viol.3}). Circles represent the true values. LP-based bounds constructed via Proposition~\ref{prop:LP1_withS} and the results in Appendix~\ref{app:sharp.LP.bounds}. Triangles at the left boundary indicate bounds that extended below 0\% but were truncated for display. No bounds are displayed when the estimated lower bound exceeds the upper bound, reflecting violation of the assumed structure.
}
    \label{fig:app:SY_SU}
\end{figure}

\subsection{Bounds and corresponding bootstrap CIs}
\begin{table}[ht]
\caption{Empirical coverage of 95\% percentile-bootstrap CIs for LP-based bounds}
\label{app:tab:coverage_lp}
\centering
\begin{tabular}{cc rr|rr|rr}
\toprule
Figure & $n$ 
  & \multicolumn{2}{c}{$VE(0)$} 
  & \multicolumn{2}{c}{$VE(1)$} 
  & \multicolumn{2}{c}{$VE_T$} \\
\cmidrule(l){3-4} \cmidrule(l){5-6} \cmidrule(l){7-8}
        &     
  & Lower    & Upper    
  & Lower    & Upper    
  & Lower    & Upper    \\
\midrule
3a      & 500  
  & 0.91     & 0.97     
  & 0.92     & 0.93     
  & 0.96     & 0.91     \\
3a      & 5000 
  & 0.96     & 0.95     
  & 0.95     & 0.94     
  & 0.94     & 0.94     \\
3d      & 500  
  & 0.94     & 0.93     
  & 0.94     & 0.93     
  & 0.93     & 0.94     \\
3d      & 5000 
  & 0.95     & 0.92     
  & 0.95     & 0.93     
  & 0.95     & 0.93     \\
\bottomrule
\end{tabular}

\end{table}

\begin{table}[ht]
\caption{Empirical coverage of 95\% percentile-bootstrap CIs for monotonicity-based bounds}
\label{app:tab:coverage_mon}
\centering
\begin{tabular}{cc rr|rr|rr}
\toprule
Figure & $n$ 
  & \multicolumn{2}{c}{$VE(0)$} 
  & \multicolumn{2}{c}{$VE(1)$} 
  & \multicolumn{2}{c}{$VE_T$} \\
\cmidrule(l){3-4} \cmidrule(l){5-6} \cmidrule(l){7-8}
        &     
  & Lower    & Upper    
  & Lower    & Upper    
  & Lower    & Upper    \\
\midrule
3a      & 500  
  & 0.93     & 0.95     
  & 0.89     & 0.95     
  & 0.87     & 0.95     \\
3a      & 5000 
  & 0.94     & 0.96     
  & 0.94     & 0.95     
  & 0.92     & 0.94     \\
  3d      & 500  
  & 0.93     & 0.94     
  & 0.94     & 0.94     
  & 0.93     & 0.94     \\
3d      & 5000 
  & 0.94     & 0.92     
  & 0.94     & 0.92     
  & 0.94     & 0.93     \\
\bottomrule
\end{tabular}

\end{table}

\clearpage
\newpage
\section{Additional results for the real data example}\label{app:real.data.example}

\subsection{The full probabilities used to generate $B$ in the real data example}\label{app:prob.B.real.data}

The probabilities were set as follows:
\[
\begin{aligned}
\Pr(B=1 \mid S=1, Y=1) &= 0.7, \\
\Pr(B=1 \mid S=1, Y=0) &= 0.5, \\
\Pr(B=1 \mid S=0, Y=1) &= 0.2, \\
\Pr(B=1 \mid S=0, Y=0) &= 0.1.
\end{aligned}
\]
\subsection{Additional results for the real data example}
\begin{figure}[h]
    \centering
    \includegraphics[width=1\linewidth]{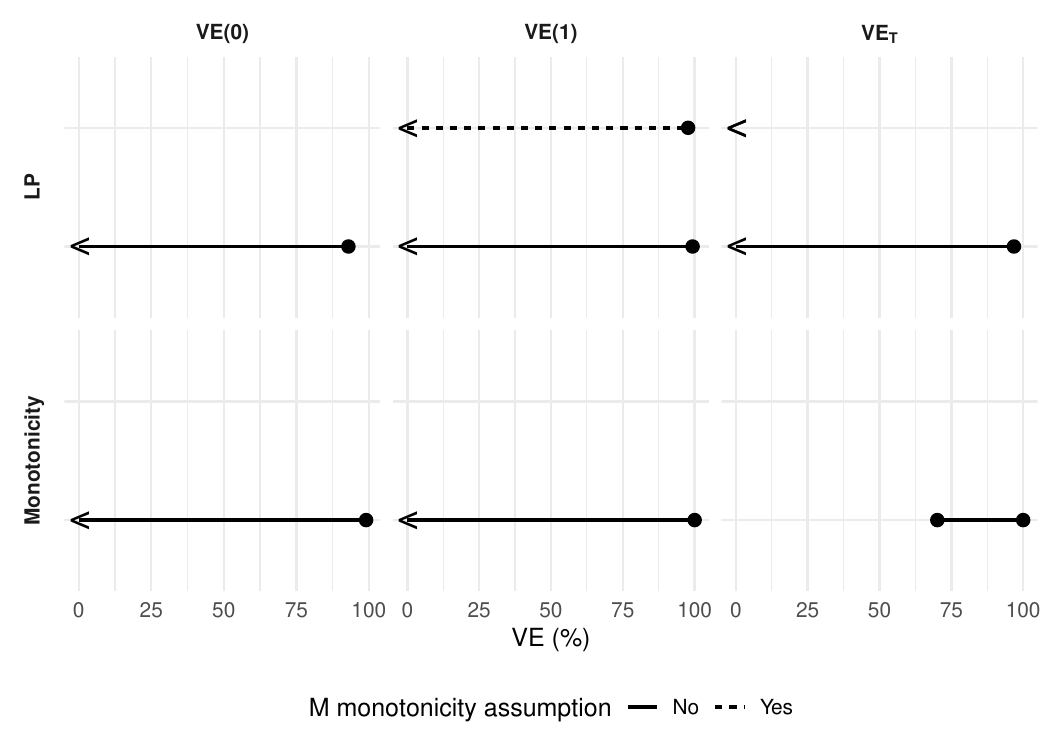}
    \caption{Estimated LP-based and monotonicity-based bounds under Figure \ref{fig:sub.viol.3}. 
Each horizontal line shows the identified interval for $VE(0)$, $VE(1)$, and $VE_{T}$, stratified by method (LP vs. Monotonicity) and whether Assumption \ref{assump:non.negative.m.general} \ref{assump:non.negative.m} was imposed (solid = no, dashed = yes). 
Triangles at the left boundary indicate bounds that extended below 0\% but were truncated for display. No bounds are displayed when the estimated lower bound exceeds the upper bound, reflecting violation of the assumed structure.}
    \label{app:fig:bounds_under_figure3}
\end{figure}

\begin{landscape}
\begin{table}[h]
\caption{Estimated VE bounds (\%) under Figure~\ref{fig:DAG_with_s_violation}. Cells show Estimate (95\% CI). Values below $-1000$ are reported as $-\infty$.}
\label{app:tab:bounds-figure4}
\raggedright
\footnotesize
\setlength{\tabcolsep}{4pt}
\begin{tabular}{ll*{6}{c}}
\toprule
\multicolumn{2}{c}{} &
\multicolumn{2}{c}{\textbf{VE(0)}} &
\multicolumn{2}{c}{\textbf{VE(1)}} &
\multicolumn{2}{c}{\textbf{VE\textsubscript{T}}} \\
\cmidrule(lr){3-4}\cmidrule(lr){5-6}\cmidrule(lr){7-8}
\textbf{Method} & \textbf{Monotonicity} & \textbf{Lower} & \textbf{Upper} & \textbf{Lower} & \textbf{Upper} & \textbf{Lower} & \textbf{Upper} \\
\midrule
LP            & No  & $-\infty$ ($-\infty$, $-\infty$) & 97.7 (96.4, 98.8)
              & $-\infty$ ($-\infty$, $-\infty$)    & 99.6 (99.3, 99.8)
              & $-\infty$ ($-\infty$, $-\infty$)       & 98.4 (97.2, 99.2) \\
Monotonicity  & No  & $-\infty$ ($-\infty$, $-\infty$) & 99.3 (98.9, 99.7)
              & $-\infty$ ($-\infty$, $-\infty$) & 100.0 (100.0, 100.0)
              & 20.4 (-49.4, 69.1)               & 100.0 (100.0, 100.0) \\
\addlinespace[2pt]
LP            & Yes & 21.7 (-35.3, 55.7)          & 64.6 (38.0, 82.8)
              & $-\infty$ ($-\infty$, $-\infty$)       & 99.0 (98.6, 99.4)
              &$-\infty$ ($-\infty$, $-\infty$)    & 39.3 (1.6, 65.5) \\
Monotonicity  & Yes & 36.5 (-7.3, 63.0)           & 47.0 (3.3, 75.4)
              & 23.9 (-39.4, 68.6)               & 48.9 (-25.7, 74.4)
              & 20.4 (-47.4, 68.4)               & 39.3 (-0.8, 63.4) \\
\bottomrule
\end{tabular}
\end{table}

\begin{table}[h]
\caption{Estimated VE bounds (\%) under Figure \ref{fig:sub.viol.3}. Cells show Estimate (95\% CI). Values below $-1000$ are reported as $-\infty$. No bounds are displayed
when the estimated lower bound exceeds the upper bound, reflecting violation of the
assumed structure.}
\label{app:tab:bounds-figure3}
\raggedright
\footnotesize
\setlength{\tabcolsep}{4pt}
\begin{tabular}{ll*{6}{c}}
\toprule
\multicolumn{2}{c}{} &
\multicolumn{2}{c}{\textbf{VE(0)}} &
\multicolumn{2}{c}{\textbf{VE(1)}} &
\multicolumn{2}{c}{\textbf{VE\textsubscript{T}}} \\
\cmidrule(lr){3-4}\cmidrule(lr){5-6}\cmidrule(lr){7-8}
\textbf{Method} & \textbf{Monotonicity} & \textbf{Lower} & \textbf{Upper} & \textbf{Lower} & \textbf{Upper} & \textbf{Lower} & \textbf{Upper} \\
\midrule
LP            & No  & $-\infty$ ($-\infty$, -591.2) & 92.9 (88.1, 96.8)
              & $-\infty$ ($-\infty$, $-\infty$)   & 99.3 (98.6, 99.7)
              & $-\infty$ ($-\infty$, $-\infty$)    & 96.8 (93.6, 98.5) \\
Monotonicity  & No  & $-\infty$ ($-\infty$, $-\infty$) & 99.1 (98.5, 99.5)
              & $-\infty$ ($-\infty$, $-\infty$) & 100.0 (100.0, 100.0)
              & 70.1 (-3.1, 94.1)               & 100.0 (100.0, 100.0) \\
\addlinespace[2pt]
LP            & Yes & ---          & ---
              & $-\infty$ ($-\infty$, $-\infty$)      & 97.7 (96.6, 98.8)
              & -4924.8 (-6785.4, -2669.4)      & -9.3 (-106.7, 43.7) \\
Monotonicity  & Yes & ---          & ---
              & ---                & ---
              & ---              & --- \\
\bottomrule
\end{tabular}
\end{table}
\end{landscape}

\end{document}